\documentclass[times,twocolumn,final,longtitle]{elsarticle}

\usepackage{pdflscape}
\usepackage{medima}
\usepackage{framed,multirow}

\usepackage{amssymb,amsmath}
\usepackage{latexsym}

\usepackage{graphicx}
\usepackage[caption=false,font=footnotesize,labelfont=sf,textfont=sf]{subfig}

\usepackage{xcolor,soul,framed} 
\PassOptionsToPackage{hyphens}{url}\usepackage{hyperref}
\colorlet{shadecolor}{yellow}

\usepackage{array}
\usepackage{mdwmath}
\usepackage{mdwtab}
\usepackage{eqparbox}
\usepackage{url}
\usepackage{algorithm}
\usepackage{algorithmic}
\usepackage{subfig}
\newcommand{\red}[1]{\textcolor{red}{#1}}
\newcommand{\green}[1]{\textcolor{green}{#1}}

\usepackage{soul}
\soulregister\cite7
\soulregister\citep7
\soulregister\ref7
\usepackage[switch]{lineno} 
\newcolumntype{M}[1]{>{\centering\arraybackslash}m{#1}}
\newcolumntype{N}{@{}m{0pt}@{}}

\usepackage{multirow}  
\usepackage{booktabs} 
\usepackage{threeparttable}
\usepackage{bbding}
\usepackage{ulem}
\newcommand{\etal}{\emph{et al.}}

\usepackage{float}
\usepackage{subfig}
\usepackage{makecell}
\makeatletter
\newcommand*\bigcdot{\mathpalette\bigcdot@{.5}}
\newcommand*\bigcdot@[2]{\mathbin{\vcenter{\hbox{\scalebox{#2}{$\m@th#1\bullet$}}}}}
\makeatother
\definecolor{newcolor}{rgb}{.8,.349,.1}
\usepackage{ragged2e}
\renewcommand{\justify}{\leftskip=0pt \rightskip=0pt plus 0cm}

\journal{Medical Image Analysis}

\begin{document}

\verso{This is the draft version for editing.}

\begin{frontmatter}

\title{Multi-site, Multi-domain Airway Tree Modeling (ATM'22): A Public Benchmark for Pulmonary Airway Segmentation}

\author[1]{Minghui \snm{Zhang}\fnref{fn1}} \author[1]{Yangqian \snm{Wu}\fnref{fn1}} \author[1]{Hanxiao \snm{Zhang}\fnref{fn1}} \author[1]{Yulei \snm{Qin}\fnref{fn1}} \author[1]{Hao \snm{Zheng}\fnref{fn1}}
\author[6]{Wen \snm{Tang}\fnref{fn2}} \author[20]{Corey \snm{Arnold}\fnref{fn2}}  \author[6]{Chenhao \snm{Pei}\fnref{fn2}} \author[6]{Pengxin \snm{Yu}\fnref{fn2}} 
\author[7]{Yang \snm{Nan}\fnref{fn2}} \author[7]{Guang \snm{Yang}\fnref{fn2}} \author[7]{Simon \snm{Walsh}\fnref{fn2}} \author[19]{Dominic C. \snm{Marshall}\fnref{fn2}} \author[19]{Matthieu \snm{Komorowski}\fnref{fn2}} 
\author[8]{Puyang \snm{Wang}\fnref{fn2}} \author[9]{Dazhou \snm{Guo}\fnref{fn2}} \author[9]{Dakai \snm{Jin}\fnref{fn2}} 
\author[10]{Ya'nan \snm{Wu}\fnref{fn2}} \author[10]{Shuiqing \snm{Zhao}\fnref{fn2}} \author[10]{Runsheng \snm{Chang}\fnref{fn2}} 
\author[11]{Boyu \snm{Zhang}\fnref{fn2}} \author[12]{Xing \snm{Lv}\fnref{fn2}} 
\author[13]{Abdul \snm{Qayyum}\fnref{fn2}} \author[21]{Moona \snm{Mazher}\fnref{fn2}} \author[14]{Qi \snm{Su}\fnref{fn2}} \author[15]{Yonghuang \snm{Wu}\fnref{fn2}} \author[16]{Ying'ao \snm{Liu}\fnref{fn2}} \author[17]{Yufei \snm{Zhu}\fnref{fn2}} \author[17,18]{Jiancheng \snm{Yang}\fnref{fn2}} 
\author[23]{Ashkan \snm{Pakzad}\fnref{fn2}} \author[24]{Bojidar \snm{Rangelov}\fnref{fn2}} \author[22]{Raul San Jose \snm{Estepar}\fnref{fn2}} \author[22]{Carlos Cano \snm{Espinosa}\fnref{fn2}} 
\author[4]{Jiayuan \snm{Sun}\fnref{fn1}}
\author[1]{Guang-Zhong \snm{Yang}\corref{cor1}%
\fnref{fn1}}\ead{gzyang@sjtu.edu.cn}
\author[1]{Yun \snm{Gu}\corref{cor1}%
\fnref{fn1}}\ead{geron762@sjtu.edu.cn}

\address[1]{Institute of Medical Robotics, Shanghai Jiao Tong University, Shanghai, 200240, China}
\address[4]{Department of Respiratory and Critical Care Medicine, Department of Respiratory Endoscopy, Shanghai Chest Hospital, Shanghai, China}
\address[6]{InferVision Medical Technology Co., Ltd., Beijing, China}
\address[7]{Imperial College London, London, UK}
\address[8]{Alibaba DAMO Academy, 969 West Wen Yi Road, Hangzhou, Zhejiang, China}
\address[9]{Alibaba DAMO Academy USA, 860 Washington Street, 8F, New York, USA}
\address[10]{College of Medicine and Biological Information Engineering, Northeastern University, Shenyang, China}
\address[11]{A.I R\&D Center, Sanmed Biotech Inc., No. 266 Tongchang Road, Xiangzhou District, Zhuhai, Guangdong, China}
\address[12]{A.I R\&D Center, Sanmed Biotech Inc., T220 Trade st. SanDiego, CA, USA}
\address[13]{ENIB, UMR CNRS 6285 LabSTICC, Brest, 29238, France}
\address[14]{Shanghai Jiao Tong University, Shanghai, China}
\address[15]{School of Information Science and Technology, Fudan University, Shanghai, China}
\address[16]{University of Science and Technology of China, Hefei, Anhui, China}
\address[17]{Dianei Technology, Shanghai, China}
\address[18]{EPFL, Lausanne, Switzerland}
\address[19]{Department of Surgery and Cancer, Imperial College London, London, UK}
\address[20]{University of California, Los Angeles, CA, USA}
\address[21]{Department of Computer Engineering and Mathematics, University Rovira I Virgili, Tarragona, Spain}
\address[22]{Brigham and Women’s Hospital, Harvard Medical School, Somerville, MA 02145, USA}
\address[23]{Medical Physics and Biomedical Engineering Department, University College London, London, UK}
\address[24]{Center for Medical Image Computing, University College London, London, UK}


\cortext[cor1]{Corresponding authors.}
\fntext[fn1]{Belongs to the ATM'22 Organizers.}
\fntext[fn2]{Belongs to the ATM'22 Participants.}

\received{1 May 2013}
\finalform{10 May 2013}
\accepted{13 May 2013}
\availableonline{15 May 2013}

\begin{abstract}
Open international challenges are becoming the de facto standard for assessing computer vision and image analysis algorithms. In recent years, new methods have extended the reach of pulmonary airway segmentation that is closer to the limit of image resolution.  Since EXACT’09 pulmonary airway segmentation, limited effort has been directed to quantitative comparison of newly emerged algorithms driven by the maturity of deep learning based approaches and clinical drive for resolving finer details of distal airways for early intervention of pulmonary diseases. Thus far, public annotated datasets are extremely limited, hindering the development of data-driven methods and detailed performance evaluation of new algorithms. To provide a benchmark for the medical imaging community, we organized the Multi-site, Multi-domain Airway Tree Modeling (ATM’22), which was held as an official challenge event during the MICCAI 2022 conference. ATM’22 provides large-scale CT scans with detailed pulmonary airway annotation, including 500 CT scans (300 for training, 50 for validation, and 150 for testing). The dataset was collected from different sites and it further included a portion of noisy COVID-19 CTs with ground-glass opacity and consolidation. Twenty-three teams participated in the entire phase of the challenge and the algorithms for the top ten teams are reviewed in this paper. Quantitative and qualitative results revealed that deep learning models embedded with the topological continuity enhancement achieved superior performance in general. ATM'22 challenge holds as an open-call design, the training data and the gold standard evaluation are available upon successful registration 
via its homepage (\url{https://atm22.grand-challenge.org/}).
\end{abstract}


\end{frontmatter}

\section{Introduction}\label{sec:introduction}
\subsection{Background}
\begin{figure*}[!t]
\centering
\includegraphics[width=1.0\linewidth]{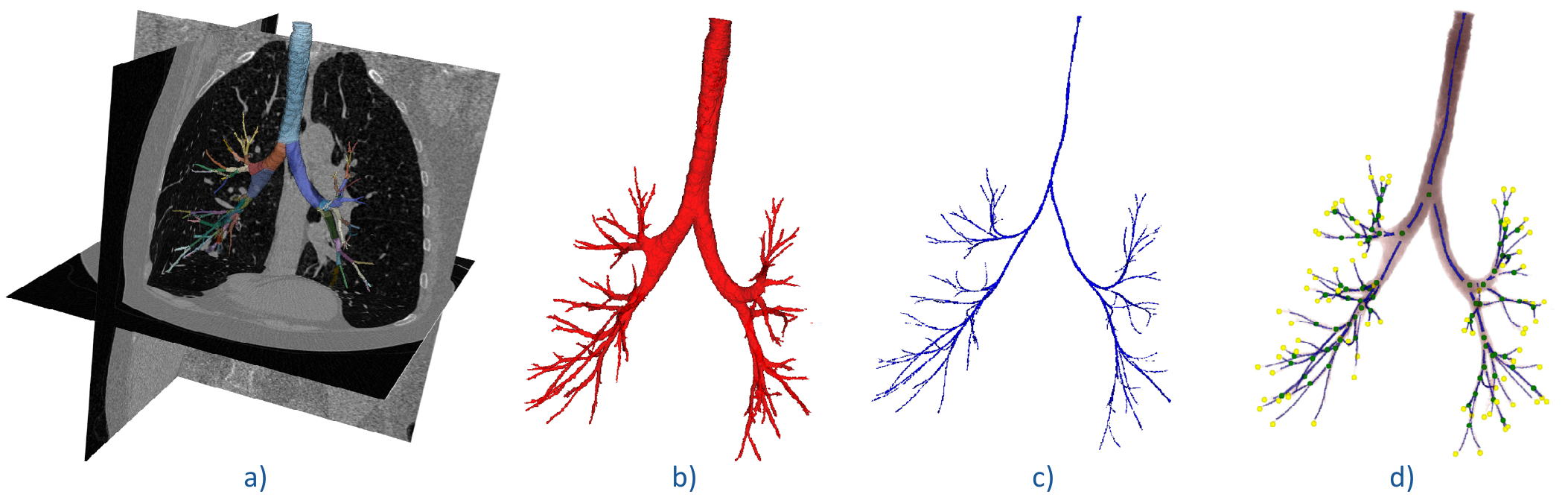}
\caption{The hierarchical illustration of the pulmonary airway structure. a) represents the branch-wise anatomical airway, overlapping with the original CT. 
b) and c) represent the binary airway and the centerline, respectively. The branch points (green) and end points (yellow) of the airway tree can be seen in d). Best viewed in color.}
\label{fig:airway_graphical_representation}
\end{figure*}
Deep learning methods are reshaping the general practice of image segmentation. In addition to novel network designs, the performance of these algorithms is largely dependent on the scale of the training data set and clinical accuracy of the annotation used.  For fair assessment of these algorithms, many grand-challenges have been organized, focusing on organs including the brain~\citep{MRBrains}, abdominal multi-organs~\citep{Abdomenct-1k}, heart~\citep{mmWHS}, skin lesion~\citep{skinlesion16} and breast cancer~\citep{Bach}.

For pulmonary airway segmentation, limited attention has been paid  since the EXACT’09  challenge~\citep{EXACT09}. Clinically, accurate segmentation of the pulmonary airway based on Computed Tomography (CT) is the prerequisite to the diagnosis and treatment of small airway diseases. It also plays an important role for pre-operative planning and intra-operative guidance of minimally invasive endobronchial interventions. With increasing miniaturization of  bronchoscopes empowered by robot assistance, small branches beyond the the 5\textsuperscript{th} generation of airways are routinely treated. Due to the fine-grained pulmonary airway structure further complicated by complex bifurcating topology, manual annotation is time-consuming, error-prone, and requires a high level of clinical skills.  

As an example, Fig.\ref{fig:airway_graphical_representation} presents a pulmonary airway structure with different levels of annotations: The branch-wise anatomical airway blended with the original CT is shown in Fig.\ref{fig:airway_graphical_representation}(a) where the generations of bronchi are also  annotated. Typical workflow involves the following steps:  The binary mask of airway is first segmented based on CT images as shown in Fig.\ref{fig:airway_graphical_representation}(b); Based on the binary mask, the skeleton or centreline of the airway in Fig.\ref{fig:airway_graphical_representation}(c) can be extracted via the morphological operations; By detecting the branching points and ending points as shown in Fig.\ref{fig:airway_graphical_representation}(d), the generations of airway can be then determined.  Since the morphology of the small distal bronchi can be fine-grained, it is challenging to delineate the airways from scratch for each patient To expedite the exploration of the airways, automatic airway segmentation algorithms are in high demand clinically.

\subsection{Challenges of Pulmonary Airway Segmentation}\label{sec:challenges_of_pulmonary_airway_segmentation}
Detailed pulmonary airway segmentation, which traditionally works on the level of trachea and bronchi and ideally reaching all the way to alveoli  should the imaging resolution permits.   However, to acquire the fine-grained airway tree structure is practically difficult. The main challenges involved the following aspects.

\begin{figure}[thbp]
  \centering
  \includegraphics[width=1.0\linewidth]{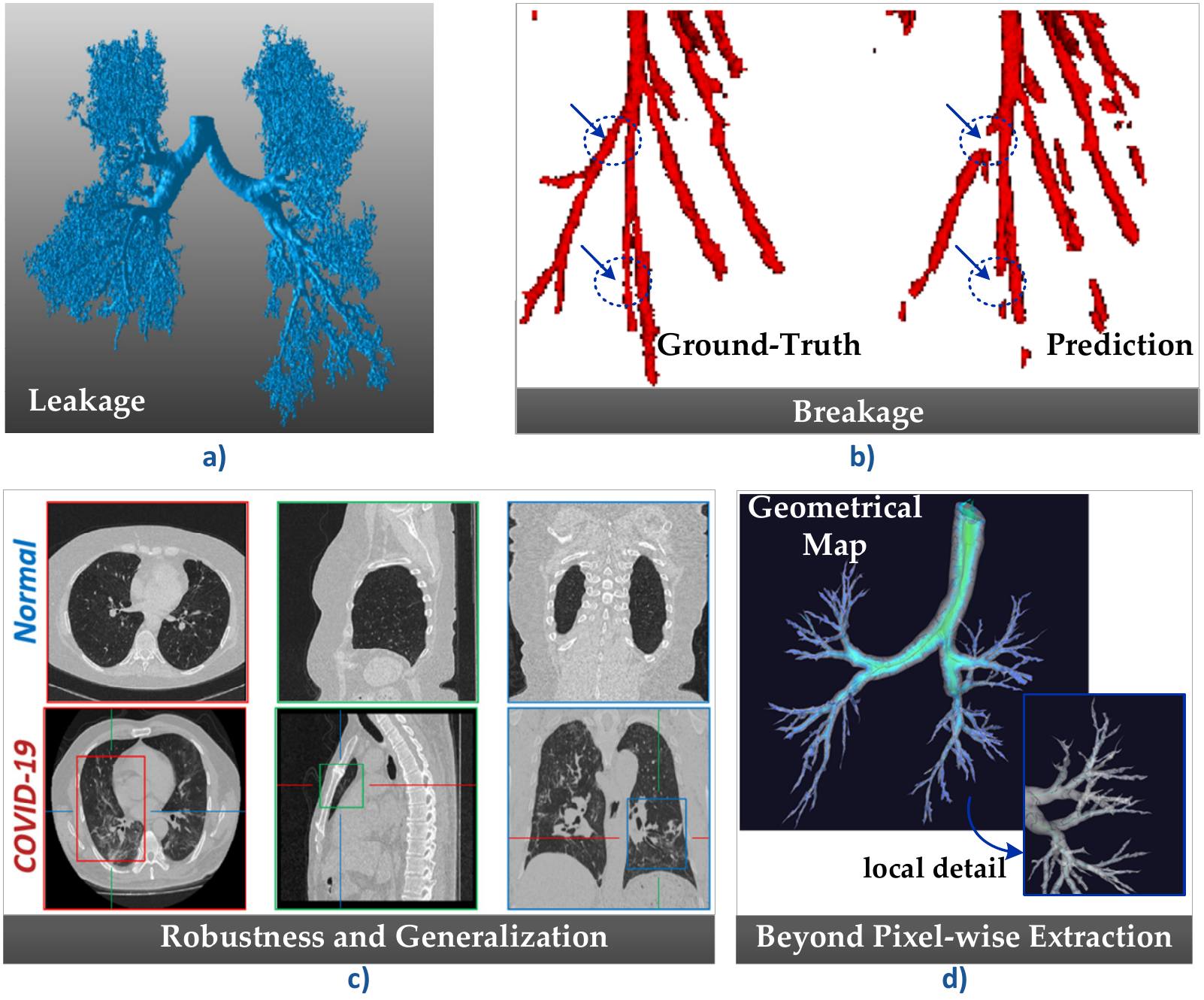}
  \caption{Four main challenges of the pulmonary airway segmentation task. a) Leakage. b) Breakage. c) Robustness and generalization. d) Beyond pixel-wise extraction.}
  \label{fig:airway_segmentation_challenge}
  \end{figure}

\textbf{Challenge 1: Leakage (C1).} The leakage phenomenon, as seen in the Figure \ref{fig:airway_segmentation_challenge}.a) (excerpted from ~\citep{charbonnier2017improving}), 
is a common challenge that exerts on the pulmonary airway segmentation algorithms. The leakage problem usually occurs on the small airway branches or 
lesion surrounding areas (e.g., emphysema and bronchiectasis.)~\citep{pu2012ct}. The segmentation methods are likely to leak into the adjacent 
lung parenchyma through blurred airway walls or soft boundaries due to the highly variable intensity levels in the lumen area. 
The traditional methods suffer the leakage problem more severely than the deep learning methods since they usually function on the low-level features of images. 
For example, the intensity-based region growing methods often leak to lung parenchyma through blurred/broken boundaries at small airways. Rule-based 
method~\citep{sonka1996rule} and morphology-based method~\citep{aykac2003segmentation} encounter similar problems as well. 
The deep learning methods can extract semantic high-level features that are discriminative to airways, which alleviates the leakage problem. 
Unlike the mass of leakage that happened in the traditional methods, the form of leakage encountered by deep learning methods 
is that the prediction are thicker than the ground-truths. This phenomenon is defined as gradient dilation ~\citep{zheng2021alleviating} when assigning 
larger weights to the peripheral bronchi. 

\textbf{Challenge 2: Breakage (C2).} The breakage phenomenon refers to the discontinuous result, which can be seen in Figure \ref{fig:airway_segmentation_challenge}.b) (adapted from ~\citep{zheng2021alleviating}). 
The breakage merely induces marginal voxel-level errors, however, the topology structure is totally changed after the largest connected component extraction. 
The breakage problem is detrimental to the airway segmentation task because only the largest connected component of the airway result is useful to the
bronchoscopic-assisted surgery. The presence of the breakages will cause interrupted trajectories. Different from the leakage problem, the deep learning method 
is more likely to generate breakages than conventional methods. The conventional methods usually rely on intensity constraints, hence, the connectivity can be 
guaranteed, especially the region-growing algorithms. However, as discussed in ~\citep{nadeem2020ct}, the intensity of the airway wall varies significantly from the proximal 
to distal sites. Consequently, the region growing methods merely function reliably in the trachea and principal bronchi. The breakages that happened in the deep learning 
methods can be ascribed to three aspects. First, the intrinsic class imbalance distribution~\citep{zheng2021alleviating,zhang2022differentiable} 
adds difficulty in extracting the whole airway tree structure with satisfactory connectivity. The class imbalance includes two type of imbalance distribution, 
the inter-class imbalance and the intra-class imbalance. Inter-class imbalance means that the number of the airway voxels is far fewer than that of background, and 
the intra-class refers to the relative total volume difference of trachea, principal bronchi, lobar bronchi, and distal segmental bronchi. Such imbalanced distribution 
influences the data-driven deep learning methods, leading to the breakages of peripheral bronchi. The uncertainty of the airway lumen is the second aspect that may 
cause breakage. The uncertainty includes low contrast, complex topological structures, and imaging noise. Third, the overlap-wise loss function, e.g., Dice ~\citep{milletari2016v}
loss function, is widely used for medical image segmentation tasks. However, it can not guarantee topological accuracy due to the severe intra-class imbalance
distribution. It is known that deep learning models trained with class-imbalanced data may perform poorly in the minor classes with scarce training data~\citep{buda2018systematic,liu2019large}.

\textbf{Challenge 3: Robustness and Generalization (C3).} The diseases, such as bronchiectasis, emphysema, and COVID-19 could influence the airway 
morphology or the characteristic of CT images. Here, we put our emphasis on the pandemic COVID-19 disease. As seen in Figure \ref{fig:airway_segmentation_challenge}.c), 
the normal CT scans can be categorized into the clean domain, where the airway lumen is relatively explicit in these clean CTs. However, the COVID-19 CT scans  
can be deemed as the noisy domain since they introduce the bias attributes, e.g., bilaterally scattered irregular patches of ground glass opacity, 
thickening of inter-lobular or intra-lobular septa, and consolidation. Preliminary experiments~\citep{zhang2021fda,zhang2022cfda} have demonstrated that the models trained on the clean domain are 
difficult to be generalized to the noisy CTs. Improve the robustness and generalization ability across domains is also critical to measure the performance of airway segmentation algorithms.

\textbf{Challenge 4: Beyond Pixel-wise Extraction (C4).} Currently, the airway tree modeling task is regarded as the pixel- or voxel-wise segmentation task. 
One of the critical purposes of airway tree modeling is for the navigation of bronchoscopic-assisted surgery. However, there exists a gap temporarily 
between these two things. The CT values are discrete signals and the airway prediction obtained by CNNs are also dense, discrete volumes. Volume rendering algorithms 
are necessary to acquire continuous results (e.g., Mesh). The topology relation in the voxel-wise structure data is weak while very strong in the continuous data. 
As seen in Figure \ref{fig:airway_segmentation_challenge}.d), the curved centerline of airways can be extracted on the Voronoi diagram via the Eikonal equation. 
More geometric attributes of the airway, including the bifurcation, radius, and centerline directions can be performed. Hence, we have a giant vision for the 
future airway tree modeling paradigm: Take the discrete CT scans as input and output the continuous airway results. This is challenging because we need to develop 
a novel pixel-wise extraction methodology.

Our organized ATM'22 focuses on the above challenges. The large-scale dataset with full annotation encourages participants to 
develop novel methods to harness the intrinsic topological airway tree knowledge and achieve remarkable segmentation performance.

\subsection{Limitation of Previous Datasets}
\makeatletter
\def\hlinew#1{%
\noalign{\ifnum0=`}\fi\hrule \@height #1 \futurelet
\reserved@a\@xhline}
\makeatother
\begin{table*}[!t]
\renewcommand\arraystretch{1.0}
\centering
\caption{\justify{Comparisons between the ATM'22 challenge and other related airway segmentation works from the dataset and 
evaluation. Commonly used metrics are Tree length detected rate (i.e., TD, \%), Branch detected rate (i.e., BD, \%), 
Dice Similarity Coefficient (i.e., DSC, \%), Precision (i.e., Pre, \%), Sensitivity (i.e., Sen, \%) and Specificity (i.e., Spe, \%). 
The reported false positive rate (FPR) is equal to reporting the specificity, and the true positive rate (TPR) 
is equal to report the sensitivity.}}\label{tab:dataset_metrics_evaluation}
\scalebox{0.6}{
\begin{threeparttable}
\begin{tabular}{cccccccccc}
\hlinew{1pt}
\multicolumn{1}{c|}{\multirow{2}{*}{\textbf{Method}} }& \multicolumn{1}{c|}{\multirow{2}{*}{\begin{tabular}[c]{@{}c@{}}\textbf{Model}\\ Description \& Characteristic\end{tabular}}} & \multicolumn{2}{c|}{\textbf{Dataset}} & \multicolumn{6}{c}{\textbf{Metrics}}   \\\cline{3-10}
\multicolumn{1}{c|}{}    & \multicolumn{1}{c|}{} & \multicolumn{1}{c|}{\green{\Checkmark\ Open-Source }} & \multicolumn{1}{c|}{\red{\XSolidBrush In-House} }   & \multicolumn{1}{c|}{TD (\%)} & \multicolumn{1}{c|}{BD (\%)} & \multicolumn{1}{c|}{DSC (\%)} & \multicolumn{1}{c|}{Pre (\%)} & \multicolumn{1}{c|}{Sen (\%)} & \multicolumn{1}{c}{Spe (\%)} \\ \hlinew{0.8pt}
\multicolumn{1}{c|}{\cite{meng2017tracking}}    & \multicolumn{1}{c|}{\scriptsize{U-Net with the tracking algorithm}}    &  \multicolumn{2}{c|}{\scriptsize{\red{\XSolidBrush} 50 scans, Tokushima University}}   & \multicolumn{1}{c|}{\scriptsize{\red{\XSolidBrush}}} & \multicolumn{1}{c|}{\scriptsize{\green{\Checkmark}}} & \multicolumn{1}{c|}{\scriptsize{\green{\Checkmark}}} & \multicolumn{1}{c|}{\scriptsize{\red{\XSolidBrush}}} & \multicolumn{1}{c|}{\scriptsize{\red{\XSolidBrush}}} & \multicolumn{1}{c}{\scriptsize{\red{\XSolidBrush}}} \\\cline{1-10}
\multicolumn{1}{c|}{\cite{jin20173d}}            & \multicolumn{1}{c|}{\scriptsize{U-Net with graph refinement}} & \multicolumn{2}{c|}{\scriptsize{\green{\Checkmark} 40 scans, EXACT'09 \green{\Checkmark} 20 scans, LTRC~\cite{karwoski2008processing}}}  & \multicolumn{1}{c|}{\scriptsize{\green{\Checkmark}}}  & \multicolumn{1}{c|}{\scriptsize{\green{\Checkmark}}} &\multicolumn{1}{c|}{\scriptsize{\red{\XSolidBrush}}}&\multicolumn{1}{c|}{\scriptsize{\red{\XSolidBrush}}} &\multicolumn{1}{c|}{\scriptsize{\red{\XSolidBrush}}}&\multicolumn{1}{c}{\scriptsize{\red{\XSolidBrush}}}\\\cline{1-10}
\multicolumn{1}{c|}{\cite{charbonnier2017improving}} & \multicolumn{1}{c|}{\scriptsize{U-Net with leak detection}}  & \multicolumn{2}{c|}{\scriptsize{\red{\XSolidBrush} 45 scans, COPDGene~\cite{regan2011genetic}}}  & \multicolumn{1}{c|}{\scriptsize{\green{\Checkmark}}} &\multicolumn{1}{c|}{\scriptsize{\red{\XSolidBrush}}}&\multicolumn{1}{c|}{\scriptsize{\red{\XSolidBrush}}}&\multicolumn{1}{c|}{\scriptsize{\red{\XSolidBrush}}}&\multicolumn{1}{c|}{\scriptsize{\red{\XSolidBrush}}}&\multicolumn{1}{c}{\scriptsize{\green{\Checkmark}}}\\\cline{1-10}
\multicolumn{1}{c|}{\cite{garcia2019joint}} & \multicolumn{1}{c|}{\scriptsize{U-Net with a gcn module}}  & \multicolumn{2}{c|}{\scriptsize{\red{\XSolidBrush} 32 scans, DLCST~\cite{pedersen2009danish}}}  &\multicolumn{1}{c|}{\scriptsize{\green{\Checkmark}}}&\multicolumn{1}{c|}{\scriptsize{\red{\XSolidBrush}}}&\multicolumn{1}{c|}{\scriptsize{\green{\Checkmark}}}&\multicolumn{1}{c|}{\scriptsize{\red{\XSolidBrush}}}&\multicolumn{1}{c|}{\scriptsize{\green{\Checkmark}}}&\multicolumn{1}{c}{\scriptsize{\red{\XSolidBrush}}}\\\cline{1-10}
\multicolumn{1}{c|}{\cite{wang2019tubular}} &  \multicolumn{1}{c|}{\scriptsize{Spatial CNN with radial distance loss}}   &\multicolumn{2}{c|}{\scriptsize{\red{\XSolidBrush} 38 private scans}}  &\multicolumn{1}{c|}{\scriptsize{\green{\Checkmark}}}&\multicolumn{1}{c|}{\scriptsize{\red{\XSolidBrush}}}&\multicolumn{1}{c|}{\scriptsize{\green{\Checkmark}}}&\multicolumn{1}{c|}{\scriptsize{\red{\XSolidBrush}}}&\multicolumn{1}{c|}{\scriptsize{\red{\XSolidBrush}}}&\multicolumn{1}{c}{\scriptsize{\green{\Checkmark}}}\\\cline{1-10}
\multicolumn{1}{c|}{\cite{nadeem2020ct}} &  \multicolumn{1}{c|}{\scriptsize{U-Net with freeze-and-grow algorithm}} & \multicolumn{2}{c|}{\scriptsize{\red{\XSolidBrush} 32 scans, SPIROMICS~\citep{couper2014design} }}  &\multicolumn{1}{c|}{\scriptsize{\red{\XSolidBrush}}}&\multicolumn{1}{c|}{\scriptsize{\red{\XSolidBrush}}}&\multicolumn{1}{c|}{\scriptsize{\red{\XSolidBrush}}}&\multicolumn{1}{c|}{\scriptsize{\red{\XSolidBrush}}}&\multicolumn{1}{c|}{\scriptsize{\red{\XSolidBrush}}}&\multicolumn{1}{c}{\scriptsize{\green{\Checkmark}}}\\\cline{1-10}
\multicolumn{1}{c|}{\cite{garcia2021automatic}} & \multicolumn{1}{c|}{\scriptsize{Efficient 3D-UNet}} & \multicolumn{2}{c|}{\scriptsize{\red{\XSolidBrush} 24 scans, CF-CT~\citep{kuo2017diagnosis}, DLCST, EXACT'09}}  &\multicolumn{1}{c|}{\scriptsize{\green{\Checkmark}}}&\multicolumn{1}{c|}{\scriptsize{\red{\XSolidBrush}}}&\multicolumn{1}{c|}{\scriptsize{\green{\Checkmark}}}&\multicolumn{1}{c|}{\scriptsize{\red{\XSolidBrush}}} &\multicolumn{1}{c|}{\scriptsize{\red{\XSolidBrush}}}&\multicolumn{1}{c}{\scriptsize{\green{\Checkmark}}} \\\cline{1-10}
\multicolumn{1}{c|}{\cite{qin2021learning}} &\multicolumn{1}{c|}{\scriptsize{UNet with attention module}} & \multicolumn{2}{c|}{\scriptsize{\green{\Checkmark} 90 scans, BAS ~\citep{qin2020learning}}} &\multicolumn{1}{c|}{\scriptsize{\green{\Checkmark}}}&\multicolumn{1}{c|}{\scriptsize{\green{\Checkmark}}}&\multicolumn{1}{c|}{\scriptsize{\green{\Checkmark}}}&\multicolumn{1}{c|}{\scriptsize{\red{\XSolidBrush}}}&\multicolumn{1}{c|}{\scriptsize{\green{\Checkmark}}}&\multicolumn{1}{c}{\scriptsize{\green{\Checkmark}}}\\\cline{1-10}
\multicolumn{1}{c|}{\cite{zheng2021alleviating}}&\multicolumn{1}{c|}{\scriptsize{WingsNet with general union loss}} &\multicolumn{2}{c|}{\scriptsize{\green{\Checkmark} BAS }}&\multicolumn{1}{c|}{\scriptsize{\green{\Checkmark}}}&\multicolumn{1}{c|}{\scriptsize{\green{\Checkmark}}}&\multicolumn{1}{c|}{\scriptsize{\red{\XSolidBrush}}}&\multicolumn{1}{c|}{\scriptsize{\green{\Checkmark}}}&\multicolumn{1}{c|}{\scriptsize{\red{\XSolidBrush}}}&\multicolumn{1}{c}{\scriptsize{\red{\XSolidBrush}}}\\ \hlinew{1pt}
\multicolumn{1}{c|}{\textbf{ATM'22 Challenge (Ours)}}& ———— &\multicolumn{2}{c|}{\scriptsize{\green{\Checkmark} \textbf{500 scans}}} &\multicolumn{1}{c|}{\scriptsize{\green{\Checkmark}}}&\multicolumn{1}{c|}{\scriptsize{\green{\Checkmark}}}&\multicolumn{1}{c|}{\scriptsize{\green{\Checkmark}}}&\multicolumn{1}{c|}{\scriptsize{\green{\Checkmark}}}&\multicolumn{1}{c|}{\scriptsize{\green{\Checkmark}}}&\multicolumn{1}{c}{\scriptsize{\green{\Checkmark}}}\\
\hlinew{1pt}
\end{tabular}
\end{threeparttable}}
\end{table*}
As reported in Table \ref{tab:dataset_metrics_evaluation}, the main drawbacks of recent representative airway segmentation 
work lie in the limited scale of the dataset and incomplete evaluation metrics. The total number of benchmark datasets 
is all smaller than one hundred, which is not sufficient for the deep learning model training. The models are prone to overfitting due to the small 
number of training set. Furthermore, the small datasets still need to split into non-overlap training/validation/test sets. 
Consequently, the validation set is inadequate to guarantee the robustness and generalization ability of the trained models 
because it owns a very small amount of samples. In addition, the in-house datasets and incomplete evaluation metrics degrade the transparency of the results. 
They also add difficulty to the fair comparison among various methods. 

Considering the public benchmarks, the  most famous airway segmentation challenge in the past few decades is the 'Extraction of Airways From CT' (EXACT’09) organized by \citep{EXACT09}. It was held at the Second International Workshop on Pulmonary Image Analysis, in conjunction with the 12\textsuperscript{th} International Conference on Medical 
Image Computing and Computer-Assisted Intervention (MICCAI 2009)\footnote{\url{https://www.lungworkshop.org/2009/index.html}}. 
The EXACT'09 provided 40 CT scans, the first 20 scans were designated as the training set and the remaining 20 scans were set as 
the testing set to evaluate different algorithms. It also provided a platform\footnote{\url{http://image.diku.dk/exact/}} 
for comparing airway extraction algorithms with standard evaluation metrics. Since the machine-learning/deep-learning methods were not the dominated methods in early 2000s, the majority of the algorithms were based on the traditional image processing methods, such as morphological filtering~\citep{irving20093d,fetita2009morphological}, and 
region growing~\citep{pinho2009robust,feuerstein2009adaptive,wiemker2009simple}. The EXACT’09 challenge aimed to develop automatic airway segmentation algorithms, however, these methods often fail in extracting the smaller peripheral bronchi due to the lack of robust features. Further, the EXACT'09 challenge 
did not publish the manual annotation of airway and the number of training samples are limited, which is not friendly for the burgeoning data-driven deep-learning methods.

Therefore, we organize the Multi-site, Multi-domain Airway Tree Modeling (\textbf{ATM'22}) Challenge, 
which was held in conjunction with MICCAI 2022. \uline{Our challenge aims to revolutionize the pulmonary 
airway segmentation task compared with the EXACT'09 from three aspects:}
\\1) \textbf{More Annotated Data}. The EXACT'09 only provided 40 CT scans without airway labels. In ATM'22, we collected 500 CT scans with elaborated airway labels, each delineated by three experienced radiologists. We believe that a large number of CT scans and airway labels could boost 
the development of robust airway segmentation algorithms based on deep neural networks. 

Compared with recent datasets adopted in deep-learning methods, as seen in Table \ref{tab:dataset_metrics_evaluation}, ATM'22 expands the scale of the dataset with record number of cases. In addition, the dataset of the 
ATM'22 is split into 300/50/150 for training, validation, and test. As well known, current deep-learning methods are mainly data-driven, 
and the large number of the training set is critical to obtain robust models. A large number of the validation set (even larger 
than whole datasets used in previous work) can avoid the over-fitting problem. The test set contains 150 CT scans which are inaccessible to the participants. 
Only Docker-based submission is acceptable for evaluating the segmentation algorithms, which guarantees the fairness and reproduciblility of the benchmark. Further, the ATM'22 challenge covers CT scans from multiple sites, which can also evaluate the generalization ability of the models. Detailed information on dataset is provided in Section \ref{sec:dataset}.\\

2) \textbf{More Comprehensive Metrics}. In the EXACT'09 challenge, the groundtruth of the airway was constructed from the results of the participants. Specifically, they first divided the airway tree into branch segments. These segments are then scored by experienced observers to determine whether it is a correctly segmented part or not. 
Finally, the reference airway trees were constructed by gathering the union of all correctly extracted branch segments. Since EXACT'09 did not acquire the fine-grained annotation of the airways, their benchmark was designed to only evaluate the depth of the predicted airway while neglecting the exact airway shape and dimensions.

To comprehensively assess the algorithms, ATM'22 considered both the depth of the airway trees and airway dimensions via the fine-grained annotations by experienced radiologists. ATM'22 aims to evaluate the airway segmentation algorithms from two perspectives including \textbf{Topological Completeness} and \textbf{Topological Correctness}.  The topological completeness is measured by the tree length detected rate (TD, \%) and branch detected rate (BD, \%), which are introduced by EXACT'09. Both TD and BD are evaluated 
on the largest component of the prediction, reflecting the topological completeness and continuity of the models. The high quality of the topological completeness is essential to the navigation usage for bronchoscopic-assisted surgery. The topological correctness represents the overlap-wise accuracy of the segmentation models. We adopted the 
Dice Similarity Coefficient (DSC, \%) and Precision (\%) for the quantitative measurement of pulmonary airways, which plays a critical role in abnormality analysis. The selection criterion and standard formula of these metrics are reported in Section \ref{sec:evaluation_metrics}.\\

3) \textbf{More Powerful Platform}.  The ATM'22 challenge is one of the Satellite Events in conjunction with MICCAI 2022\footnote{MICCAI 2022 challenge list: \url{https://conferences.miccai.org/2022/en/MICCAI2022-CHALLENGES.html}}, and hosted on \textit{grand-challenge.org}\footnote{ATM'22 website: \url{https://atm22.grand-challenge.org/}}, which allows the flexible and extendible management of benchmarks. Compared with the EXACT'09 challenge, our website not only 
provides the registration and dataset access but also supports the submission of prediction results and prompt feedback. 
The submissions will be evaluated automatically by the executable docker and the metrics can be presented on the leaderboard in a few minutes. The improvement of the evaluation procedure has significantly accelerated the research, as the researchers do not need to wait for the official result reply via e-mails like EXACT'09.
In addition, the ATM'22 owns a live leaderboard that presents all valid results from different teams. The public leaderboard ensures the fairness in evaluating various algorithms. In conclusion, the ATM'22 challenge has deployed on a more effective platform, which is beneficial to the research community. A detailed evaluation procedure can be seen in Section \ref{sec:participation_procedure}.

\subsection{Contributions}
Our challenge was accepted as a Satellite Event of the MICCAI 2022 challenge, 
and our official challenge website is constructed and maintained via the platform of grand-challenge.org. The contribution of our organized challenge 
can be briefly summarized below:
\begin{itemize}
  \item ATM'22 is a critical milestone that establishes the standard norm of the airway segmentation field in this deep learning 
  era. To our best knowledge, ATM'22 is the first challenge to provide the large-scale dataset, 500 CT scans with full pulmonary airway annotation. 
  A large amount of the dataset is beneficial to the development of the deep-learning based algorithms. Further, our challenge was deployed on the 
  public platform that executes the evaluation in time and then presents the results online. Hence, it is convenient to compare with different algorithms 
  and speed up the research procedure. 
  \item ATM'22 arouses the reflection that airway segmentation should be a beyond pixel-wise segmentation task. Unlike other common segmentation tasks, 
  the overlap based and the surface distance-based measures are enough to evaluate the performance of the algorithms. 
  However, these measures only consider the topological correctness of the airway segmentation methods. The topological completeness is another significant 
  aspect to measure the performance of airway segmentation algorithms. ATM'22 first establishes the most comprehensive evaluation system, including both 
  topological correctness and topological completeness, to determine the performance of the algorithms. Combined with the large-scale datasets, the intrinsic 
  topological features of airways are expected to be harnessed.
  \item ATM'22 focus on the generalization ability of automatic airway segmentation algorithms. 
  ATM'22 contains divergent data from multi-site and multi-domain. The deep learning models are expected to explore more 
  substantive characteristics of the pulmonary airway to perform well across different sites and domains. In addition, ATM'22 provides 
  a valuable prerequisite database for various clinical centers worldwide. They could leverage this database for the pre-training of models 
  and then apply it to their in-house data.
\end{itemize}
The rest of the paper is organized as follows: Section \ref{sec:relatedwork} summarizes the previous work related to pulmonary airway segmentation. Section \ref{sec:challengesetup} provides the details of the materials, evaluation framework, and participation procedure in our challenge. Section \ref{sec:methodologies} introduces and compares the top 10 methods ranked in this challenge, along with our insights. Section \ref{sec:results} presents the quantitative and qualitative results of the validation phase and the final test phase, 
followed by the discussion in Section \ref{sec:discussion}. Finally, we conclude our work in Section \ref{sec:conclusion}.

\section{Related Work}\label{sec:relatedwork}
\subsection{EXACT'09 Challenge}
To compare different airway segmentation algorithms using a standard dataset and performance evaluation method, the Extraction of Airways From CT (EXACT’09)~\citep{EXACT09} is successfully hosted in 2009. The EXACT'09 dataset provided 40 CT scans including 20 scans for the training usage and 20 scans for the test stage. They evaluated 15 airway tree extraction algorithms from different research groups. In that pre-deep learning era, most of the participants adopted region-growing and vessel filters to 
address this problem. The results of the participants were further used by the organizers to construct the golden 
standard of the airway reference. Specifically, the airway prediction of different participants was first 
subdivided into individual branches, and then visually scored by the trained observers. The correctly segmented 
branches were retained while the incorrect branches were rejected. Finally, all accepted branches were aggregated to 
acquire the final reference standard. However, the training observers merely decided whether the individual branches 
were acceptable or not while they did not annotate the original CT scans for those branches neglected by 
all the algorithms. In addition, due to the lack of precise voxel-wise annotation, the evaluation of EXACT'09 
was designed to only take the extracted airway tree length into consideration without the shape and dimension. 

EXACT'09 contributed to the field of pulmonary airway segmentation as they established a framework to evaluate the 
airway extraction algorithms in a standard manner. They had established their own website\footnote{EXACT'09 Website: \url{http://image.diku.dk/exact/}}, 
where detailed information and challenge results are presented. Although this website is maintained manually 
for registration and submission, their feedback period is extremely long, which is inappropriate for the 
current scientific research.

\begin{figure*}[thbp]
  \centering
  \includegraphics[width=0.95\linewidth]{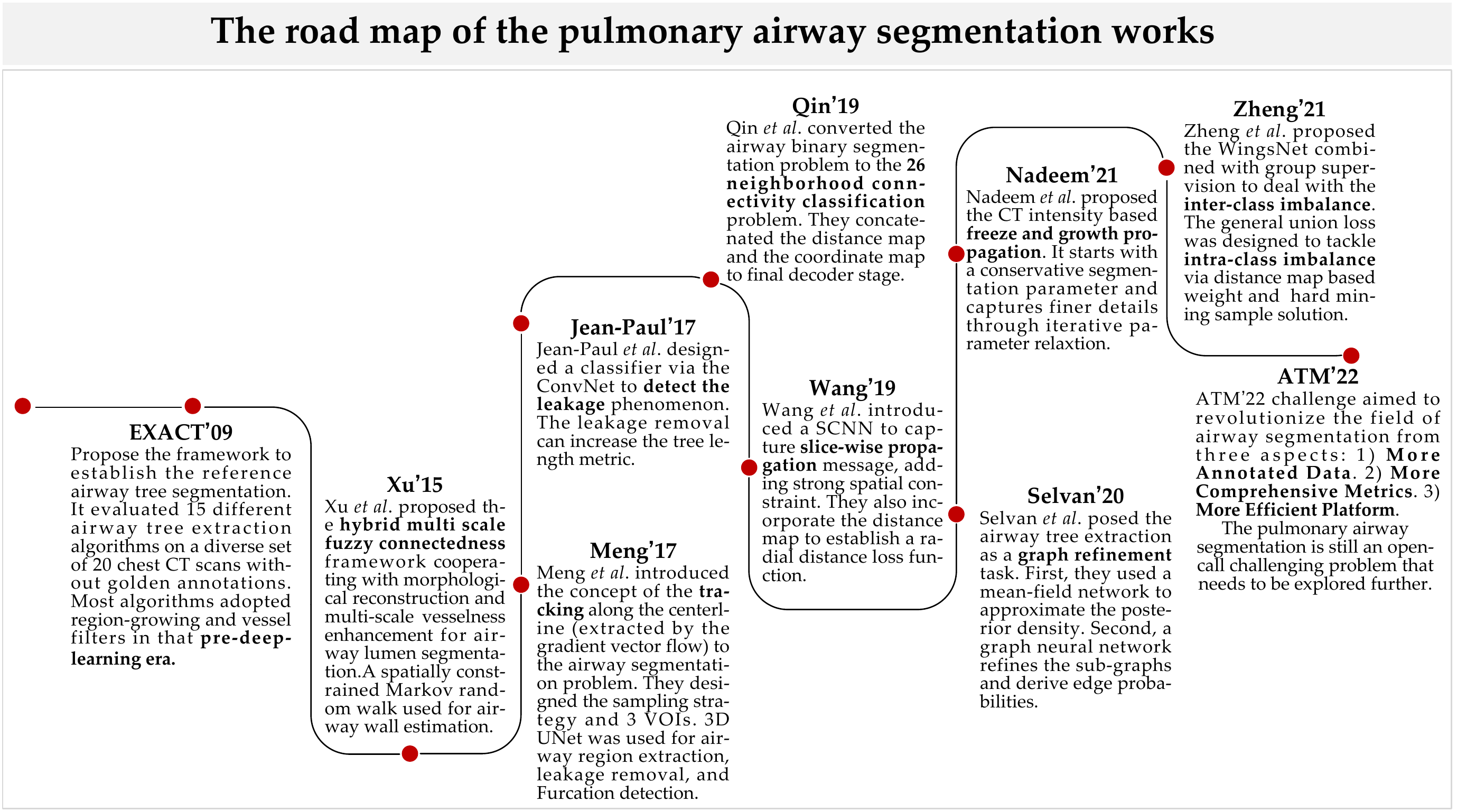}
  \caption{The road map of the representative airway segmentation works from EXACT'09 challenge to ATM'22 challenge.}
  \label{fig:airway_roadmap}
  \end{figure*}
\subsection{Deep Learning Methods for Airway Segmentation}
Since EXACT'09, several methods that employed techniques such as adaptive thresholding, region growing, and filtering-based enhancement were proposed. These methods successfully segmented the trachea and main bronchi but often failed to extract peripheral bronchi because the intensity contrast between the airway lumen and wall weakens as airways bifurcate into thinner branches. \cite{xu2015hybrid} proposed the hybrid multi-scale fuzzy connectedness framework cooperating with morphological reconstruction and multi-scale vessel enhancement for airway lumen segmentation. As presented in Fig.\ref{fig:airway_roadmap}, the recent progress of deep learning, especially Convolutional Neural Networks (CNNs) have promoted the research on airway segmentation~\citep{charbonnier2017improving,jin20173d,meng2017tracking,meng2017automatic,selvan2018extraction,
nadeem2018topological,garcia2018automatic,zhao2019bronchus, yun2019improvement,qin2019airwaynet,wang2019tubular,garcia2019joint,nadeem2020ct,selvan2020graph,qin2021learning,
zheng2021alleviating,zheng2021refined,garcia2021automatic,wu2022ltsp,yu2022break,nan2022fuzzy,zhang2022differentiable}. 

To reduce the mass of false positives and increase the length of the detected airway tree length, 2-D CNN~\citep{yun2019improvement} and 2.5-D CNN~\citep{charbonnier2017improving} were respectively applied to the coarse segmentation to  reduce false positives and increase the length of the detected airway tree.
3D CNNs were developed to handle the airway segmentation task via either the fixed-stride patch-wise sliding window 
fashion~\citep{garcia2018automatic} or a dynamic VOI-based tracking way~\citep{meng2017tracking}. 
To further extract discriminative features, specific designs of neural networks were also incorporated into the 3D UNet. Graph refinement~\citep{selvan2020graph,garcia2019joint} was explored to incorporate neighborhood knowledge of airways in feature aggregation. \cite{wang2019tubular} proposed a spatial propagation layer and 
radial distance loss for tubular topology perception. \cite{qin2021learning} designed a feature calibration and attention distillation module to force the 3D UNet to share superiority to tenuous peripheral bronchioles. \cite{zhao2019bronchus} proposed a linear-programming tracking method to combine the results of 3D CNNs and 2D CNNs. 

Meanwhile, the importance of the connectivity of the airway prediction also raised attention. AirwayNet~\citep{qin2019airwaynet} was proposed to transform the binary airway segmentation task into 26-neighborhood connectivity prediction problem. \cite{wu2022ltsp} utilized the long-range slice continuity information to enhance the connectedness of airway prediction. The connectivity attribute was further explored by \cite{zheng2021alleviating} and \cite{zhang2022differentiable}.
Zheng \etal put forward the class imbalance problem that existed in the airway segmentation task while Zhang \etal pointed out that 
a satisfactory trade-off between the topological completeness and correctness should be achieved. 
The WingsNet was adopted by \cite{zheng2021alleviating} and \cite{yu2022break} as the backbone for a multi-stage training solution. Zheng \etal designed a general union loss (GUL) to alleviate the intra-class imbalance problem. Yu \etal resolved the problem via a breakage-sensitive loss. To further tackle the topology-preserving challenge, \cite{zhang2022differentiable} proposed a convolutional distance transform (CDT) module to refine the fractured areas that are critical to the topological structures. \cite{nan2022fuzzy} designed a continuity and accumulation mapping (CAM) loss,  which enhanced the continuity degree and minimized projection errors of airway predictions.

The EXACT'09 challenge has been hosted over a decade, it is time to promote the airway segmentation task to a new level for the next generation of 
medical image analysis and bronchoscopic-assisted surgery. The ATM'22 challenge aimed to revolutionize this field via providing more annotated data, more comprehensive evaluation, and more efficient feedback for the research community. A promising trend of the pulmonary airway segmentation is to harness the intrinsic topological features from the significant annotated data.

\subsection{Topological Prior Knowledge}
The most relevant task to airway segmentation is tubular object segmentation, where topological prior knowledge plays a critical role. A typical class of tubular objects shares a tree-like structures~\citep{li2022human}, such as blood vessel ~\citep{lyu2022reta}, coronary artery~\citep{kong2020learning}, neuron images~\citep{li20193d}, and the airway. 
Despite the powerful data-fitting ability of the deep learning models, they barely can learn the extrinsic topological features. For example, it is extremely difficult for 
deep learning models to represent the characteristic that "An object shares one single connected domain". The poor representation of the topology leads to the discontinuity problem that often happens in tubular object segmentation tasks. To alleviate this problem, previous works could be categorized into three dimensions: 
1) Enhancing the representation ability of the deep learning models. 2) Designing surrogate objective functions to increase the topological accuracy. 3) Adding the explicit topological restriction to the optimization procedure. 

As for the first aspect, Mosinska~\citep{agarap2018deep} discriminated the higher-order topological features of linear structures by adding the restriction term to minimize the differences between the VGG19 descriptor of the ground-truth images and the corresponding prediction delineations. The Local Intensity Order Transformation (LIOT) ~\citep{shi2022local} was dedicated to representing the tubular structure, which is invariant to the increasing change of the contrast. LIOT transformed the original image into a  feature map with four channels, reinforcing the network to learn more discriminative features. A Joint Topology-preserving and Feature-refinement Network (JTFN) ~\citep{cheng2021joint} 
was designed to jointly handle the global topology and refined features via an iterative feedback learning strategy. 

In addition to enhancing the representation ability of deep learning models, other works endeavored to achieve this goal by designing surrogate objective functions to increase topological accuracy. Distance transform is a natural alternative ~\citep{ma2020distance} used in medical image analysis to uncover topology information. \cite{kervadec2019boundary} focused on the boundary of the distance map and designed the boundary loss to minimize the boundary variations between prediction and ground-truth via an integral approach. \cite{xue2020shape} directly regressed the signed distance map (SDM), 
followed by the least absolute error loss to penalize the output SDM with the wrong sign. To repair the fractured areas, 
a convolutional distance transform (CDT) module~\citep{zhang2022differentiable} was proposed to be perceptible to the breakage.
Other topological elements were also investigated in the tubular object segmentation. The centerline, bifurcation, local 
radius, curvature, normal, and so on are the ponderable characteristics for the representation of tubular structures. 
Wang \etal ~\citep{wang2020deep} presented tubular shapes as the envelope of a family of spheres with continuously changing center points and radii. 
They rephrased the distance map prediction as a quantified classification based on the center points and radii. 
Shit \etal ~\citep{shit2021cldice} proposed a differentiable measurement, CenterlineDice (clDice), to simultaneously handle the 
over- or under-segmentation phenomenon. However, the centerline ground-truth of volumetric data is not easily acquired. Despite 
it can be approximately computed via the 3D skeletonization method~\citep{lee1994building}, the curve-skeleton/medial axis extraction 
from 3D mesh representation itself is an open challenging and unsolved problem~\citep{au2008skeleton,dey2006defining,cornea2005computing}.

As for the third perspective, critical properties in the algebraic topology were gradually applied to 
add the explicit topological restriction to the optimization procedure. Persistent homology~\citep{edelsbrunner2000topological,cornea2005computing} is a topological 
data analysis method for calculating the robustness of topological features of a dataset at different scales. 
Persistent homology involves counting the number of topological features from different dimensions, 
termed \textit{Betti numbers}. The Betti numbers are crucial topological invariants that count the number of features of 
dimension $k$, where $\beta_{0}$, $\beta_{1}$, and $\beta_{2}$ represent the number of connected components, 
the number of loops or holes, and the number of hollow voids, respectively. Clough \etal ~\citep{clough2020topological} 
first analyzed the Betti numbers as a set of \textit{birth} and \textit{death} threshold values for each topological feature, which 
can be represented in a barcode diagram. They then specified the desired topology of the segmented objects and adopted the 
Persistent homology upon the candidate segmentation to reinforce it to share the specified topological features. 
Similarly, Hu \etal ~\citep{hu2019topology} optimized the persistence diagram to emphasize one-dimensional topological features, 
i.e., the connected components. The Morse theory ~\citep{milnor2016morse}, which captures the 
singularities of the gradient vector field of the likelihood function, was also investigated to 
identify critical global structures, including 1D skeletons and 2D patches~\citep{hu2021topology}. 
Zhang \etal explored several unsupervised geometry-based methods for tubular object reconstruction.
The divergence prior ~\citep{zhang2019divergence} and confluence property ~\citep{zhang2021confluent}
were incorporated as the explicit constraints to improve reconstruction accuracy.

\begin{table*}[th]
  \centering
  \caption{The summarized properties of the training, validation and test sets.}\label{tab:dataset_property}
  \scalebox{0.9}{
  \begin{tabular}{@{}cccccc@{}}
  \toprule
  Dataset     & Scanner               & Slice Number          & Slice thickness (mm)       & Resolution (mm)  \\ 
  \midrule
  Training    & Philips iCT 256, GE LightSpeed16                      & 157-1125       & 0.500-1.000                 & 0.514-0.919        \\
  Validation  & Philips iCT 256, GE LightSpeed16                      & 408-803       & 0.500-0.750                 & 0.531-0.822        \\
  Test        & Philips iCT 256, GE LightSpeed16, TOSHIBA Aquilion    & 257-830       & 0.450-0.801                 & 0.500-0.859        \\ 
  \bottomrule      
  \end{tabular}}
  \end{table*}
  
\section{Challenge Setup}\label{sec:challengesetup}
\subsection{Dataset}\label{sec:dataset}
\subsubsection{Dataset Information}
We collected and annotated 500 chest CT scans from multi sites. The CT scans were collected from the public LIDC-IDRI dataset ~\citep{armato2011lung} and 
the Shanghai Chest hospital. The chest CT scans were acquired with three vendors including Philips iCT 256, GE MEDICAL SYSTEMS LightSpeed16, 
TOSHIBA Aquilion. The health conditions of the scanned subjects are diverse, ranging from healthy people to patients with severe pulmonary disease. 
The information of patients and scanners were manually anonymized. We then annotated the selected 500 CT scans by three experienced radiologists. 
The annotation details are carefully elaborated in the Section \ref{sec:annotation_details}.

Each chest CT scan consisted of varying number of slices, ranging from 157 to 1125 with a slice thickness of 0.450-1.000 mm. The axial size of all slices is $512 \times 512$ pixels with a spatial resolution of 0.500-0.919 mm. The training set consists of 300 chest CT scans, while 50 and 150 CT scans for the validation set and test set, respectively. The properties of the training, validation, and test sets are summarized in Table \ref{tab:dataset_property}.

\subsubsection{Annotation Details}\label{sec:annotation_details}
To acquire the fine-grained annotations of the airway from chest CT scans, each CT scan was firstly preprocessed by the models by \citep{zheng2021alleviating,3DU-Net,yu2022break} trained on BAS dataset~\citep{qin2021learning}. The results are then ensembled by majority voting strategy to acquire the preliminary segmentation result. Theses preliminary annotations were carefully delineated and manually double-checked by three radiologists with more than five years of professional experience to acquire the final refined airway tree structure, 
which took 60-90 minutes for each CT scan. The organizers spent almost one year to collect the 500 chest CT scans from different sites and carefully delineating the refined airway annotations for each scan. In the annotation process, we tried our best to ensure that each radiologist stuck to the same annotation 
principle and thus guaranteed the consistency of airway annotation. 

\subsection{Evaluation Metrics}\label{sec:evaluation_metrics}
As presented in Table \ref{tab:dataset_metrics_evaluation}, previous works adopted incomplete metrics to measure the performance. 
In this challenge, we established a comprehensive evaluation system. Followed by ~\citep{maier2022metrics}, we chose two types of metrics 
to evaluate the airway segmentation algorithms. The first is the \textit{common segmentation task metric} and the second is the \textit{specific property-related metric}. 
Specifically, we adopted the Dice Similarity Coefficient (DSC, \%), Precision (\%) to measure the overlap-based and voxel-wise segmentation accuracy. 
Let $\mathcal{Y}$ and $\mathcal{\hat{Y}}$ denote the binary ground-truth label and the prediction result. The calculation of DSC and Precision can be formulated as below:
\begin{align}
  DSC &= \frac{2|\mathcal{\hat{Y}} \bigcap \mathcal{Y}|}{|\mathcal{Y}+\mathcal{\hat{Y}}|}, \\
  Precision &= \frac{|\mathcal{\hat{Y}} \bigcap \mathcal{Y}|}{|\mathcal{\hat{Y}}|},
\end{align}
where the $|\cdot|$ denotes the sum operation that returns the number of voxels. In addition, we also supplemented the evaluation of the voxel-wise segmentation accuracy 
with the Sensitivity (Sen, \%) and Specificity (Sep, \%). The Sen and Spe are respectively associated with the true positive (TP) volume fractions and the true negative (TN) 
volume fractions:
\begin{align}
  \textrm{Sen} &= \frac{|TP|}{|TP+FN|} = \frac{|\mathcal{\hat{Y}} \bigcap \mathcal{Y}|}{|\mathcal{Y}|},\\
  \textrm{Spe} &= \frac{|TN|}{|TN+FP|} = \frac{|I|-|\mathcal{\hat{Y}} \bigcup \mathcal{Y}|}{|I| - |\mathcal{Y}|},
\end{align}
where the $FN$ denotes the false negative volume fractions and $FP$ is the false positive volume fractions. $I$ represents the image to segment.

As for the specific property-related metric, topological completeness is the most critical attribute in the airway segmentation challenge. 
Following~\citep{EXACT09}, we defined the Tree length detected rate (TD, \%) and Branch detected rate (BD, \%) to measure the performance of algorithms in detecting the airway. TD is defined as the fraction of the tree length that is detected appropriately with regard to the length of the airway tree in the ground-truth:
\begin{align}
  TD = \frac{T_{det}}{T_{ref}},
\end{align}
where $T_{det}$ denotes the total length of all branches detected in the prediction, and $T_{ref}$ represents the whole tree length in the ground-truth. 
BD denotes the percentage of the airway branches that are detected correctly with association to the whole number of branches in the ground-truth:
\begin{align}
  BD = \frac{B_{det}}{B_{ref}},
\end{align}
where $B_{det}$ denotes the total correct branches detected in the prediction, and $B_{ref}$ represents the whole number of branches in the ground-truth. 
Note that a branch in the prediction is identified as 'correct' only if more than $80\%$ of centerline voxels extracted from the certain branch are within the ground-truth. 
  \makeatletter
  \def\hlinew#1{%
  \noalign{\ifnum0=`}\fi\hrule \@height #1 \futurelet
  \reserved@a\@xhline}
  \makeatother
  \begin{table*}[thbp]
  \renewcommand\arraystretch{1.1}
  \centering
  \caption{\justify{The list and details of the participant teams who successfully participated in the \textbf{validation phase}
  (Full submission including 50 binary predictions and a qualified short paper was received before 17, Aug, 2022)  and \textbf{testing phase}(Full submission including an executable docker and a qualified short paper was received before 31, Aug, 2022). For simplicity, the short team index is used in 
  the main text for the reference of the different teams, e.g., V1 means Validation team 1, representing for the team of Sanmed\_AI. Index is random. T1 means Test phase team 1, representing for the team of Sanmed\_AI. Index depends on the successful submission order.}}\label{tab:validation-participants-info}
  \scalebox{0.6}{
  \begin{threeparttable}
  \begin{tabular}{ccccc}
  \hlinew{1pt}
  \textbf{Validation index}    &   \textbf{Test index}    &\textbf{Team name }& \textbf{Affiliation} & \textbf{Location} \\ \hline
  V1 & T1&Sanmed\_AI & A.I. R\&D Center, Sanmed Biotech Inc. & Guangdong, China \\ 
  V2 & T4 &YangLab & National Heart and Lung Institute, Imperial College London & London, UK \\ 
  V3 & T20 & notbestme & School of information science and technology, Fudan University & Shanghai, China \\ 
  V4 & - &xiaqi & Hygea Medical Technology Corporation & Beijing, China \\ 
  V5 & T22 &cvhthreedee & Department of Informatics, Karlsruhe Institute of Technology & Karlsruhe, Germany \\ 
  V6 & T3 &LinkStartHao & College of Physics and Information Engineering, Fuzhou University & Fujian, China \\ 
  V7 & T7 &neu204 & College of Medicine and Biological Information Engineering, Northeastern University & Liaoning, China \\ 
  V8 & T12 &miclab & Department of Computer Engineering and Industrial Automation, University of Campinas & Campinas, Brazil \\ 
  V9 & T8 &blackbean & Shanghai AI Lab & Shanghai, China \\ 
  V10 & T19 & Median & Median Technologies & Valbonne, France \\ 
  V11 & T9 &lya & University of Science and Technology of China & Hefei, China \\ 
  V12 & T18 &satsuma & Centre for Medical Image Computing, University College London & London, UK \\ 
  V13 & - &ailab & Shanghai AI Lab & Shanghai, China \\ 
  V14 & T6 & timi & InferVision Medical Technology Co., Ltd. & Beijing, China\\ 
  V15 & T17 &suqi & School of Electronic Information and Electrical Engineering, Shanghai Jiao Tong University & Shanghai, China \\ 
  V16 & - &MibotTeam & Smart surgery, Alg Department, Microport & Shanghai, China \\ 
  V17 & T13 &CITI-SJTU & School of Biomedical Engineering, Shanghai Jiao Tong University & Shanghai, China \\ 
  V18 & - &SEU & Key Laboratory of Computer Network and Information Integration, Southeast University & Nanjing, China \\ 
  V19 & T14 &deeptree\_damo & Alibaba DAMO Academy & Hangzhou, China \\ 
  V20 & T15 &CBT\_IITDELHI & Indian Institute of Technology Delhi(IITD) & Delhi, India\\ 
  V21 & T5 &dolphins & Computer Science Department, National Engineering School of Brest & Brest, France\\ 
  V22 & T11 &bms410 & National Yang Ming Chiao Tung University Yangming Campus & Taipei, Taiwan, China\\ 
  V23 & - &airwayseg & Center of Product Research\&Development, Keya Medical & Shenzhen, China\\
  V24 & - &atmmodeling2022 & Pittsburgh Institute, Sichuan University & Sichuan, China\\
  V25 & T16 &bwhacil & Applied Chest Imaging Laboratory, Brigham and Women's Hospital, Harvard Medical School & Boston, USA\\
  V26 & T10 &dnai & Diannei Technology & Shanghai, China\\
  V27 & - &mlers & R\&D, Microport & Shanghai, China\\
  V28 & T2 &fme & Fraunhofer Institute for Digital Medicine MEVIS & Bremen, Germany\\
  V29 & T21 &biomedia & Mohamed bin Zayed University of Artificial Intelligence, UAE & Abu Dhabi, United Arab Emirates\\
  V30 & - &ntflow & Mathematics, Nanjing University & Nanjing, China\\
  \hlinew{1pt}
  \end{tabular}
  \end{threeparttable}}
  \end{table*}
  
TD and BD are adopted to measure the topological completeness of segmentation algorithms, meanwhile, DSC and Precision are chosen as the 
topological correctness measurements. Since all metrics are normalized into $[0\%,100\%]$, the mean score calculation is adopted as the ranking criterion: 
\begin{align}
  \text{Mean Score} = 0.25 * TD + 0.25 * BD \notag \\  + 0.25 * DSC + 0.25 * Precision.
\end{align}  
The implementation of the evacuation code can be found in our official ATM'22 repository\footnote{\url{https://github.com/Puzzled-Hui/ATM-22-Related-Work/tree/main/evaluation}}.

\subsection{Participants}\label{sec:participants}

As an open-call challenge, the ATM'22 received \textbf{305} requests of registration before the MICCAI 2022 conference (September 22, 2022), among which 
\textbf{30} teams had successfully participated in the validation phase before the validation phase submission deadline (August 17, 2022). 
\textbf{22} teams had submitted the algorithm dockers successfully in the test phase before the test phase submission deadline (August 31, 2022). 
\begin{figure}[!htbp]
\centering
\includegraphics[width=0.6\linewidth]{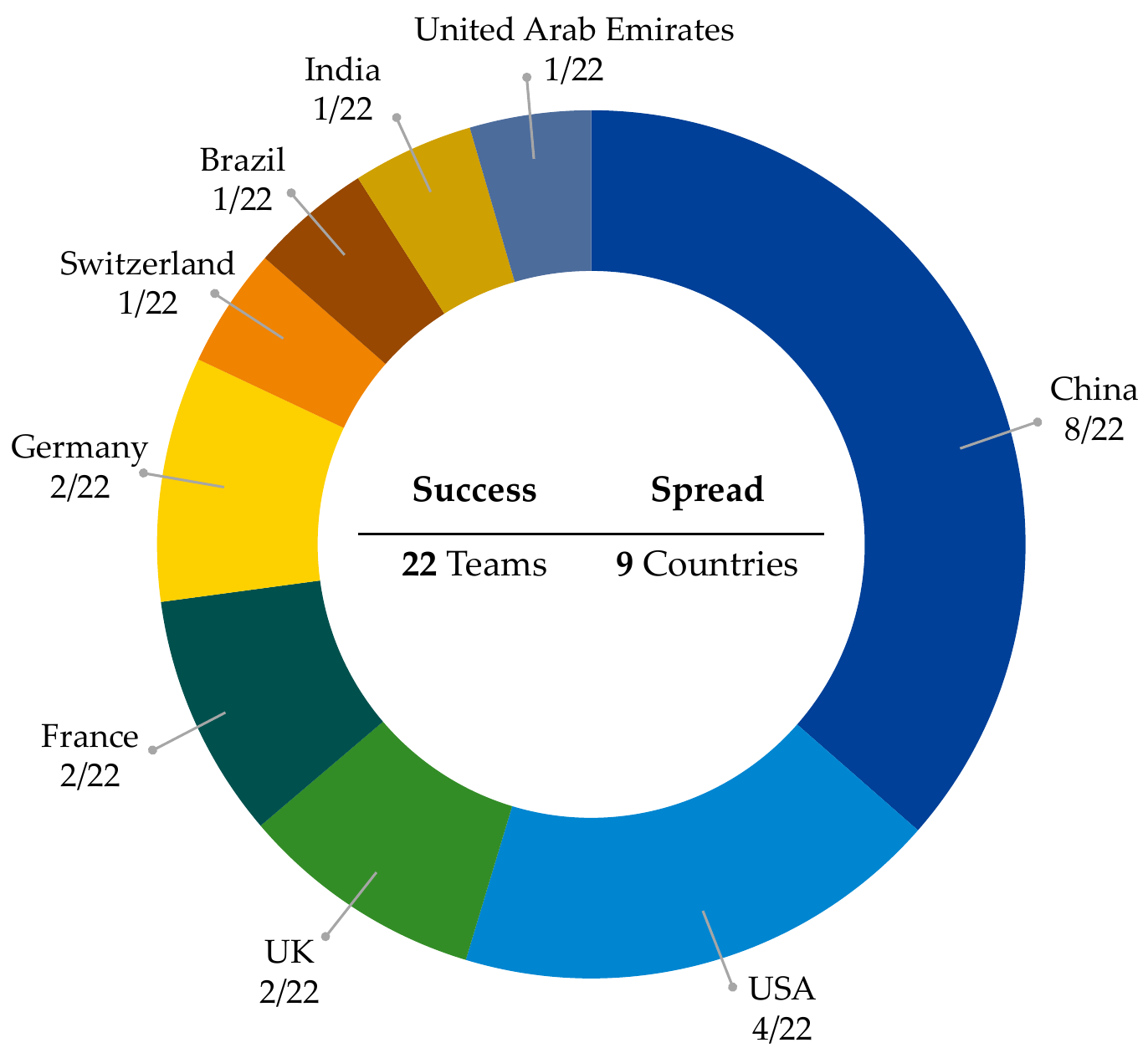}
\caption{The individual team statistics that fully participated successfully in ATM'22 challenge.}
\label{fig:team_distribution}
\end{figure}
As detailed in Figure \ref{fig:team_distribution}, these 22 teams are from 9 different countries.
In this paper, the information of all teams are reported in Table~\ref{tab:validation-participants-info}. \textit{Note that we assign a unique Team index for different teams in the validation phase and test phase respectively.} We adopted the Team index to describe their methodologies and results for simplicity. 
\textbf{10} representative algorithms were selected to be reported in detail. All teams agreed to include their methods and results for this publication. The selected algorithms were under the consideration of both novelty and evaluation performance. 

\subsection{Challenge Phases}\label{sec:participation_procedure}
The challenge includes three phases. First, to complete the registration and get access to the dataset,
the participants should register on the official challenge website, sign the data agreement file then send the scanned 
file via e-mail to organizers, and keep their promise to abide by the challenge rules. Second, the participants should take part 
in the validation phase, where the binary predictions are required to submit. The evaluation is automatically 
executed on the platform of grand-challenge.org. The leaderboard is also presented online and updated promptly
\footnote{The leaderboard of the validation phase: \url{https://atm22.grand-challenge.org/evaluation/validation-phase-1-live-leaderboard/leaderboard/}}.
Third, the participants should take part in the final test phase to complete the full participation in this challenge. 
To guarantee the fairness of the competition, the packaged docker is the only valid submission in the test stage. 
The instructions for the preparations of dockers are provided in the ATM'22 challenge repository
\footnote{Docker submission guideline of the test phase: \url{https://github.com/Puzzled-Hui/ATM-22-Related-Work/tree/main/baseline-and-docker-example}}. 
In this repository, we also provide the basic pipeline to package your models to the docker image, which is helpful to those 
who have little expertise with docker.

\begin{table}[h]
\centering
\caption{Descriptions of the notation.}\label{tab::notation}
\begin{tabular}{@{}cc|cc@{}}
\toprule
Notation & Description & Notation & Description \\ \midrule
$\mathbf{x}$        & Input          & $\mathbf{y}$        & Label           \\
$\mathcal{X}$        & Feature space           & $\mathcal{Y}$        & Label space           \\
$\mathcal{L(\cdot,\cdot)}$        & Loss function    & $\mathbf{\hat{y}}$   & Likelihood map \\ \bottomrule      
\end{tabular}
\end{table}

\makeatletter
\def\hlinew#1{%
\noalign{\ifnum0=`}\fi\hrule \@height #1 \futurelet
\reserved@a\@xhline}
\makeatother
\begin{table*}[thbp]
\renewcommand\arraystretch{1.3}
\centering
\caption{Characteristics of the top 10 models. Abbr: Lung Window (LW), Lung Region Extraction (LRE), Intensity Normalization (Norm), 
Rotation (R), Flip (F), Scale (S), Jitter (J), Gaussian Noise (GN), Brightness (B), Gamma (GA). 
It is noted that the largest connected component extraction is executed on 
all methods by the official organizers to evaluate the final metrics.}\label{tab:characteritic_of_model}
\scalebox{0.8}{
\begin{tabular}{cc|ccc|cccccccc|c}
\hlinew{1pt}
\multicolumn{1}{c|}{\multirow{3}{*}{\textbf{Team}}} & \multirow{3}{*}{\textbf{Backbone}} & \multicolumn{3}{c|}{\multirow{2}{*}{\textbf{Pre-Process}}} & \multicolumn{8}{c|}{\textbf{Data Augmentation}} & \multirow{3}{*}{\textbf{Post-Process}} \\
\multicolumn{1}{c|}{}                      &                           & \multicolumn{3}{c|}{}                             & \multicolumn{3}{c|}{Spatial-based} & \multicolumn{4}{c|}{Intensity-based} & \multirow{2}{*}{Others} &                               \\
\multicolumn{1}{c|}{}                      &                           & LW   & LRE  & Norm                      &R     &F   & \multicolumn{1}{c|}{S}   &J    &GN    &B   & \multicolumn{1}{c|}{GA}  &        &                               \\ \hlinew{1pt}
\multicolumn{1}{c|}{T6}                    & WingsNet~\citep{zheng2021alleviating} & [-1000,500] & \green{\Checkmark}  & \green{\Checkmark} &     &    & \multicolumn{1}{c|}{}   &    &    &   & \multicolumn{1}{c|}{}  & N/A & fill holes\\
\multicolumn{1}{c|}{T4}                    & Attention UNet~\citep{oktay2018attention} & [-1200,600] & \green{\Checkmark}  & \green{\Checkmark}  &\green{\Checkmark}     &\green{\Checkmark}    & \multicolumn{1}{c|}{}   &    &    &   & \multicolumn{1}{c|}{}  & N/A        & region grow\\
\multicolumn{1}{c|}{T14}                   & nnUNet~\citep{isensee2021nnu} &  N/A              &                &  \green{\Checkmark}  & \green{\Checkmark}&\green{\Checkmark}& \multicolumn{1}{c|}{\green{\Checkmark}}   &    &\green{\Checkmark}&\green{\Checkmark}& \multicolumn{1}{c|}{\green{\Checkmark}}  & N/A & TTA  \\
\multicolumn{1}{c|}{T7}                    & 3D UNet~\citep{3DU-Net}  & [-1000,600]  & \green{\Checkmark}  &  \green{\Checkmark}  &     &\green{\Checkmark}    & \multicolumn{1}{c|}{}   &\green{\Checkmark}    &    &   & \multicolumn{1}{c|}{}  &N/A & ensemble \\
\multicolumn{1}{c|}{T1}                    & 3D UNet~\citep{3DU-Net}  & [-1400,200]  & \green{\Checkmark}  &  \green{\Checkmark}  &     &\green{\Checkmark}    & \multicolumn{1}{c|}{}   &\green{\Checkmark}    &    &   & \multicolumn{1}{c|}{}  &N/A &N/A \\
\multicolumn{1}{c|}{T5}                    & 3DResNet~\citep{tran2018closer} & N/A & &\green{\Checkmark} &     &\green{\Checkmark}& \multicolumn{1}{c|}{}   &    & \green{\Checkmark}&   & \multicolumn{1}{c|}{\green{\Checkmark}}  & pseudo label & N/A \\
\multicolumn{1}{c|}{T17}                   & nnUNet~\citep{isensee2021nnu}  & N/A   &  &\green{\Checkmark} &\green{\Checkmark}  &\green{\Checkmark}    & \multicolumn{1}{c|}{\green{\Checkmark}}   &    &\green{\Checkmark}    &\green{\Checkmark}   & \multicolumn{1}{c|}{\green{\Checkmark}}  &N/A &N/A  \\
\multicolumn{1}{c|}{T20}                   & Transformer  & N/A &\green{\Checkmark} & \green{\Checkmark} &     &    & \multicolumn{1}{c|}{}   &    &    &   & \multicolumn{1}{c|}{}  & N/A & Resize \\
\multicolumn{1}{c|}{T9}                    & nnUNet~\citep{isensee2021nnu} & [-1200,600] &  &\green{\Checkmark} &\green{\Checkmark} &\green{\Checkmark} & \multicolumn{1}{c|}{\green{\Checkmark}} &    &    &\green{\Checkmark} & \multicolumn{1}{c|}{}  & deformation  & N/A   \\
\multicolumn{1}{c|}{T10}                   & 3D UNet~\citep{3DU-Net}   & [-1028,266] &  & \green{\Checkmark} &\green{\Checkmark} &\green{\Checkmark}    & \multicolumn{1}{c|}{}   &    &    &   & \multicolumn{1}{c|}{}  &N/A & ensemble \\ \hlinew{1pt}
\end{tabular}}
\end{table*}

\section{Methodologies}\label{sec:methodologies}
In this section, the overall comparison and analysis of different methods are first reported. 
Then, we elaborate on the top 10 methods ranked in the final test phase. 
For each method, we summarize the main contributions and report the implementation details. 
The potential directions of improvements are finally discussed. For simplicity, Table \ref{tab::notation} lists the 
frequently-used notation. The order of method description is in accordance with the performance ranking of the final test stage.

\subsection{Overall Comparison}
In this section, we focus on the overall comparison of the top 10 methods. Table \ref{tab:characteritic_of_model} summarizes the main characteristics 
of the top 10 models, including the backbone architectures, the pre-process procedures, data augmentation strategies, and the post-process procedure. 
3D UNet~\citep{3DU-Net} and nnUNet~\citep{isensee2021nnu} are the common choices for the backbones. nnUNet (used by T14, T17) 
adopted the percentage clipping instead of the lung window to truncate CT values. 
Generally speaking, nnUNet conducts more comprehensive data augmentation compared with other methods. All top 10 methods perform the intensity normalization 
and a part of methods (T6, T4, T7, T1, T20) adopt the lung region extraction as the pre-process to crop the unrelated regions. 

Table \ref{tab:method_brief_summary} presents a brief summary of the top 10 methods. In this table, we compare the key components and training strategies among 
these methods. 

\subsection{Participants Methods}
Next, we will report the top 10 ranked methods and highlight the key novelty or component of each method.
\subsubsection{A.timi}
\begin{figure}[!htbp]
\centering
\includegraphics[width=1.0\linewidth]{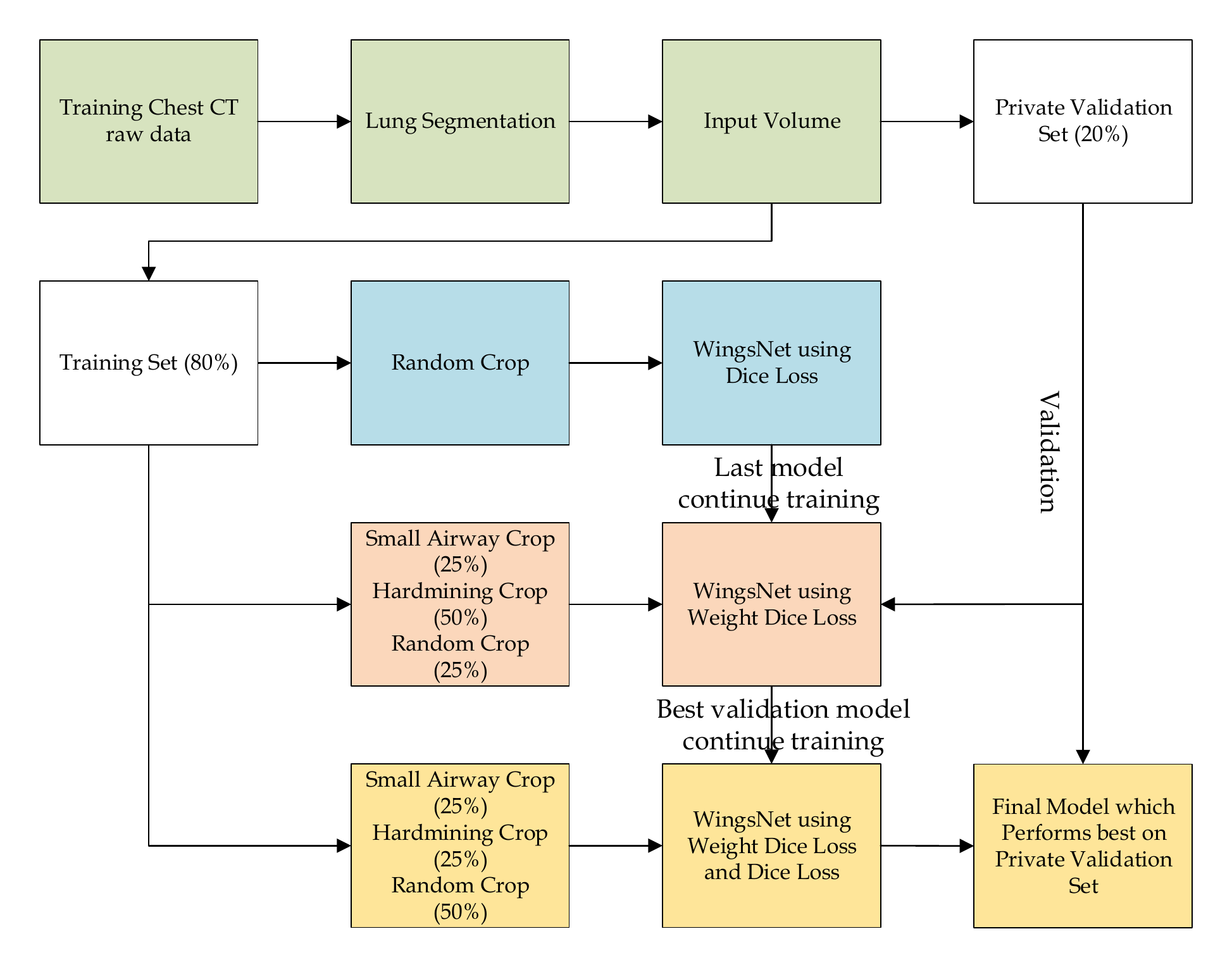}
\caption{The three-stage deep learning pipeline for the airway segmentation by the team of \textit{timi}(T6).}
\label{fig:timi_method}
\end{figure}

The team of timi (T6) proposed a well-designed three-stage deep learning pipeline for the airway segmentation, as seen in Figure \ref{fig:timi_method}. 
The WingsNet~\citep{zheng2021alleviating} was adopted as the backbone architecture. 
In the first stage, the network was trained with only the dice loss and the random crop sampling strategy. 
Their contribution is concentrated on the second stage, where the loss function and training procedure were carefully designed.
Inspired by the local-imbalance-based weight~\citep{zheng2021refined}, 
they designed a variant of the general union loss (GUL)~\citep{zheng2021alleviating}, which adjusted 
the weight factor to focus on the small airways according to the different sizes of branches. They derived the $w_p$ from the local 
foreground rate within the pre-defined neighborhood space.
Furthermore, similar to ~\citep{zheng2021alleviating,zhang2022differentiable}, 
the voxels near the centerline of the airway were assigned more attention. This weight ratio, $w_d$, was defined as inverse square to the 
Euclidean distance from the current position of the voxel to its nearest voxel on the centerline. 
In conclusion, the final weight of each voxel could be defined as $w = w_p + w_d$, and the loss function was defined as below:
\begin{align}
  \mathcal{L(\mathbf{y},\mathbf{\hat{y}})} = 1 - \frac{\sum_{i = 1}^{N}w_i{\mathbf{\hat{y}}_{i}}^{\gamma}\mathbf{y}_{i}}{\sum_{i = 1}^{N}w_i(\alpha\mathbf{\hat{y}}_{i}+\beta\mathbf{y}_{i})},
\end{align}
where the $\gamma, \alpha, \beta$ were set to 0.7, 0.2, and 0.8 respectively.
To improve the efficiency of the training procedure, the small airway over-sampling and the skeleton-based hard-mining were adopted in 
different stages. The small airway over-sampling strategy represents that the cropped patches around the small airways 
(diameter less than 2 pixels) were densely over sampled. The prediction of the first stage was the prerequisite of the skeleton-based hard-mining strategy. 
The misclassified voxels on the skeleton were defined as the hard-mining voxels, from where the cropped patches were densely extracted for training usage.
In the third stage, the variant of GUL, combined with a weighted common Dice loss was adopted to fine-tune the model. 
\uline{The main novelty of T6 method can be summarized as: 1) Adopt the local-imbalance and centerline-distance based weight to dynamically re-weight each voxel.
2) Design the small airway oversampling and skeleton-based hard-mining strategies.}
\begin{figure}[!t]
  \centering
  \includegraphics[width=0.8\linewidth]{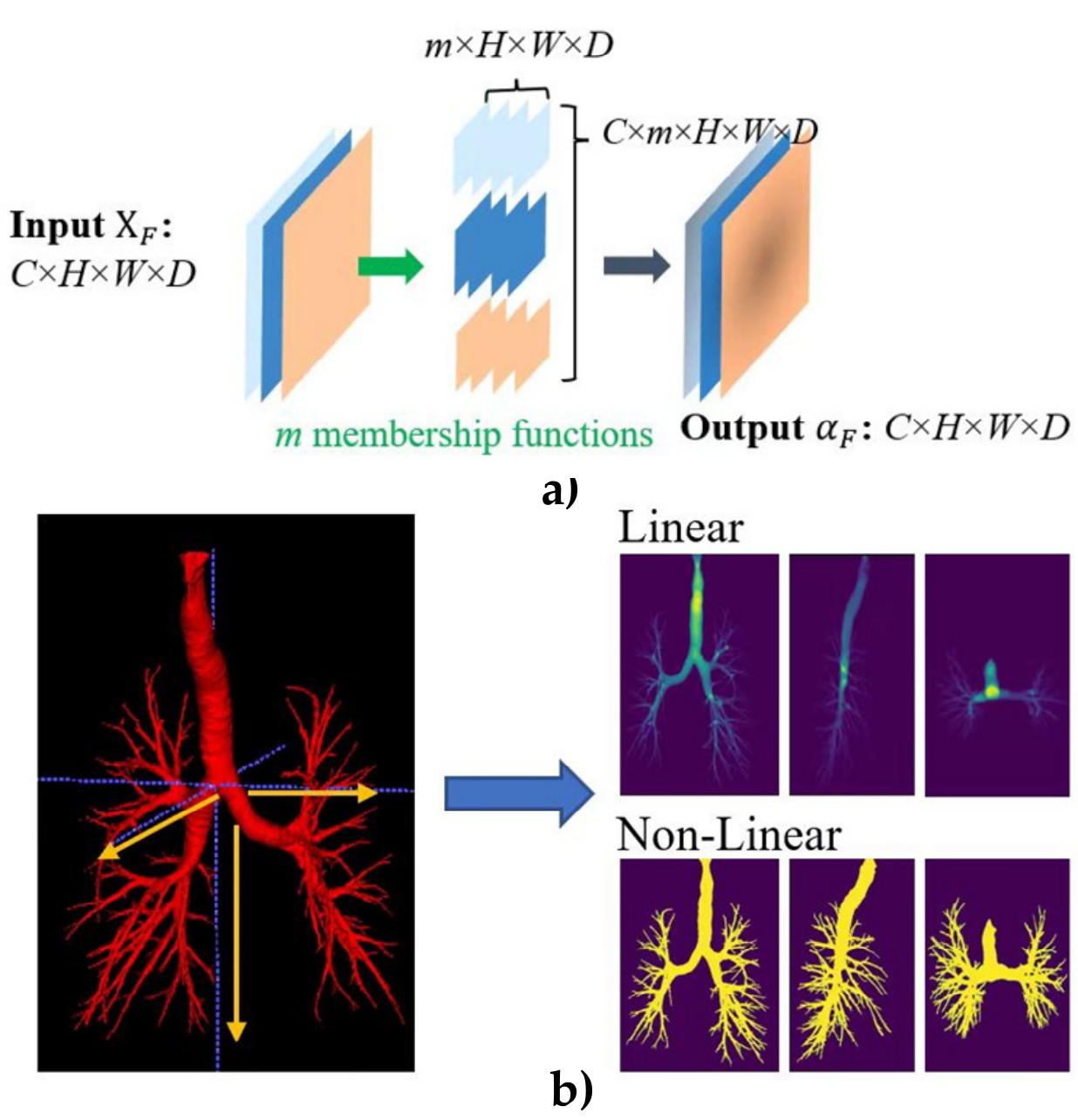}
  \caption{The proposed two modules by the team \textit{YangLab} (T4). a) fuzzy attention gate (FAG). b) Jaccard continuity and accumulation mapping (JCAM) loss.}
  \label{fig:yang-method}
  \end{figure}
 
\subsubsection{B.YangLab}
 
The team of YangLab (T4) designed a novel fuzzy attention gate (FAG) and the Jaccard continuity and accumulation mapping (JCAM) loss for pulmonary airway segmentation. 
The fuzzy attention gate was designed to tackle with the uncertainty of annotations and the inhomogeneous intensity within the airway regions.

They followed the paradigm of the attention gate~\citep{oktay2018attention} while replacing the sigmoid function with the trainable Gaussian membership functions. 
The Gaussian membership functions are favored to specify the deep fuzzy sets due to the smoothness and concise notation.
Moreover, they advocated designing the channel-specific attention gate instead of assigning the same coefficient to all channels that belong to the same spatial 
feature point. This way aimed to extract reliable feature representations in different channels since they are processed by different kernels. Motivated by the 
strength of the uncertainty reduction in original data by fuzzy logic and neural networks~\citep{deng2016hierarchical}, they applied the fuzzy logic with the FAG 
using trainable Gaussian membership functions to assist the neural networks to focus on the regions of interests. The diagram of the FAG is demonstrated in 
Figure \ref{fig:yang-method}.a). Specifically, assume that $\mathcal{X}$ shares the shape of $C \times D \times H \times W$, each feature map was filtered by $M$ 
Gaussian membership functions with the trainable mean $\mu_{m,c}$, and standard deviation $\sigma_{m,c}$:
\begin{align}
  f_{m,c}(\mathcal{X},\mu,\sigma) = e^{\frac{-(\mathcal{X}_{c}-\mu_{m,c})^2}{2\sigma_{m,c}^{2}}},
\end{align}
where $m = 1, 2, ..., M,$ and $c=1, 2, ..., C$. The operator '$\mathrm{OR}$' was adopted to aggregate the fuzzy sets. To guarantee differentiability, they 
used the $\mathrm{max}$ operation instead. The overall fuzzy attention gate upon the $c - th$ channel can be finally derived as:
\begin{align}
  f_{c}(\mathcal{X},\mu,\sigma) = \bigvee_{m = 1}^{M}e^{\frac{-(\mathcal{X}_{c}-\mu_{m,c})^2}{2\sigma_{m,c}^{2}}} = \mathrm{max}(e^{\frac{-(\mathcal{X}_{c}-\mu_{m,c})^2}{2\sigma_{m,c}^{2}}})
\end{align}

The Jaccard continuity and accumulation mapping (JCAM) loss was another contribution that proposed to pay more attention to the continuity of the airway predictions. 
As seen in Figure \ref{fig:yang-method}.b), The JCAM estimated two topological types of errors between the prediction and the ground-truth. The first is the projection error, which was executed through the 
coronal, sagittal, and axial planes. The second error, termed $\mathcal{L_{C}}$ measures the difference of centerlines extracted from the prediction and the ground-truth, respectively. 
The projection error was split into two parts, the linear accumulation maps (LAM), and the nonlinear transformation of the linear accumulation maps (nLAM) performed by 
the $\mathrm{tanh}$ operation. The overall loss function they used can be summarized as: 
\begin{align}
  \mathcal{L}(\mathbf{x},\mathbf{y}) = \alpha\mathcal{L}_{J}(\mathbf{x},\mathbf{y}) + \beta\mathcal{L}_{C}(\mathbf{x},\mathbf{y}) + \varphi\mathcal{L}_{CE}(\mathbf{x},\mathbf{y}) + \notag \\
  \gamma\mathcal{L}_{LAM}(\mathbf{x},\mathbf{y}) + \delta\mathcal{L}_{nLAM}(\mathbf{x},\mathbf{y}),
\end{align}
where the $\mathcal{L}_{J}$ denotes the Jaccard loss function, the $\mathcal{L}_{CE}$ denotes the Cross-Entropy loss function. The $\alpha, \beta, \gamma$ was set to 1, and 
the $\varphi, \delta$ was set to 0.3 in all experimental settings. In addtion, they adopted the region growing method to fine-tune the trachea part. 
\uline{In conclusion, the main novelty of T4 method lies in the channel-specific fuzzy attention layer and the JCAM loss that designed to enhance the continuity of airways.}

\subsubsection{C.deeptree\_damo}
\begin{figure}[!htbp]
\centering
\includegraphics[width=1.0\linewidth]{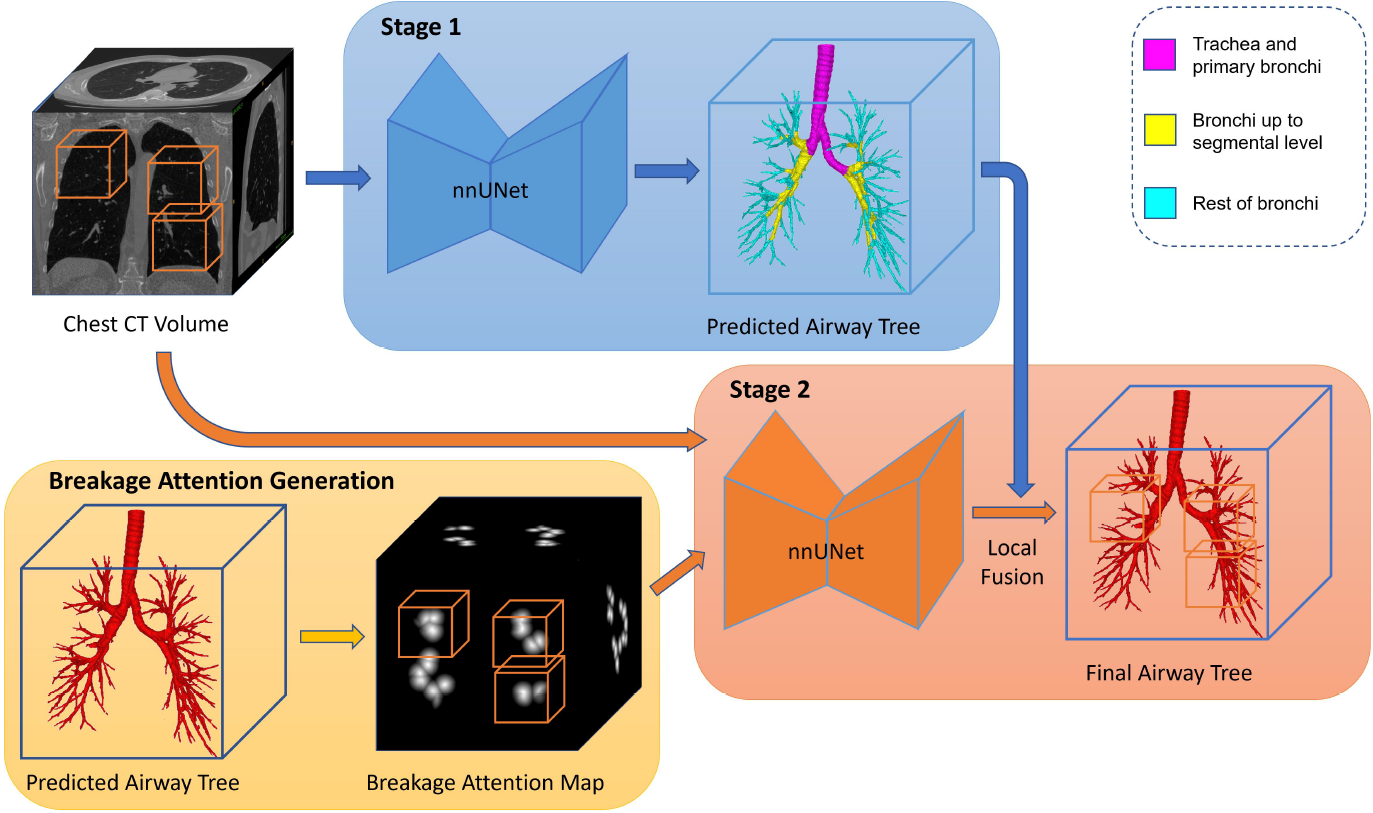}
\caption{Overall workflow of the proposed two-stage airway segmentation by team \textit{deeptree} (T14). The coarse airway is first extracted by the anatomy-aware deep network. Secondly, the breakage 
map is calculated by the morphological operations to connect the breaking branches.}
\label{fig:puyang-method}
\end{figure}

The team of deeptree\_damo (T14) proposed a two-stage framework for airway segmentation, as demonstrated in Figure \ref{fig:puyang-method}. 
In the first stage, to tackle the intra-class imbalance 
between the different levels of airway branches, they formulated the binary segmentation task to the multi-class segmentation 
task in accordance with the airway branch size.
Specifically, they preliminarily decomposed the ground-truth of the pulmonary airway label into three classes: 1) The trachea 
and two main bronchi are classified as large-level airways, $\mathcal{Y^{L}}$. 2) From the bronchial up to the segmental airways are 
considered as the middle-level airways, $\mathcal{Y^{M}}$. 3) The rest of the peripheral airways, whose average lumen 
diameter $<$ 2 mm, are small-level airways, $\mathcal{Y^{S}}$. The anatomy-aware multi-class (AMC) airway segmentation was formulated 
as follows:
\begin{align}
  \mathcal{L}_{first}(\mathcal{\hat{Y}},\mathcal{Y}) = \mathcal{L}(\mathcal{\hat{Y^{L}}},\mathcal{Y^{L}}) + \mathcal{L}(\mathcal{\hat{Y^{M}}},\mathcal{Y^{M}}) +\mathcal{L}(\mathcal{\hat{Y^{S}}},\mathcal{Y^{S}}),
\end{align}
where they applied the general union loss function~\citep{zheng2021alleviating} in the AMC framework. The AMC framework assisted in 
explicitly differentiating the anatomic context of different branches in the model training procedure. Thus, each class 
owned a distinguished airway branch size range and the class-specific features could be naturally learned.

Secondly, to deal with the breakage phenomenon that happened in the first stage, they calculated the breakage attention maps and simulated 
the domain-specific breakage training data. These preparations aimed to accomplish the deep breakage connection. 
The breakage attention map, termed $\mathcal{H}$, was designed to highlight the breaking area 
via the second-shortest distance calculation between background points to all separate connected components in a prediction. 
$\mathcal{H}$ was further normalized by the parameterized Sigmoid function $\overline{\mathcal{H}} = \mathrm{Sigmoid}(5-\mathcal{H})$, where 
$\overline{\mathcal{H}}$ formed a normal 3D ball-like intensity distribution at a breakage location. Further, the domain-specific 
breakage simulation was performed to acquire sufficient breakage condition data $\mathcal{Y^{B}}$ from the ground-truth 
for the 2nd-stage breakage-connection network training. This network was fed with the fusion of $\mathcal{X}$ and $\overline{\mathcal{H}}$ 
and predict the breakage $\mathcal{\hat{Y^{B}}}$:
\begin{align}
  \mathcal{\hat{Y^{B}}} &= \mathcal{F(\mathcal{X},\overline{\mathcal{H}}; \textbf{W})},\\
  \mathcal{L}_{second} &= \mathcal{L}(\mathcal{\hat{Y^{B}}},\mathcal{Y^{B}}),
\end{align}
where the $\mathcal{F(\cdot)}$ and $\textbf{W}$ denote the 2nd-stage breakage-connection network and the corresponding network parameters, respectively. 
Finally, the output of the 1st- and 2nd-stage are merged to generate the whole airway tree prediction.
\uline{In summary, the breakage-connection network based on breakage attention maps is the main novelty of T14 method.}

\subsubsection{D.neu204}
\begin{figure}[!htbp]
\centering
\includegraphics[width=1.0\linewidth]{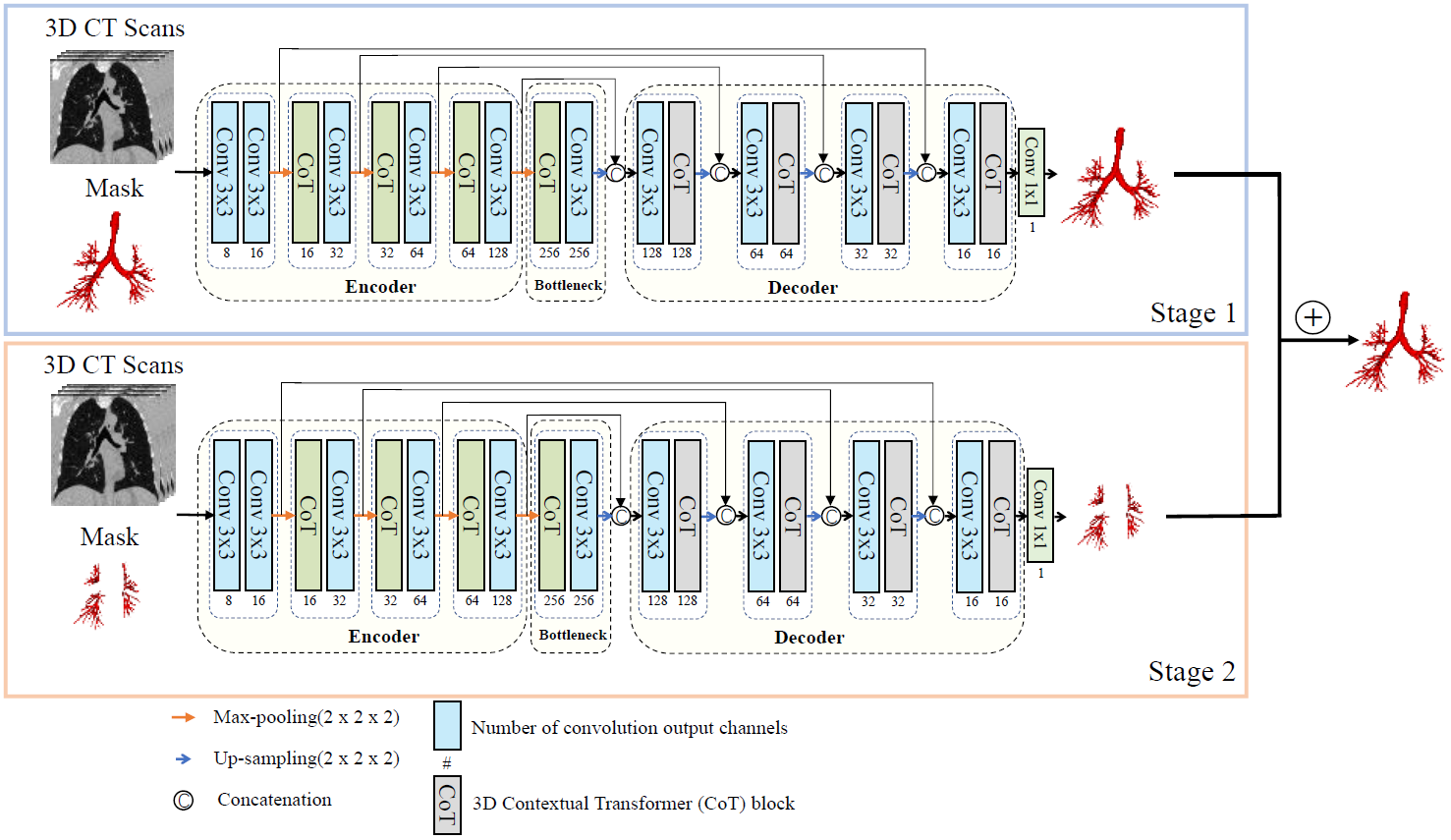}
\caption{The proposed two-stage network for the airway segmentation by team \textit{neu204} (T7). The first stage trains the whole airway while the second stage refine the airways inside the lungs. 
The CoT refers to the contextual transformer block.}
\label{fig:yanan-method}
\end{figure}

The team of neo204 (T7) developed a two-stage network for airway segmentation as described in Figure \ref{fig:yanan-method}. 
In stage 1, the 3D computed tomography (CT) scans and the full airway annotation was fed into the proposed network, 
and the 3D computed tomography (CT) scans and partial intra-pulmonary airway annotation were fed in stage 2. 
Then the results of the two stages were merged as the final prediction.
3D UNet~\citep{3DU-Net} was chosen as the basic neural network architecture in both two stages. They replaced one of 
the $3 \times 3 \times 3$ convolutional kernel layers with the emerging contextual transformer (CoT)~\citep{li2022contextual} 
module in both encoder and decoder parts. The design of the CoT capitalizes on the contextual information among input keys to 
guide the learning of the dynamic attention matrix and thus strengthens the capacity of visual representation. 
The CoT module aimed to exploit the rich contexts among the neighbor keys, which is beneficial to highlighting 
the topological connection in the airway tree structure.
\uline{In conclusion, the independent processing of airways based on their locations and introduction of the CoT are the main contribution.}

\subsubsection{E.Sanmed\_AI}
The team of Sanmed\_AI (T1) designed a modified attention UNet for pulmonary airway tree modeling. First, a channel- and spatial-wise 
attention module, the Project \& Excite (PE)~\citep{rickmann2019project} module was embedded into each layer, following the common convolution operations. 
PE squeezes the feature maps along different axes of slices separately to retain more spatial information rather than perform global average pooling. 
The extracted spatial information is further used in the excitation step. It helps the network to learn the important feature information of the airways 
and improve the generalization ability of the model. 

Secondly the coordinate attention mechanism was applied on the last decoder layer. It recorded the local information 
of its corresponding patch in the whole image. Due to the GPU memory limit, 3D CT images were cropped into sub-volumes as model 
inputs, and such patch-based training strategy caused a loss of position and context information. 
The coordinate map was introduced to make up such information loss. It was inserted to the high-dimension feature maps of the last decoder 
because they share the same spatial dimension. Similar to other airway segmentation works~\citep{qin2020learning,zhang2021fda,zhang2022cfda}, 
the Dice with Focal loss was applied in all experiments.
\uline{To sum up, the fusion of attention map and the coordinate is the key design of T1 method.}

\makeatletter
\def\hlinew#1{%
\noalign{\ifnum0=`}\fi\hrule \@height #1 \futurelet
\reserved@a\@xhline}
\makeatother
\begin{table*}[thbp]
\renewcommand\arraystretch{2.0}
\centering
\caption{\justify{Brief summary and comparison of the top 10 methods. It includes the concise description of the method and the training strategy. 
The order of ROI size is (depth, height, width).}}\label{tab:method_brief_summary}
\scalebox{0.5}{
\begin{threeparttable}
\begin{tabular}{lll}
\hlinew{1pt}
\multicolumn{1}{c}{\textbf{Team name}}& \multicolumn{1}{c}{\textbf{Main novelty/contribution of the method}} & \multicolumn{1}{c}{\textbf{Training strategy}}\\ \hline
timi, T6 & \makecell[l]{$\bigcdot$ Use WingsNet~\citep{zheng2021alleviating} as the backbone, adopt the local-imbalance-based~\citep{zheng2021refined} and \\ 
the centerline-distance-based weight~\citep{zheng2021alleviating,zhang2022differentiable} to dynamically re-weight each voxel. \\
$\bigcdot$ Design the small airway oversampling and skeleton-based hard-mining strategies.}& 
\makecell[l]{$\bigcdot$ ROI size: 128 $\times$ 128 $\times$ 128. Batchsize is set to 24. \\ AdamW optimizer with learning rate 0.0001 is used.\\ 
$\bigcdot$ Dice loss and random crop were used in the training stage 1 for \\
100 epochs. The variant of the GUL and the designed sampling \\
strategies were adopted in the training stage 2 for 50 epochs. The \\ 
training stage 3 continued for 30 epochs with the combination of \\
the variant of GUL and 0.5 * Dice loss.} \\ \hline
YangLab, T4 & \makecell[l]{$\bigcdot$ Take the Attention U-Net~\citep{oktay2018attention} structure as the backbone, propose the 
\\channel-specific fuzzy attention layer combined with fuzzy logic.\\ 
$\bigcdot$ The JCAM loss is proposed to enhance the continuity and completeness of airways. \\ Correspondingly, a CCF-score is designed for the measurement.}& 
\makecell[l]{$\bigcdot$ Adopt the average size of the 3D minimum bounding box of \\ the ground-truth as the patch size. Total epoch is set to 200.\\
Initial learning rate is 0.001 and a decay of 0.5 at the 20th, \\ 50th, 80th, 110th and 150th epoch. \\
$\bigcdot$ The online smart patch sampling strategy is used in the \\ training procedure. It ensures the cropped patches \\ own enough centerline or foreground voxels.} \\ \hline
deeptree\_damo, T14 & \makecell[l]{$\bigcdot$ Take the nnU-Net~\citep{isensee2021nnu} as backbones for both two stages. Formulate the \\ 
anatomy-aware multi-class segmentation task for airways that share \\ large context variation of different branches.\\ 
$\bigcdot$ Introduce a breakage attention map that highlights the breaking regions. \\ Train a breakage-connection network with the simulated data.}& 
\makecell[l]{$\bigcdot$ A modified nnUNet is adopted. Reduce the downsampling \\ operation to 3 times, and enlarge the width of the convolutional \\ 
layers at the deeper blocks to increase capacity.\\
$\bigcdot$ A linear time distance transform algorithm~\citep{maurer2003linear} is adopted to \\ calculate the breakage attention map. 
The curve skeleton and \\ skeleton-to-volume propagation algorithm~\citep{jin2016robust} is applied to 
\\ create simulated training samples for deep breakage connection.} \\ \hline
neu204, T7 & \makecell[l]{$\bigcdot$ Take the 3D-UNet~\citep{3DU-Net} as basic architectures for both two stages.\\
$\bigcdot$ The 1st stage processes the full airway tree while the 2nd stage only handle \\
the airway inside the lungs. The contextual transformer (CoT)~\citep{li2022contextual} module \\ is embedded in both encoders and decoders.}& 
\makecell[l]{$\bigcdot$ ROI size: 64 $\times$ 192 $\times$ 192 and 64 $\times$ 128 $\times$ 128 for the 1st- and \\ 
2nd- stage training, respectively. The CoT replaces one of the \\ 3 $\times$ 3 $\times$ 3 convolution, followed by the IN~\citep{ulyanov2016instance} and ReLU~\citep{agarap2018deep}.\\
$\bigcdot$ Adam optimizer with an initial learning rate of 0.01 is adopted. \\ Exponential decay solution (rate:0.9) is used after each epoch.} \\ \hline
Sanmed\_AI, T1 & \makecell[l]{$\bigcdot$ Take the 3D-UNet~\citep{3DU-Net} as the backbone and add the attention mechanisms.\\
$\bigcdot$ Project \& Excite (PE)~\citep{rickmann2019project} module was embedded into each layer to \\ recalibrate feature maps. The coordinate map is 
applied to compensate for \\ the information loss due to the patch-wise training procedure.}& 
\makecell[l]{$\bigcdot$ ROI size: 128 $\times$ 128 $\times$ 128. The coordinate map is normalized 
\\to [-1, 1] in three axes. The warm-up cosine annealing learning 
\\ strategy is used. Learning rate ranges from 1e-5 to 0.01 with \\ cycle period of 20 epochs and decay ratio is 0.5.\\
$\bigcdot$ Choose Dice with Focal loss function in all experiments.} \\ \hline
dolphins, T5 & \makecell[l]{$\bigcdot$ Take the 3DResNet~\citep{tran2018closer} as the backbone. The deep supervision is introduced\\ 
to generate the aggregated loss. Residual block is inserted via the skip connection.\\
$\bigcdot$ Use nnUNet to perform one-fold cross-validation of training volumes \\ with ground-truth and validation volumes with pseudo labels.}& 
\makecell[l]{$\bigcdot$ ROI size: 16 $\times$ 256 $\times$ 256. Batchsize is set to 2. The learning \\ 
rate of 0.0004 with Adam optimizer is used. Total epoch is \\ set to 200 with an early stop solution of 20 epochs.} \\ \hline
suqi, T17 & \makecell[l]{$\bigcdot$ Take the nnUNet~\citep{isensee2021nnu} as backbone with the introduction of the dense block~\citep{huang2017densely}.\\
$\bigcdot$ Adopt some immediate results to conduct deep supervision that \\ is conducive to the convergence of the network.}& 
\makecell[l]{$\bigcdot$ ROI size: 96 $\times$ 160 $\times$ 160. Batchsize 2. Resample pixel spacing.\\
Total epoch is 1000, SGD optimizer with the initial learning rate \\ 
0.01 is used. The weighted Dice and binary cross entropy loss is \\ applied in all experiments. } \\ \hline
notbestme, T20 & \makecell[l]{$\bigcdot$ Adopt the transformer structure as backbone. To reduce \\ 
computational cost, a 2.5 D Compute-cheap Gated Global Attention is designed.\\
$\bigcdot$ A multi-resolution network is designed to enhance multi-scale mining ability.}& 
\makecell[l]{$\bigcdot$ ROI size: 32 $\times$ 160 $\times$ 160. The final 5362 image patches are \\ 
extracted for training. Batchsize is 2 and total epoch is 50. \\
$\bigcdot$ CE loss for the first 5 epochs and then use pixel-wise weighted \\ 
CE loss derived from categorical information distribution. \\ Adam optimizer with a learning rate of 0.0005 is used for training.} \\ \hline
lya, T9 & \makecell[l]{$\bigcdot$ Take the nnUNet~\citep{isensee2021nnu} as the backbone. Besides original data \\ augmentation, Elastic 
and brightness transformation are introduced.\\ $\bigcdot$
$\bigcdot$ The small branches receive more attention in the sampling procedure, \\ and a combination of TopK and Dice loss 
are designed to conduct hard mining.}& 
\makecell[l]{$\bigcdot$ The Adam optimizer with an initial learning rate of 3e-4 is used. \\ 
Total epoch is 1000 and Batchsize is set to 2.\\ 
$\bigcdot$ The TopK loss is intractable, thus the combined loss function \\ 
is only used to fine-tune the network.} \\ \hline
dnai, T10 & \makecell[l]{$\bigcdot$ Apply the 3D UNet~\citep{3DU-Net} for coarse segmentation and \\ Attention UNet~\citep{oktay2018attention} for the refine usage. }& 
\makecell[l]{$\bigcdot$ ROI size: 96 $\times$ 160 $\times$ 160 for the coarse stage, \\ 48 $\times$ 80 $\times$ 80 for the refining stage.\\ 
$\bigcdot$ 75\% labeled patches and 25\% random patches are sampled for the \\ coarse stage training while 25\% random patches and 75\% \\ patches contained 
peripheral airways are extracted for fine part.} \\
\hlinew{1pt}
\end{tabular}
\end{threeparttable}}
\end{table*}

\subsubsection{F.dolphins}
The team of dolphins (T5) proposed a 3DResNet~\citep{tran2018closer} with deep supervision model for the segmentation of pulmonary airways.
The convolutional block consisted of convolutional layers with Batch Normalization~\citep{ioffe2015batch} and ReLU activation function to extract the different 
feature maps from each block on the encoder side. The residual block was inserted at each encoder block with skip connection. 
The feature concatenation was executed at each encoder and decoder block except the last 1 $\times$ 1 convolutional layer. 
The three-level deep-supervision technique was applied to generate the aggregated loss between ground-truth and prediction. 
In addition, they used the nnUNet for one-fold cross-validation of training volumes with ground-truth and validation volumes with pseudo labels.
\uline{In summary, the introduction of deep supervision and the leverage of pseudo labels are the key components of T5 method.}

\subsubsection{G.suqi}
The team of suqi (T17) proposed a Dense-UNet based on the nnUNet for airway segmentation. As the airway is the fine-grained structure, 
to prevent the network from losing too much information during upsampling and downsampling, they used transposed convolution to realize upsampling, 
and used convolution with a step size of 2 to realize downsampling. Further, to enhance the feature embedding and alleviate the feature forgetting, 
the dense block~\citep{huang2017densely} was introduced into the nnUNet. Specifically, the encoder mapped the features into the hidden space with 
the size of (6,5,5) and all 1 $\times$ 1 $\times$ 1 convolutions in each dense block had 256 channels.
They still used the output results of the encoder except the lowest two layers to predict the probability map and obtain the supervision signal, 
which is conducive to the convergence of the network.
\uline{In conclusion, the integration of the dense block into nnUNet is the key solution proposed by T17 method.}

\subsubsection{H.notbestme}
The team of notbestme (T20) developed a multi-resolution network for airway segmentation. 
It implemented a three-axis fusion, computationally inexpensive self-attention mechanism. The multi-resolution network was designed to 
enhance the multi-scale mining ability of the model and adapt to the segmentation task of different objects due to the significant difference between 
the trachea and small airways. Specifically, they used the interpolation algorithm to resize the original input to different resolution sizes and fed
them into subnetworks whose weights are not shared. 

As transformers had expanded into the field of computer vision~\citep{dosovitskiy2020image, liu2021swin}, 
the shortcomings of CNNs in capturing global dependencies have been paid more and more attention by researchers. However, it is impractical to 
directly transfer current transformer structures to the volumetric medical images due to the limitation of computational resources. To deal with this 
problem, they designed the 2.5D Compute-cheap Gated Global Attention for 3D medical images. The self-attention calculation among three 
matrices (Q,K,V) followed the standard criterion~\citep{vaswani2017attention} while they adopted the $\mathrm{Pooling}$ operation to reduce feature dimension. 
In addition, they used the attention map to enhance the expression of the Value matrix (V) via the Gated Linear Unit~\citep{dauphin2017language} mechanism.
\uline{The main novelty of T20 method is the proposed 2.5D compute-cheap gated global attention that introduced into the transformer.}

\subsubsection{I.lya}
The team of lya (T9) applied an improved nnUNet for the airway segmentation. More data augmentation, a specified voxel sampling strategy, and a modified loss function 
were incorporated into the nnUNet to improve the segmentation performance of the small peripheral bronchi. Besides the transformation by nnUNet config, they 
adopted the elastic transformation and brightness transformation to conduct data augmentation. Further, they replaced the percentage clipping with a fixed CT window. 
The window was set to [-1200,600], and the maximum HU value was randomly selected from 400 to 600 for data augmentation in the training stage.

To handle the intra-class imbalance problem of the airways, they made efforts from two aspects. For one thing, they discarded the random sample solution and 
located the sampling central points more on the small branches. For another, the deep neural networks intend to fit the major class, thus the small peripheral 
bronchi are easily missed. They applied a combination of the TopK loss function and the dice loss function:
\begin{align}
  \mathcal{L(\mathbf{y},\mathbf{\hat{y}})} &= \mathcal{L(\mathbf{y},\mathbf{\hat{y}})}_{TopK} + \mathcal{L(\mathbf{y},\mathbf{\hat{y}})}_{dice},\\
  \mathcal{L(\mathbf{y},\mathbf{\hat{y}})}_{TopK} &= -\frac{1}{K}\sum_{i = 1}^{K}\mathbf{y}_{i}log(\mathbf{\hat{y}}_{i})+(1-\mathbf{y}_{i})log(1-\mathbf{\hat{y}}_{i}),
\end{align}
where the TopK loss aimed to force the network to focus on the hard samples. 
\uline{In summary, the key design of T9 method lies in paying more attention on small branches in the sampling procedure and the compound loss function.}

\subsubsection{J.dnai}
The team of dnai (T10) designed a two-stage coarse-to-fine framework for airway segmentation. In the coarse stage, they chose the 3D UNet~\citep{3DU-Net} as 
the backbone and used the Instance Normalizaiton~\citep{ulyanov2016instance}, ConvTranspose operation instead of the original components. The patch size was 
96 $\times$ 160 $\times$ 160 for the coarse stage training, with 75\% of patches labeled and 25\% patches sampled randomly. 

In the refining stage, they adopted a relatively shallow network based on the Attention UNet~\citep{oktay2018attention}. The patch size also a shared smaller scale, 
48 $\times$ 80 $\times$ 80. However, These patches were sampled with 25\% randomly and 75\% contained high-level airway branches, which demonstrated that in 
the refining stage, they aimed to increase the segmentation performance of the peripheral airways. 
The combination of cross entropy and Dice loss was used in all experiments.
\uline{The two-stage coarse-to-fine framework is the main design of T10 method.}

\subsection{Consensus on Effective Methods}\label{consensus_of_effective_methods}
After introducing the main contributions on individual methods, we conclude some consensus of effective methods 
to deal with the challenges of pulmonary airway segmentation.  

\textbf{Solution 1: Multi-stage Solution (S1).} The multi-stage training pipeline has demonstrated the advantage of pulmonary 
airway segmentation. First, the lung region extraction is a simple yet effective hard attention mechanism to 
focus on related regions, which can deal with the leakage challenge (C1). Secondly, the initial training stage can obtain the 
preliminary predictions, which provide useful information for the following training stage to acquire a more complete airway 
tree structure, such as hard sample mining (T6) and breakage attention map calculation (T14).  

\textbf{Solution 2: Improve Intra-class Discrimination (S2).} Improving the intra-class discrimination ability is a reasonable 
choice to tackle the C2, breakage challenge. The extra information can be extracted from the CT scans and binary airway annotation, 
such as the centerline points (T4, T6, T14), radius (T14), and spatial location (T7) of the branches. 
These additional knowledges can be leveraged from several aspects to improve intra-class discrimination: 
1) Over-sampling. T4 proposed a smart patch sampling strategy to put more emphasis on peripheral airways 
based on the centerline points ratio. 2) Differentiate the training procedure between airway branches. 
T14 formulated a multi-class task between the different levels 
of airway branches, hence, the multi-level discriminative features are extracted from different branches of airways.
T7 designed a two-stage framework, the first stage was for the whole airway tree segmentation while the second stage 
was only trained with the partial intra-pulmonary airways. 

\textbf{Solution 3: Novel Objective Functions (S3).} Designing novel loss functions that emphasize topology 
completeness and topology correctness is beneficial to deal with the challenge of Leakage (C1) and Breakage (C2), 
Robustness and Generalization (C3). For example, T4 proposed a JCAM loss function that focuses on topological errors.
The JCAM measures the projection error and centerline detected ratio error. T14 proposed the breakage attention map to 
construct the objective function used in the breakage-connection network. T6 adopted the variant of the general union loss to 
force the network to enjoy superiority to the continuity. The objective functions that pay attention to the topology 
could harness the high-level feature of an airway tree structure, which may improve the robustness and generalization ability 
of the algorithms.

\makeatletter
\def\hlinew#1{%
\noalign{\ifnum0=`}\fi\hrule \@height #1 \futurelet
\reserved@a\@xhline}
\makeatother
\begin{table*}[thbp]
\renewcommand\arraystretch{1.4}
\centering
\caption{\justify{Quantitative results on the \textbf{validation set} of ATM'22 challenge achieved by participants. 
The results are reported in the format of mean $\pm$ standard deviation. In the validation set, The number of sample CT scans is 50, 
and the total number of airway branches is 212.48$\pm$52.10. The results are first post-processed by the largest component extraction 
and eventually reported by the Tree length detected rate (i.e., TD, \%), Branch detected rate (i.e., BD, \%), Dice Similarity Coefficient (i.e., DSC, \%),
Precision (\%), Sensitivity (i.e., Sen, \%) and Specificity (i.e., Spe, \%). }}\label{tab:validation-phase-1-result}
\scalebox{0.9}{
\begin{threeparttable}
\begin{tabular}{lcccccc}
\hlinew{1.25pt}
\textbf{Team name}     & \textbf{TD (\%) $\uparrow$}            & \textbf{BD (\%) $\uparrow$}            & \textbf{DSC (\%) $\uparrow$ }         & \textbf{Precision (\%) $\uparrow$}     & \textbf{Sen (\%) $\uparrow$}  & \textbf{Spe (\%) $\uparrow$}   \\ \hline
Sanmed\_AI (V1)    & 89.874\scriptsize{$\pm$6.609} & 85.102\scriptsize{$\pm$10.085} & \underline{95.555\scriptsize{$\pm$1.376}} & 95.551\scriptsize{$\pm$2.385} & \underline{95.644\scriptsize{$\pm$2.483}} & 99.986\scriptsize{$\pm$0.008} \\
YangLab (V2)          & 94.406\scriptsize{$\pm$3.798} & 91.302\scriptsize{$\pm$6.439}  & \textbf{95.926\scriptsize{$\pm$1.249}} & \underline{97.180\scriptsize{$\pm$1.941}} & 94.766\scriptsize{$\pm$2.270} & 99.991\scriptsize{$\pm$0.006} \\
notbestme (V3)      & 85.756\scriptsize{$\pm$7.560} & 79.181\scriptsize{$\pm$11.514} & 95.212\scriptsize{$\pm$1.893} & 95.706\scriptsize{$\pm$1.986} & 94.824\scriptsize{$\pm$3.511} & 99.987\scriptsize{$\pm$0.007} \\
suqi (V4) & 80.680\scriptsize{$\pm$7.475} & 70.555\scriptsize{$\pm$10.284} & 94.713\scriptsize{$\pm$1.178} & 96.191\scriptsize{$\pm$1.431} & 93.302\scriptsize{$\pm$1.705} & 99.989\scriptsize{$\pm$0.004} \\
cvhthreedee (V5) & 87.856\scriptsize{$\pm$8.482} & 80.291\scriptsize{$\pm$13.996} & 94.939\scriptsize{$\pm$1.597} & 96.839\scriptsize{$\pm$1.948} & 93.188\scriptsize{$\pm$2.890} & 99.990\scriptsize{$\pm$0.006} \\
LinkStartHao (V6) & 89.764\scriptsize{$\pm$7.611} & 83.439\scriptsize{$\pm$12.780} & 94.392\scriptsize{$\pm$1.674} & 95.758\scriptsize{$\pm$2.091} & 93.140\scriptsize{$\pm$2.860} & 99.987\scriptsize{$\pm$0.007} \\
neu204 (V7) & 94.441\scriptsize{$\pm$4.008} & 92.279\scriptsize{$\pm$5.987} & 95.800\scriptsize{$\pm$1.142} & 93.451\scriptsize{$\pm$1.929} & 98.302\scriptsize{$\pm$1.221} & 99.979\scriptsize{$\pm$0.007} \\
miclab (V8) & 82.865\scriptsize{$\pm$5.861} & 74.223\scriptsize{$\pm$40.124} & 95.501\scriptsize{$\pm$1.007} & 96.557\scriptsize{$\pm$1.575} & 94.507\scriptsize{$\pm$1.785} & 99.990\scriptsize{$\pm$0.005} \\
blackbean (V9) & 89.422\scriptsize{$\pm$7.675} & 83.210\scriptsize{$\pm$12.401} & 94.554\scriptsize{$\pm$1.778} & 95.461\scriptsize{$\pm$2.219} & 93.730\scriptsize{$\pm$2.747} & 99.986\scriptsize{$\pm$0.008} \\
Median (V10) & 88.765\scriptsize{$\pm$7.669} & 82.441\scriptsize{$\pm$12.040} & 94.667\scriptsize{$\pm$1.699} & 95.642\scriptsize{$\pm$2.088} & 93.769\scriptsize{$\pm$2.579} & 99.987\scriptsize{$\pm$0.007} \\
lya (V11)  & 89.613\scriptsize{$\pm$ 7.395} & 83.583\scriptsize{$\pm$12.306} & 94.371\scriptsize{$\pm$1.566} & 95.146\scriptsize{$\pm$2.661} & 93.689\scriptsize{$\pm$2.339} & 99.985\scriptsize{$\pm$0.009} \\
satsuma (V12) & 89.783\scriptsize{$\pm$7.734} & 83.571\scriptsize{$\pm$12.822} & 94.649\scriptsize{$\pm$1.588} & 95.574\scriptsize{$\pm$2.233} & 93.817\scriptsize{$\pm$2.623} & 99.986\scriptsize{$\pm$0.008} \\
ailab (V13)    & 90.989\scriptsize{$\pm$6.912} & 86.102\scriptsize{$\pm$11.011} & 94.624\scriptsize{$\pm$1.737} & 95.211\scriptsize{$\pm$2.334} & 94.111\scriptsize{$\pm$2.572} & 99.985\scriptsize{$\pm$0.008} \\
timi (V14)   & \underline{95.866\scriptsize{$\pm$3.366}}  & \underline{94.921\scriptsize{$\pm$4.399}} & 93.987\scriptsize{$\pm$2.337} & 94.041\scriptsize{$\pm$2.837} & 94.008\scriptsize{$\pm$3.075} & 99.981\scriptsize{$\pm$0.010} \\
sen (V15)   & 89.295\scriptsize{$\pm$7.587} & 83.269\scriptsize{$\pm$12.300} & 93.614\scriptsize{$\pm$2.306} & 95.000\scriptsize{$\pm$2.473} & 92.386\scriptsize{$\pm$3.879} & 99.985\scriptsize{$\pm$0.009} \\
MibotTeam (V16)    & 89.293\scriptsize{$\pm$8.577} & 81.184\scriptsize{$\pm$13.734} & 94.772\scriptsize{$\pm$1.275} & 95.596\scriptsize{$\pm$2.353} & 94.044\scriptsize{$\pm$2.354} & 99.986\scriptsize{$\pm$0.008} \\
CITI-SJTU (V17)   & 91.840\scriptsize{$\pm$6.000} & 87.239\scriptsize{$\pm$9.724} & 92.943\scriptsize{$\pm$1.595} & 91.132\scriptsize{$\pm$2.526} & 94.891\scriptsize{$\pm$1.981} & 99.971\scriptsize{$\pm$0.011} \\
SEU (V18) & 84.704\scriptsize{$\pm$10.573} & 76.672\scriptsize{$\pm$16.599} & 93.776\scriptsize{$\pm$1.885} & 95.240\scriptsize{$\pm$2.561} & 92.468\scriptsize{$\pm$3.325} & 99.985\scriptsize{$\pm$0.009} \\
deeptree\_damo (V19) & \textbf{97.369\scriptsize{$\pm$2.957}} & \textbf{96.717\scriptsize{$\pm$3.711}} & 92.812\scriptsize{$\pm$1.488} & 87.324\scriptsize{$\pm$2.652} & \textbf{99.090\scriptsize{$\pm$0.449}} & 99.955\scriptsize{$\pm$0.013} \\
CBT\_IITDELHI (V20) & 73.928\scriptsize{$\pm$10.847} & 65.672\scriptsize{$\pm$12.250} & 94.336\scriptsize{$\pm$1.897} & 97.127\scriptsize{$\pm$1.505} & 91.812\scriptsize{$\pm$3.768} & \underline{99.992\scriptsize{$\pm$0.004}} \\
dolphins (V21) & 83.478\scriptsize{$\pm$12.616} & 77.496\scriptsize{$\pm$15.540} & 93.228\scriptsize{$\pm$2.160} & 95.961\scriptsize{$\pm$1.652} & 90.814\scriptsize{$\pm$4.478} & 99.988\scriptsize{$\pm$0.006} \\
bms410 (V22)    & 60.705\scriptsize{$\pm$8.762} & 46.924\scriptsize{$\pm$6.555} & 88.394\scriptsize{$\pm$2.562} & \textbf{99.515\scriptsize{$\pm$1.464}} & 79.635\scriptsize{$\pm$4.252} & \textbf{99.999\scriptsize{$\pm$0.004}} \\
airwayseg (V23)   & 78.128\scriptsize{$\pm$16.110} & 72.714\scriptsize{$\pm$16.955} & 92.828\scriptsize{$\pm$3.555} & 95.136\scriptsize{$\pm$2.000} & 90.939\scriptsize{$\pm$6.606} & 99.985\scriptsize{$\pm$0.007} \\
atmmodeling2022 (V24)   & 63.540\scriptsize{$\pm$20.759} & 56.034\scriptsize{$\pm$21.56} & 92.416\scriptsize{$\pm$4.053} & 97.156\scriptsize{$\pm$2.503} & 88.556\scriptsize{$\pm$7.644} & 99.992\scriptsize{$\pm$0.008} \\
bwhacil (V25)    & 72.524\scriptsize{$\pm$9.621} & 58.391\scriptsize{$\pm$9.181} & 87.628\scriptsize{$\pm$2.105} & 83.076\scriptsize{$\pm$3.826} & 92.963\scriptsize{$\pm$3.532} & 99.941\scriptsize{$\pm$0.018} \\
dnai (V26)  & 87.596\scriptsize{$\pm$5.529} & 79.467\scriptsize{$\pm$9.031} & 91.341\scriptsize{$\pm$1.409} & 91.234\scriptsize{$\pm$2.159} & 91.473\scriptsize{$\pm$1.205} & 99.973\scriptsize{$\pm$0.009} \\
mlers (V27)   & 74.207\scriptsize{$\pm$12.406} & 67.412\scriptsize{$\pm$14.011} & 90.702\scriptsize{$\pm$1.589} & 89.926\scriptsize{$\pm$3.309} & 91.636\scriptsize{$\pm$2.490} & 99.967\scriptsize{$\pm$0.014} \\
fme (V28)    & 74.785\scriptsize{$\pm$6.849} & 55.643\scriptsize{$\pm$7.580} & 87.053\scriptsize{$\pm$1.408} & 87.978\scriptsize{$\pm$1.965} & 86.213\scriptsize{$\pm$2.401} & 99.963\scriptsize{$\pm$0.010} \\
biomedia$\ddagger$ (V29)    & 60.598\scriptsize{$\pm$12.005} & 51.359\scriptsize{$\pm$11.437} & 74.778\scriptsize{$\pm$10.255} & 91.687\scriptsize{$\pm$1.495} & 64.424\scriptsize{$\pm$14.166} & 99.982\scriptsize{$\pm$0.005} \\
ntflow$\ddagger$ (V30)    & 28.372\scriptsize{$\pm$6.008} & 21.930\scriptsize{$\pm$5.304} & 86.452\scriptsize{$\pm$3.299} & 95.834\scriptsize{$\pm$1.595} & 78.914\scriptsize{$\pm$5.320} & 99.990\scriptsize{$\pm$0.004} \\
\hlinew{1.25pt}
\end{tabular}
\begin{tablenotes}
\scriptsize
\item[$\ddagger$] Their submissions before the deadline cannot be correctly evaluated by grand-challenge.org. We downloaded their results and evaluated on local devices.
\end{tablenotes}
\end{threeparttable}}
\end{table*}

\section{Results}\label{sec:results}
In this section, we reported the obtained results in the validation phase (Table~\ref{tab:validation-phase-1-result}) and the test phase (Table~\ref{tab:test-phase-result}, Table~\ref{tab:test-phase-covid19-result}), respectively. The results of the validation phase were stated from an overall statistical perspective while the analysis of the 
test phase results focused on the top 10 algorithms, which is in accordance with Section \ref{sec:methodologies}. 
We conducted an elaborated comparison among the top 10 algorithms, including quantitative and qualitative analysis, 
model complexity analysis, deep analysis of the relationships among metrics, and ranking stability analysis. 
\uline{It should be noticed that the airway segmentation task itself is challenging. 
Our analysis only focused on the results of the top 10 methods to derive critical observation and effective methods, then provide insights for the research community. 
That means even the tenth-place produced better results than other successful participation teams (i.e., better than the average of all valid results).}

\makeatletter
\def\hlinew#1{%
\noalign{\ifnum0=`}\fi\hrule \@height #1 \futurelet
\reserved@a\@xhline}
\makeatother
\begin{table*}[thbp]
\renewcommand\arraystretch{1.4}
\centering
\caption{\justify{Quantitative results on the full hidden \textbf{test set} of ATM'22 challenge achieved by participants. 
The results are reported in the format of mean $\pm$ standard deviation. In the hidden test set, The number of sample CT scans is 150, 
and the total number of airway branches is 178.91$\pm$48.81. The results are first post-processed by the largest component extraction 
and eventually reported by the Tree length detected rate (i.e., TD, \%), Branch detected rate (i.e., BD, \%), Dice Similarity Coefficient (i.e., DSC, \%),
Precision (\%), Sensitivity (i.e., Sen, \%) and Specificity (i.e., Spe, \%).}}\label{tab:test-phase-result}
\scalebox{0.9}{
\begin{threeparttable}
\begin{tabular}{lcccccc}
\hlinew{1.25pt}
\textbf{Team name}     & \textbf{TD (\%) $\uparrow$}            & \textbf{BD (\%) $\uparrow$}            & \textbf{DSC (\%) $\uparrow$ }         & \textbf{Precision (\%) $\uparrow$}     & \textbf{Sen (\%) $\uparrow$}  & \textbf{Spe (\%) $\uparrow$}   \\ \hline
Sanmed\_AI (T1)    & 88.843\scriptsize{$\pm$7.250} & 83.350\scriptsize{$\pm$10.900} & \textbf{94.969\scriptsize{$\pm$1.800}} & 95.055\scriptsize{$\pm$3.210} & 95.047\scriptsize{$\pm$3.349} & 99.984\scriptsize{$\pm$0.011} \\
fme (T2)  & 70.695\scriptsize{$\pm$12.393} & 54.615\scriptsize{$\pm$12.946} & 87.986\scriptsize{$\pm$10.177} & 87.137\scriptsize{$\pm$10.730} & 90.460\scriptsize{$\pm$4.874} & 99.698\scriptsize{$\pm$2.418} \\
LinkStartHao (T3) & 81.721\scriptsize{$\pm$10.295} & 71.140\scriptsize{$\pm$15.614} & 92.938\scriptsize{$\pm$2.133} & 96.140\scriptsize{$\pm$2.414} & 90.128\scriptsize{$\pm$4.438} & 99.988\scriptsize{$\pm$0.008} \\
YangLab (T4) & 94.512\scriptsize{$\pm$8.598} & 91.920\scriptsize{$\pm$9.435} & \underline{94.800\scriptsize{$\pm$7.925}} & 94.707\scriptsize{$\pm$8.302} & 95.015\scriptsize{$\pm$8.240} & 99.985\scriptsize{$\pm$0.010} \\
dolphins (T5) & 90.134\scriptsize{$\pm$6.477} & 84.201\scriptsize{$\pm$11.151} & 92.734\scriptsize{$\pm$2.094} & 94.656\scriptsize{$\pm$3.434} & 91.122\scriptsize{$\pm$4.273} & 99.983\scriptsize{$\pm$0.012} \\
timi (T6) & \underline{95.919\scriptsize{$\pm$5.234}} & \underline{94.729\scriptsize{$\pm$6.385}} & 93.910\scriptsize{$\pm$3.682} & 93.553\scriptsize{$\pm$3.420} & 94.500\scriptsize{$\pm$5.168} & 99.979\scriptsize{$\pm$0.012} \\
neu204 (T7) & 90.974\scriptsize{$\pm$10.409} & 86.670\scriptsize{$\pm$13.087} & 94.056\scriptsize{$\pm$8.021} & 93.027\scriptsize{$\pm$8.410} & \underline{95.284\scriptsize{$\pm$8.581}} & 99.979\scriptsize{$\pm$0.013} \\
blackbean (T8) & 82.103\scriptsize{$\pm$10.719} & 71.418\scriptsize{$\pm$16.435} & 93.153\scriptsize{$\pm$2.284} & 96.146\scriptsize{$\pm$2.380} & 90.545\scriptsize{$\pm$4.748} & 99.988\scriptsize{$\pm$0.008} \\
lya (T9) & 85.215\scriptsize{$\pm$9.146} & 75.705\scriptsize{$\pm$14.887} & 93.758\scriptsize{$\pm$2.174} & \underline{96.501\scriptsize{$\pm$2.908}} & 91.412\scriptsize{$\pm$4.795} & \underline{99.989\scriptsize{$\pm$0.010}} \\
dnai (T10) & 86.733\scriptsize{$\pm$5.393} & 77.888\scriptsize{$\pm$8.703} & 90.871\scriptsize{$\pm$1.748} & 91.674\scriptsize{$\pm$2.787} & 90.871\scriptsize{$\pm$1.748} & 99.974\scriptsize{$\pm$0.011} \\
bms410 (T11)\textsuperscript{$\star$}  & 3.898\scriptsize{$\pm$6.481} & 2.812\scriptsize{$\pm$5.499} & 16.965\scriptsize{$\pm$19.429} & 77.583\scriptsize{$\pm$36.470} & 10.920\scriptsize{$\pm$14.278} & 99.997\scriptsize{$\pm$0.005} \\
miclab (T12) & 75.408\scriptsize{$\pm$14.094} & 65.994\scriptsize{$\pm$17.667} & 93.493\scriptsize{$\pm$2.678} & 96.440\scriptsize{$\pm$2.565} & 91.035\scriptsize{$\pm$5.907} & \textbf{99.989\scriptsize{$\pm$0.009}} \\
CITI-SJTU (T13) & 83.545\scriptsize{$\pm$9.942} & 73.012\scriptsize{$\pm$15.854} & 92.443\scriptsize{$\pm$2.195} & 94.756\scriptsize{$\pm$2.911} & 90.445\scriptsize{$\pm$4.384} & 99.984\scriptsize{$\pm$0.010} \\
deeptree\_damo (T14) & \textbf{97.853\scriptsize{$\pm$2.275}} & \textbf{97.129\scriptsize{$\pm$3.411}} & 92.819\scriptsize{$\pm$2.191} & 87.928\scriptsize{$\pm$4.181} & \textbf{98.448\scriptsize{$\pm$1.402}} & 99.957\scriptsize{$\pm$0.018} \\
CBT\_IITDELHI (T15) & 66.588\scriptsize{$\pm$26.624} & 59.044\scriptsize{$\pm$24.793} & 81.280\scriptsize{$\pm$30.103} & 94.865\scriptsize{$\pm$2.810} & 79.892\scriptsize{$\pm$29.790} & 99.984\scriptsize{$\pm$0.010} \\
bwhacil (T16) & 75.556\scriptsize{$\pm$24.091} & 68.478\scriptsize{$\pm$25.843} & 81.380\scriptsize{$\pm$13.376} & 80.076\scriptsize{$\pm$8.127} & 87.180\scriptsize{$\pm$18.607} & 99.927\scriptsize{$\pm$0.040} \\
suqi (T17) & 89.209\scriptsize{$\pm$7.338} & 82.164\scriptsize{$\pm$12.264} & 93.646\scriptsize{$\pm$2.102} & 95.777\scriptsize{$\pm$3.318} & 91.839\scriptsize{$\pm$4.378} & 99.987\scriptsize{$\pm$0.012} \\
satsuma (T18)& 81.565\scriptsize{$\pm$11.017} & 70.819\scriptsize{$\pm$16.828} & 93.307\scriptsize{$\pm$2.196} & 96.181\scriptsize{$\pm$2.411} & 90.813\scriptsize{$\pm$4.745} & 99.988\scriptsize{$\pm$0.008} \\
Median (T19)& 78.653\scriptsize{$\pm$10.365} & 68.314\scriptsize{$\pm$14.529} & 93.119\scriptsize{$\pm$2.095} & 96.159\scriptsize{$\pm$2.305} & 90.443\scriptsize{$\pm$4.361} & 99.988\scriptsize{$\pm$0.008} \\
notbestme (T20)& 87.518\scriptsize{$\pm$9.028} & 81.343\scriptsize{$\pm$13.560} & 94.515\scriptsize{$\pm$2.270} & \textbf{96.590\scriptsize{$\pm$2.673}} & 92.701\scriptsize{$\pm$4.325} & \textbf{99.989\scriptsize{$\pm$0.009}} \\
biomedia (T21)& 64.254\scriptsize{$\pm$11.578} & 53.988\scriptsize{$\pm$12.679} & 80.370\scriptsize{$\pm$11.816} & 93.533\scriptsize{$\pm$2.953} & 71.986\scriptsize{$\pm$15.532} & 99.984\scriptsize{$\pm$0.007} \\
\hlinew{1.25pt}
\end{tabular}
\begin{tablenotes}
\scriptsize
\item[$\star$] The results are abnormal hence were excluded in the final ranking.
\end{tablenotes}
\end{threeparttable}}
\end{table*}

\subsection{Validation Phase}
\begin{figure}[!htbp]
\centering
\includegraphics[width=1.0\linewidth]{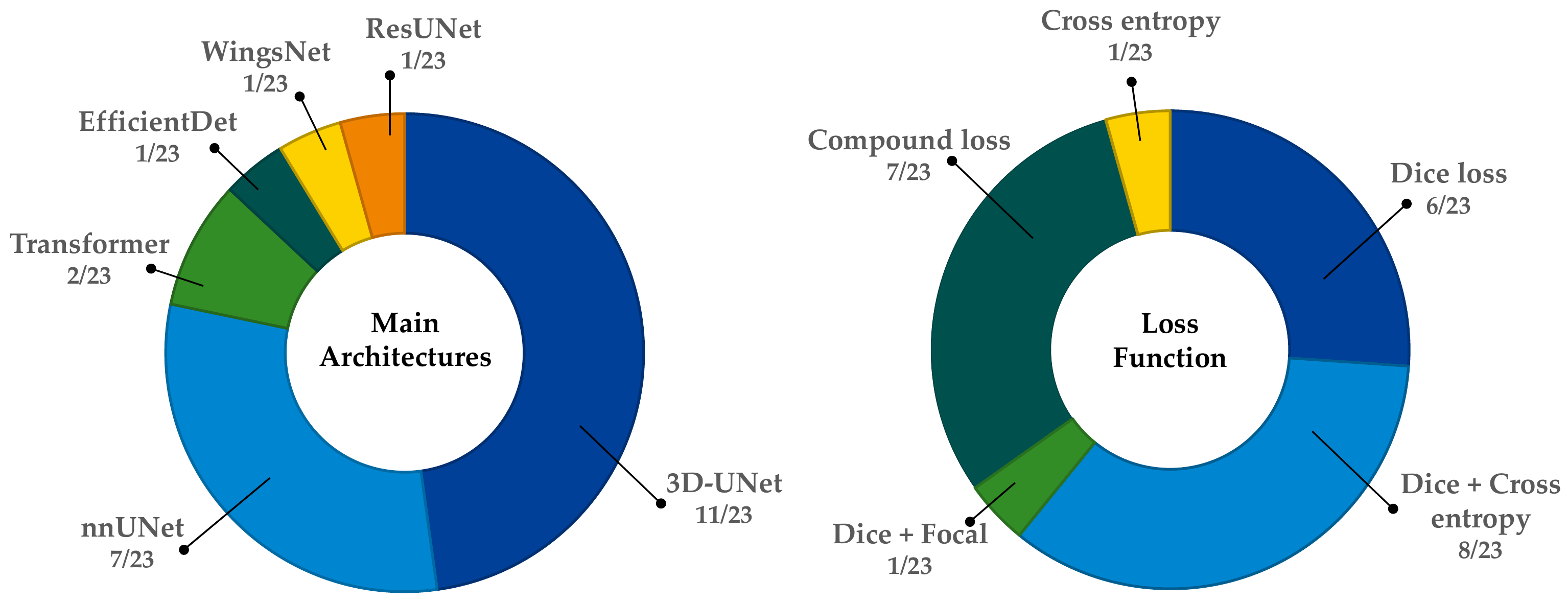}
\caption{Main network architectures (left) and loss function (right) adopted by the participants in the validation phase (n = 23 teams).}
\label{fig:validation-arch-loss}
\end{figure}

\textbf{Overall Outcome:}
50 CT scans without the pulmonary airway ground-truth are provided for evaluation in the validation phase. 
The participants were required to submit the binary prediction results to the platform of grand-challenge.org, where the evaluation was automatically executed. 
In the validation phase\footnote{Full ranking results of the validation phase (Time period: 1 Jun 2022 -- 17 Aug 2022): 
\url{https://atm22.grand-challenge.org/evaluation/validation-phase/leaderboard/}}, we received 30 valid submissions from different teams, 
23 submissions of them are detailed enough to report their main architectures and loss function, which can be seen in Figure~\ref{fig:validation-arch-loss}. 
\makeatletter
\def\hlinew#1{%
\noalign{\ifnum0=`}\fi\hrule \@height #1 \futurelet
\reserved@a\@xhline}
\makeatother
\begin{table*}[thbp]
\renewcommand\arraystretch{1.4}
\centering
\caption{\justify{Quantitative results on the \textbf{noisy domain (i.e., COVID-19 CT scans) of the hidden test set} achieved by participants. 
The results are reported in the format of mean $\pm$ standard deviation. In the hidden test set, The number of COVID-19 CT scans is 58, 
and the total number of airway branches is 167.17$\pm$34.97. The results are first post-processed by the largest component extraction 
and eventually reported by the Tree length detected rate (i.e., TD, \%), Branch detected rate (i.e., BD, \%), Dice Similarity Coefficient (i.e., DSC, \%),
Precision (\%), Sensitivity (i.e., Sen, \%) and Specificity (i.e., Spe, \%).}}\label{tab:test-phase-covid19-result}
\scalebox{0.9}{
\begin{threeparttable}
\begin{tabular}{lcccccc}
\hlinew{1.25pt}
\textbf{Team name}     & \textbf{TD (\%) $\uparrow$}            & \textbf{BD (\%) $\uparrow$}            & \textbf{DSC (\%) $\uparrow$ }         & \textbf{Precision (\%) $\uparrow$}     & \textbf{Sen (\%) $\uparrow$}  & \textbf{Spe (\%) $\uparrow$}   \\ \hline
Sanmed\_AI (T1)    & 83.517\scriptsize{$\pm$6.686} & 74.562\scriptsize{$\pm$9.299} & 94.615\scriptsize{$\pm$1.202} & 97.533\scriptsize{$\pm$0.913} & 91.898\scriptsize{$\pm$2.204} & 99.991\scriptsize{$\pm$0.004} \\
fme (T2) & 57.623\scriptsize{$\pm$5.934} & 42.319\scriptsize{$\pm$4.507} & 88.217\scriptsize{$\pm$1.819} & 91.703\scriptsize{$\pm$1.110} & 85.036\scriptsize{$\pm$3.001} & 99.973\scriptsize{$\pm$0.004} \\
LinkStartHao (T3) & 72.648\scriptsize{$\pm$7.033} & 56.880\scriptsize{$\pm$8.482} & 91.310\scriptsize{$\pm$1.626} & 97.398\scriptsize{$\pm$0.656} & 85.993\scriptsize{$\pm$2.945} & 99.991\scriptsize{$\pm$0.003} \\
YangLab (T4) & 92.358\scriptsize{$\pm$4.174} & 88.218\scriptsize{$\pm$5.802} & \underline{94.982\scriptsize{$\pm$0.998}} & 97.052\scriptsize{$\pm$1.309} & 93.036\scriptsize{$\pm$1.919} & 99.990\scriptsize{$\pm$0.006} \\
dolphins (T5) & 84.061\scriptsize{$\pm$5.356} & 73.740\scriptsize{$\pm$8.478} & 91.984\scriptsize{$\pm$1.544} & 97.336\scriptsize{$\pm$0.898} & 87.241\scriptsize{$\pm$2.803} & 99.991\scriptsize{$\pm$0.004} \\
timi (T6) & \underline{94.251\scriptsize{$\pm$3.541}} & \underline{92.049\scriptsize{$\pm$4.898}} & \textbf{95.063\scriptsize{$\pm$1.227}} & 96.026\scriptsize{$\pm$1.307} & \underline{94.147\scriptsize{$\pm$1.978}} & 99.986\scriptsize{$\pm$0.005} \\
neu204 (T7)& 86.021\scriptsize{$\pm$6.614} & 77.724\scriptsize{$\pm$9.825} & 94.672\scriptsize{$\pm$1.055} & 96.186\scriptsize{$\pm$1.860} & 93.280\scriptsize{$\pm$2.426} & 99.986\scriptsize{$\pm$0.008} \\
blackbean (T8)& 72.231\scriptsize{$\pm$7.783} & 56.094\scriptsize{$\pm$9.758} & 91.176\scriptsize{$\pm$1.785} & 97.390\scriptsize{$\pm$0.656} & 85.768\scriptsize{$\pm$3.198} & 99.992\scriptsize{$\pm$0.003} \\
lya (T9)& 76.528\scriptsize{$\pm$5.575} & 61.400\scriptsize{$\pm$8.787} & 91.932\scriptsize{$\pm$1.602} & \underline{98.355\scriptsize{$\pm$0.794}} & 86.350\scriptsize{$\pm$2.901} & \underline{99.995\scriptsize{$\pm$0.003}} \\
dnai (T10)& 84.181\scriptsize{$\pm$5.538} & 73.619\scriptsize{$\pm$8.026} & 90.191\scriptsize{$\pm$1.482} & 92.911\scriptsize{$\pm$1.844} & 87.664\scriptsize{$\pm$2.129} & 99.976\scriptsize{$\pm$0.009} \\
bms410 (T11)\textsuperscript{$\star$}  & 2.508\scriptsize{$\pm$2.631} & 1.663\scriptsize{$\pm$2.247} & 8.869\scriptsize{$\pm$8.785} & 71.356\scriptsize{$\pm$36.422} & 4.895\scriptsize{$\pm$5.198} & 99.997\scriptsize{$\pm$0.004} \\
miclab (T12)& 59.650\scriptsize{$\pm$6.603} & 46.486\scriptsize{$\pm$5.433} & 90.814\scriptsize{$\pm$1.906} & \textbf{98.427\scriptsize{$\pm$0.389}} & 84.350\scriptsize{$\pm$3.242} & \textbf{99.995\scriptsize{$\pm$0.002}} \\
CITI-SJTU (T13)& 73.647\scriptsize{$\pm$6.963} & 57.194\scriptsize{$\pm$8.783} & 90.908\scriptsize{$\pm$1.559} & 96.476\scriptsize{$\pm$0.777} & 85.998\scriptsize{$\pm$2.834} & 99.988\scriptsize{$\pm$0.004} \\
deeptree\_damo (T14)& \textbf{96.242\scriptsize{$\pm$2.634}} & \textbf{94.947\scriptsize{$\pm$4.104}} & 93.990\scriptsize{$\pm$1.240} & 91.049\scriptsize{$\pm$2.239} & \textbf{97.165\scriptsize{$\pm$0.866}} & 99.965\scriptsize{$\pm$0.012} \\
CBT\_IITDELHI (T15)& 53.197\scriptsize{$\pm$33.505} & 45.729\scriptsize{$\pm$29.034} & 66.297\scriptsize{$\pm$40.944} & 96.003\scriptsize{$\pm$2.967} & 64.358\scriptsize{$\pm$39.798} & 99.986\scriptsize{$\pm$0.011} \\
bwhacil (T16)& 53.486\scriptsize{$\pm$23.100} & 42.926\scriptsize{$\pm$20.643} & 75.861\scriptsize{$\pm$17.975} & 84.930\scriptsize{$\pm$8.197} & 73.230\scriptsize{$\pm$21.150} & 99.950\scriptsize{$\pm$0.031} \\
suqi (T17)& 83.259\scriptsize{$\pm$5.518} & 71.429\scriptsize{$\pm$8.825} & 92.493\scriptsize{$\pm$1.389} & 98.040\scriptsize{$\pm$1.081} & 92.493\scriptsize{$\pm$1.389} & 99.993\scriptsize{$\pm$0.005} \\
satsuma (T18)& 71.235\scriptsize{$\pm$7.414} & 55.133\scriptsize{$\pm$9.328} & 91.316\scriptsize{$\pm$1.682} & 97.537\scriptsize{$\pm$0.630} & 85.894\scriptsize{$\pm$3.013} & 99.992\scriptsize{$\pm$0.003} \\
Median (T19)& 70.054\scriptsize{$\pm$7.891} & 55.809\scriptsize{$\pm$8.596} & 91.367\scriptsize{$\pm$1.671} & 97.388\scriptsize{$\pm$0.637} & 86.099\scriptsize{$\pm$2.946} & 99.992\scriptsize{$\pm$0.003} \\
notbestme (T20)& 81.283\scriptsize{$\pm$8.344} & 70.723\scriptsize{$\pm$11.425} & 93.175\scriptsize{$\pm$1.668} & 97.917\scriptsize{$\pm$0.885} & 88.940\scriptsize{$\pm$3.199} & 99.993\scriptsize{$\pm$0.004} \\
biomedia (T21)& 59.444\scriptsize{$\pm$8.351} & 47.130\scriptsize{$\pm$7.753} & 82.915\scriptsize{$\pm$11.107} & 94.014\scriptsize{$\pm$1.158} & 82.915\scriptsize{$\pm$11.107} & 99.983\scriptsize{$\pm$0.006} \\
\hlinew{1.25pt}
\end{tabular}
\begin{tablenotes}
\scriptsize
\item[$\star$] The results are abnormal hence were excluded in the final ranking.
\end{tablenotes}
\end{threeparttable}}
\end{table*}
All the results are derived from the best submission of each team before the deadline of the validation phase.\\
\textbf{Architectures and Loss Functions:} As seen in Figure~\ref{fig:validation-arch-loss}, the 3D UNet~\citep{3DU-Net} and the nnUNet~\citep{isensee2021nnu} 
were the popular network architectures adopted 
by the participants. Two teams adopted the transformer architectures~\citep{liu2021swin,tang2022self} as the backbone. The WingsNet~\citep{zheng2021alleviating} 
EfficientDet~\citep{tan2020efficientdet}, and the ResUNet~\citep{diakogiannis2020resunet} were used by one team, respectively. 
The 3D UNet and the nnUNet are demonstrated to be the most effective architectures for the medical image segmentation task, hence, they are the most popular 
options for the participants. The WingsNet was adopted to explicitly tackle the inter-class imbalance problem that existed in airway segmentation. For easy and 
fast multi-scale feature fusion, one team used the EfficientDet. The ResUNet was designed to alleviate the problem of vanishing
and exploding gradients, thus achieving consistent training as the depth of the network increases. As for the aspect of the loss functions, the 
Dice loss~\citep{milletari2016v} with its variants (e.g., Dice loss with Focal loss~\citep{zhu2019anatomynet}, Dice loss with Cross Entropy 
loss~\citep{taghanaki2019combo}) dominate the majority. Some other teams designed the compound loss functions. For simplicity, we used the team index defined in 
the Table \ref{tab:validation-participants-info}. V9 designed the centerline weighted loss function. In addition to the centerline weighted loss function, V14 
further used the local-imbalanced-based loss function to dynamically re-weight each voxel. V19 proposed a breakage-sensitive loss function in the second stage 
to repair the breakage regions within the airways. 
V25 explored the clDice loss~\citep{shit2021cldice} combined with the multi-weighted loss to preserve the topology of the airway. V2 designed a 
Jaccard Continuity and Accumulation Mapping (JCAM) loss to tackle the discontinuity problem. V14, V2, V19 achieved the leading performance in the validation phase 
and the later test phase, which demonstrated that the reasonable design of loss function is effective for the pulmonary airway segmentation task. This experimental 
observation is in line with the consensus of effective methods finding, the novel objective functions that emphasize the topology completeness and the topology 
correctness are beneficial to deal with challenge of the Leakage (C1) and Breakage (C2), Robustness and Generalization (C3). The details of all qualified papers in the validation phase 
could be found in our official repository collection\footnote{ATM'22 Validation Phase Papers: \url{https://drive.google.com/drive/folders/1FTrc1AGqEqNDHvfCtEpQG3O2agxHXu46?usp=share_link}}.\\
\textbf{Quantitative Results:} Six metrics were reported in both the validation phase and test phase. Four metrics, including the tree length detected rate (TD, \%), branch detected 
rate (BD, \%), Dice similarity coefficient (DSC, \%) and Precision, are used for the score calculation. In addition, the sensitivity (Sen, \%) and the specificity (Spe, \%) were also covered to present a more comprehensive report. The results are presented in Table~\ref{tab:validation-phase-1-result}. It was observed that no single team could achieve the best performance on all six metrics. 
As for the topological completeness of airway segmentation results, 
teams V19 and V14 achieved remarkably higher performance than other teams. Specifically, V19 achieved 97.369\% TD and 96.717\% BD, 
and V14 attained 95.866\% TD and 94.921\% BD, which substantially exceeded the average results (84.696\% TD and 77.680\% BD). As for the topological 
correctness, V2 achieved the best performance on the DSC (95.926\%) and V22 achieved the highest Precision (99.515\%). However, the other metrics of V22 were 
quite low, only 60.705\% TD, 46.924\% BD, and 88.394\% DSC. The underlying reason is that all metrics were calculated on the largest component of the airway 
prediction, the precision can be very high whereas the breakage (C2) is severe (e.g., only the trachea and the main bronchi were preserved). 
On the contrary, V2 obtained the second-best performance of the Precision (97.180\%) while achieving the best DSC. It is worthwhile noting that in the validation phase, 
V2 achieved satisfactory DSC and Precision while maintaining competitive TD (94.406\%) and BD (91.302\%). Similarly, V14 preserved compelling results of DSC (93.987\%) 
and Precision (94.041\%) under the circumstance that they achieved the highest performance on topological completeness. Another important observation is that the 
high performance of topological correctness can not consistently guarantee the topological completeness of the airways and vice versa. 
For one thing, take V1 as an example, they achieved the competitive DSC (95.555\%) and Precision (95.551\%), 
higher than the average (93.383\% DSC and 94.275\% Precision), however, their TD and BD were under 90\%. More strikingly, V20 obtained 94.336\% DSC and 97.127\% Precision 
while the TD and BD were as low as 73.928\% and 65.672\% respectively. For another, V19 over-emphasized the topological completeness, consequently, 
the topological correctness was inevitably affected. Specifically, they achieved the superior performance of the TD and BD while the DSC and Precision suffered a decrease 
to 92.812\% and 87.324\% respectively. The above findings demonstrated that the pulmonary airway segmentation task differs from other common medical segmentation tasks. 
DSC usually dominates in the evaluation of many medical segmentation tasks due to its superiority in measuring overlap-wise accuracy. However, the overlap-wise accuracy is 
not sufficient to evaluate the airway extraction algorithms because the topology is intrinsically embedded in the voxel-wise airway annotation data.

\subsection{Test Phase}
\begin{figure*}[thbp]
\centering
\subfloat[Tree length detected rate]{
    \includegraphics[width=0.48\linewidth]{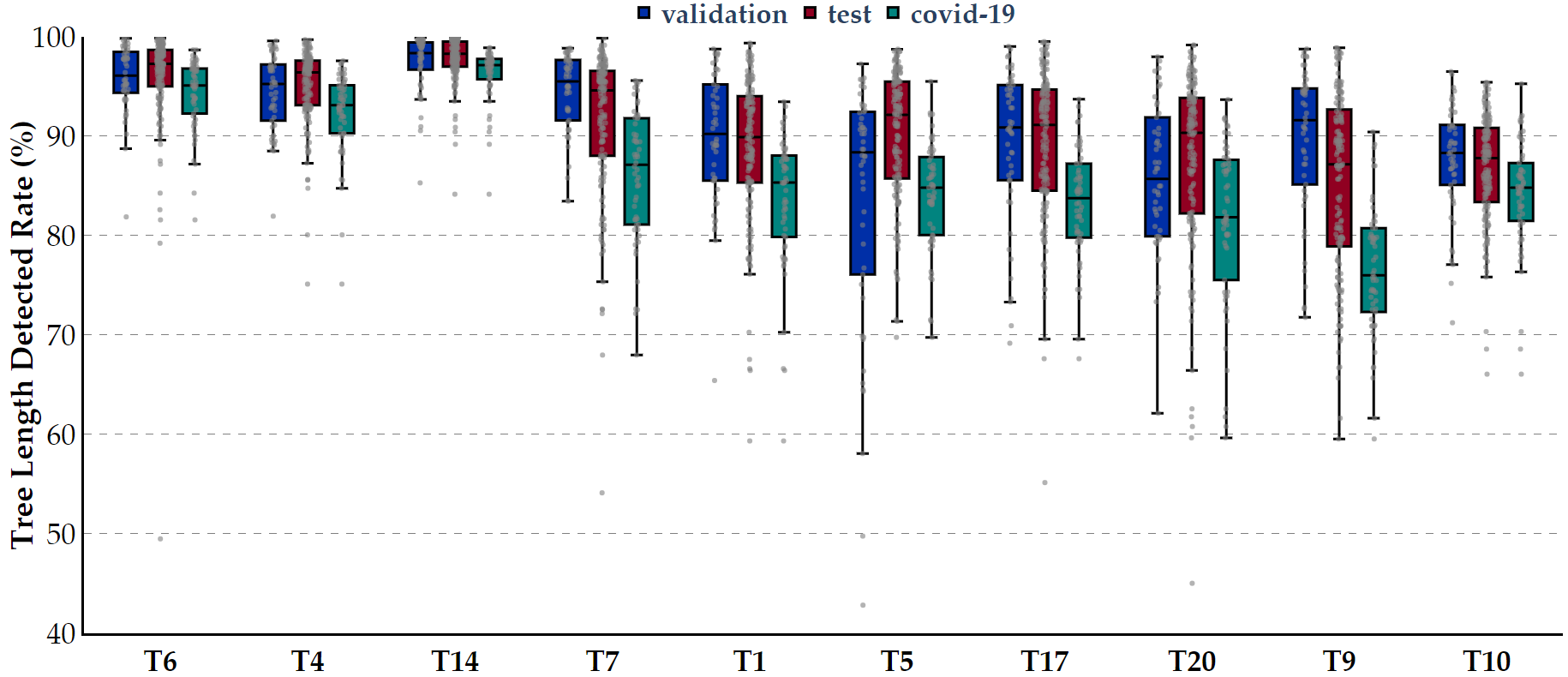}
}
\subfloat[Branch detected rate]{
\includegraphics[width=0.48\linewidth]{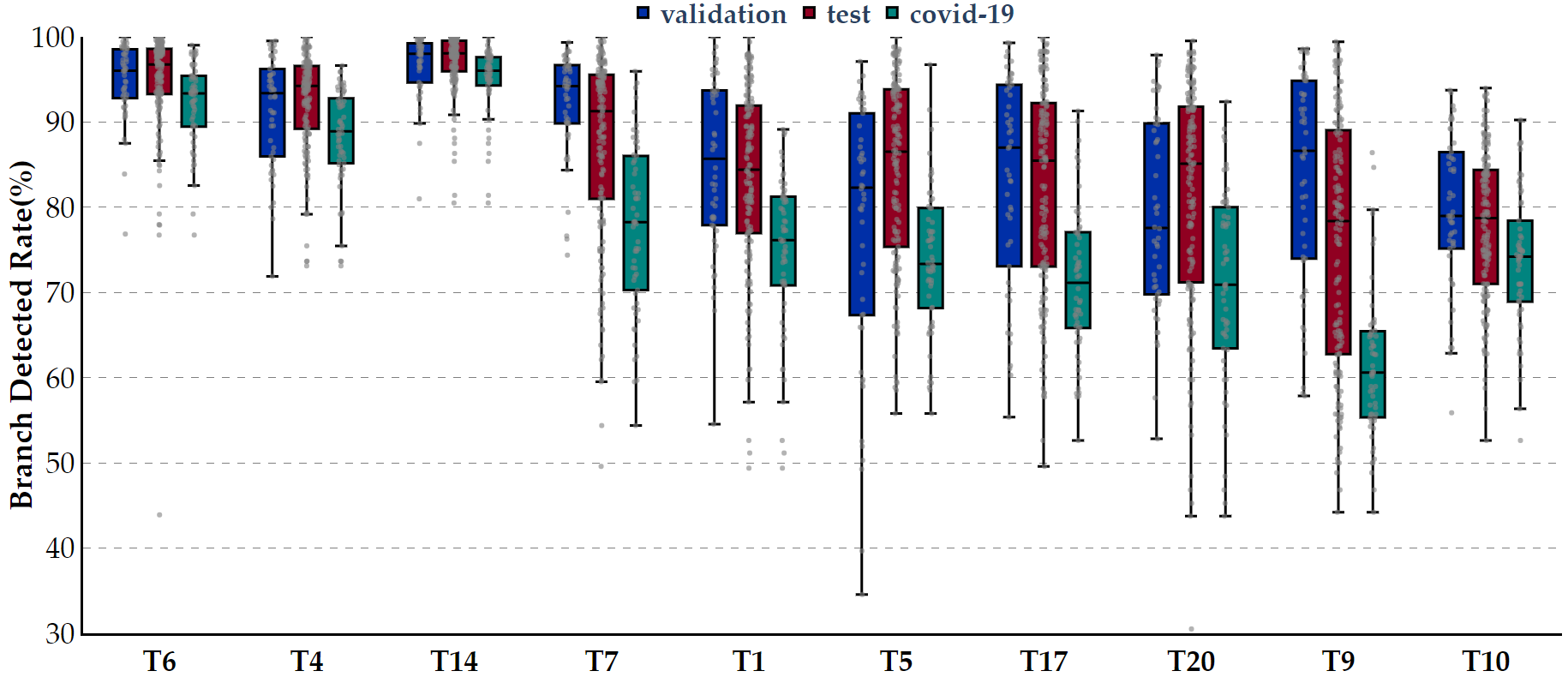}
}
\quad    
\subfloat[Dice Similarity Coefficient]{
  \includegraphics[width=0.48\linewidth]{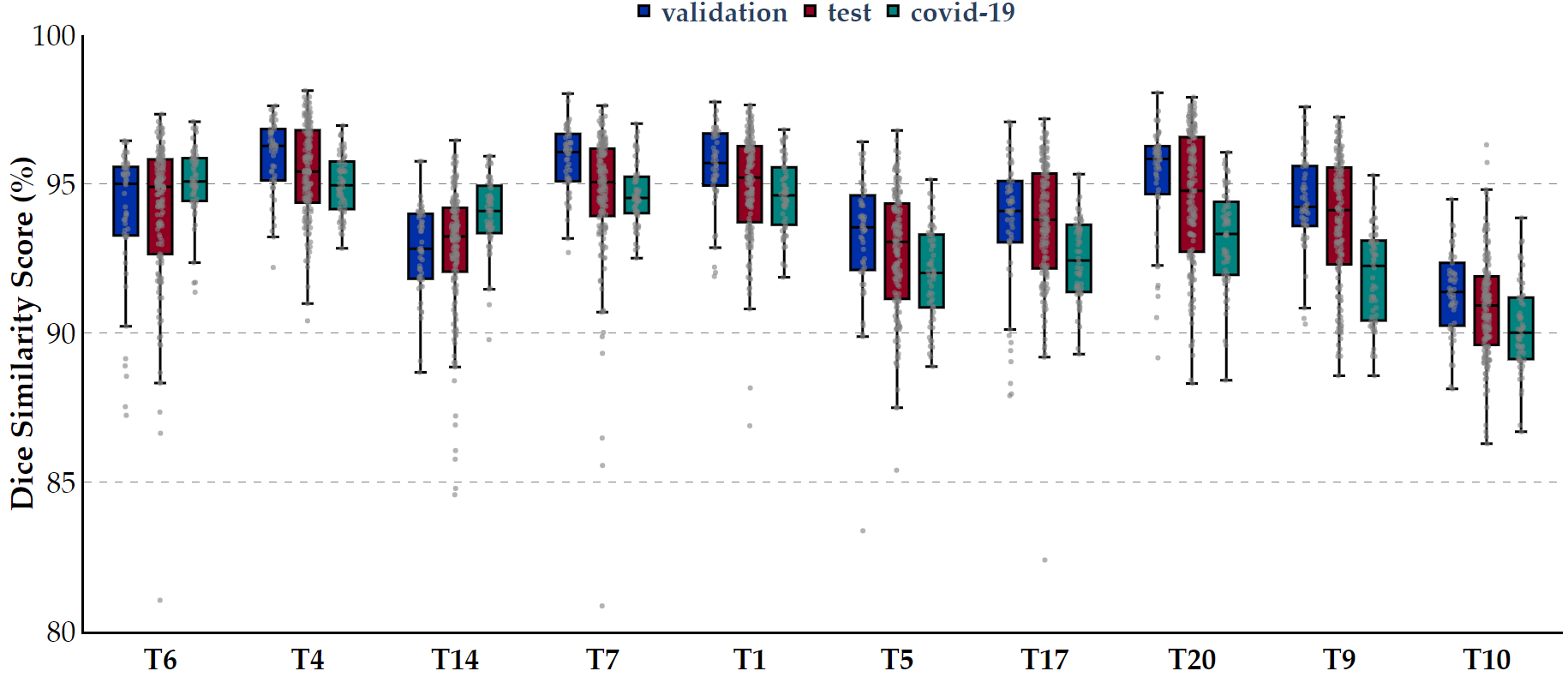}
}
\subfloat[Precision]{
\includegraphics[width=0.48\linewidth]{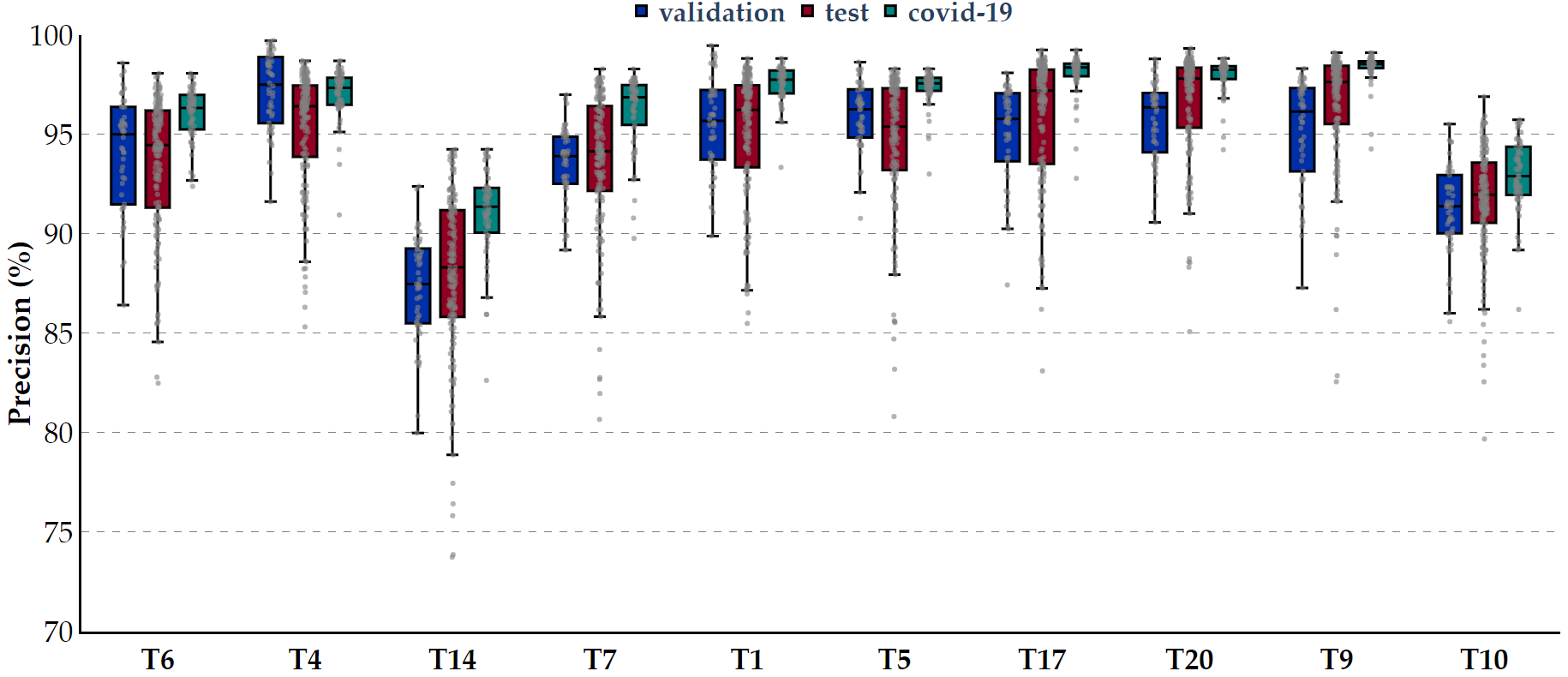}
}
\caption{Box plots of the quantitative results achieved by the top 10 algorithms on the validation set, test set, and COVID-19 set separately. 
The quantitative results include the (a) tree length detected rate, (b) branch detected rate, (c) dice similarity coefficient and (d) precision. Team index 
is adopted to represent different algorithms, and the order in the x-axis is dependent on the final rank (descend from left to right).}
\label{fig:box-plot}
\end{figure*}

\subsubsection{Overall Outcome}
150 CT scans are kept entirely hidden by the organizers, which means that the input images are inaccessible to the participants. The participants are 
required to package their algorithms into the docker images and followed the official instructions to execute these dockers. 
21 docker submissions were received and successfully executed on our server to generate the final results.
The standard format and instructions of the docker are provided in our official repository\footnote{The Docker Tutorial for the ATM'22 Test Phase Submission: \url{https://github.com/Puzzled-Hui/ATM-22-Related-Work/tree/main/baseline-and-docker-example}}. 
Owning a high number of registrations however only a fraction of the fully-completed participants is a typical phenomenon that takes place in 
biomedical image analysis challenges. (e.g., the Medical Segmentation Decathlon~\citep{antonelli2022medical} towards a multitude of both tasks and
modalities 2019 challenge with 19/180 successful submissions, the Skin lesion analysis detection~\citep{codella2018skin} 2017 challenge with 46/593 submissions or 
the Multi-Center, Multi-Vendor, and Multi-Disease Cardiac Segmentation~\citep{campello2021multi} challenge 2020 with 16/80 submissions). 
Some challenge participants usually register for the data access while they cannot participate in the validation and test phase within the deadline due to other commitments. 
Furthermore, the dissatisfied training and validation results may frustrate them to step back from the final submission. 
All qualified papers in the final test phase could be found in our official repository 
collection\footnote{ATM’22 Test Phase Papers:\url{https://drive.google.com/drive/folders/1T9OQ552TZUK5bxKm4D9wPt83AzSgqLAr?usp=sharing}}.
\begin{figure*}[htbp]
\centering
\includegraphics[width=1.0\linewidth]{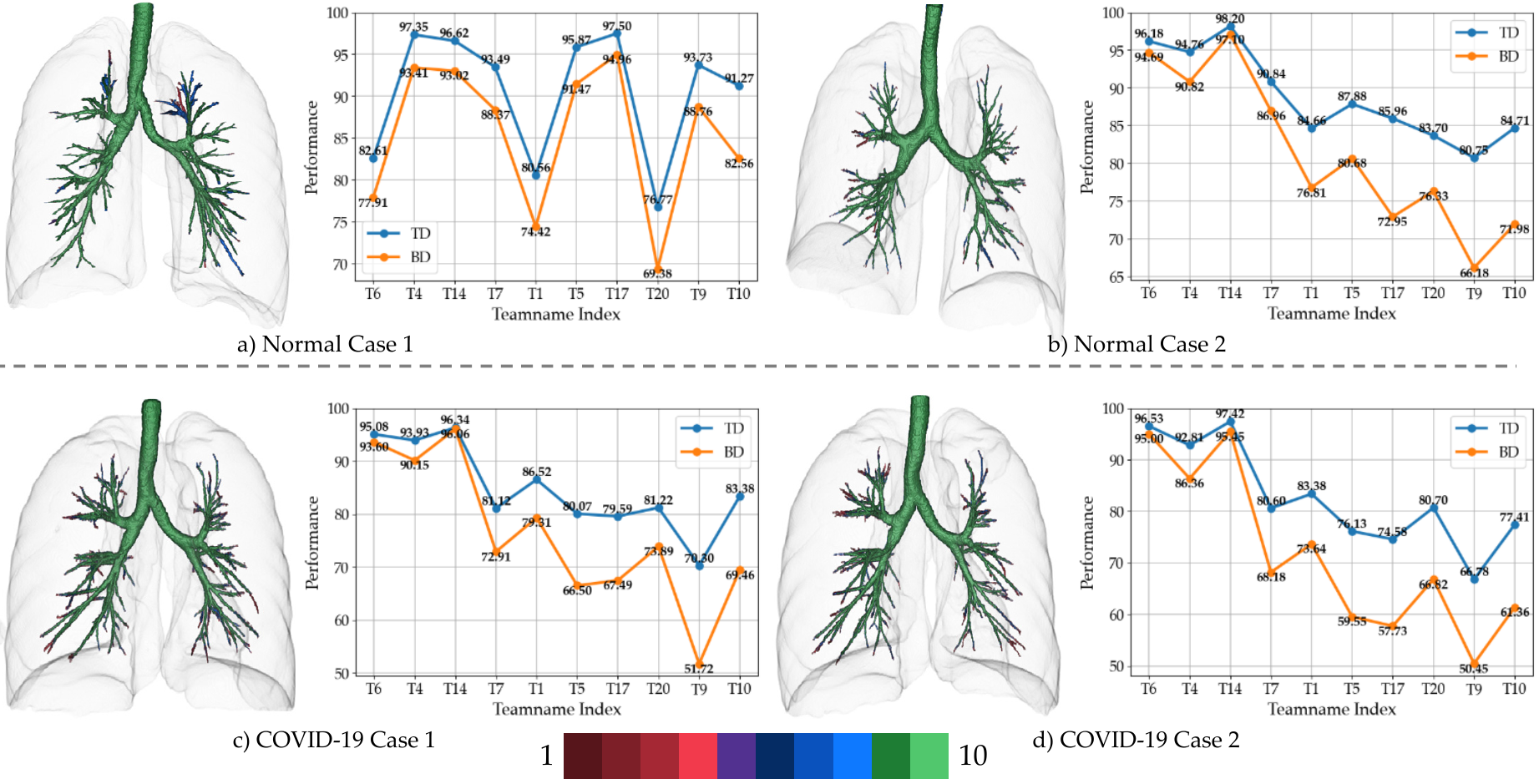}
\caption{The visualization of the color-encoded airway prediction by the Top 10 algorithms. Only the true positive part is presented, and the 
airway branches are assigned with a unique color from dark red (detected by only one team) to green (detected by all ten teams). Two normal 
cases and two COVID-19 cases from the test set are chosen for illustration. Individual line chart describes the TD and BD values of each team 
is also included beside the branch-wise airway visualization. Best viewed in color.}
\label{fig:airway_pred_encoded}
\end{figure*}
\subsubsection{Quantitative and Qualitative Comparisons}
\textbf{Performance across Domains:} Table \ref{tab:test-phase-result} and Table \ref{tab:test-phase-covid19-result} respectively
reported the quantitative results of the full set and the partial COVID-19 set of the hidden test set. The overall results were broadly similar 
to those in the validation phase, especially the top 5 algorithms. The ranking result (seen in Table \ref{tab:rank_mean_score_val_test}) revealed that 
the same 5 teams occupied the top 5 positions in both the validation and test phase, which demonstrated their generalization ability was 
superior to the rest of the teams. T6 and T4 ranked in the top 2 of the whole test phase, which proved their strong capacity under the comprehensive 
evaluation system. Compared to the average results (TD: 83.350\%, BD: 75.596\%, DSC: 91.277\%, Precision: 93.669\%), T6 and T4 achieved better performance.
Specifically, T6 achieved the 95.919\% TD, 94.729\% BD, 93.910\% DSC, 93.53\% Precision, and T4 obtained the 94.512\% TD, 91.920\% BD, 
94.800\% DSC, 94.707\% Precision. The overall compelling performance achieved by T6 can be ascribed to their elaborated optimization procedure tailored for 
the pulmonary airway segmentation task. Furthermore, the large batchsize (24 in their experiments) may be helpful to explore the features of airway datasets. 
The proposed fuzzy attention layer and continuity and accumulation mapping loss by T4 could assist to preserve the topological structure of airways. In addition, 
they used the region growing in the post-process procedure, which is beneficial to improving the accuracy of the trachea while maintaining the fidelity of the airway 
structure. T14 ranked third place and achieved the highest TD (97.853\%) and BD (97.129\%), while the Precision declined to 87.928\% that lower than the average. 
This observation implied that their proposed breakage attention map may over-emphasize the topology completeness while leading to the dilation problem. 
Figure \ref{fig:visualization_FP} corroborated this finding as T14 generated thicker prediction surrounding the boundary of the airway ground-truth.
T7 and T1 ranked forth and fifth place, respectively. They achieved the leading performance in the DSC (94.056\%, 94.696\%) and Precision (93.027\%, 95.055\%). 
However, their TD and BD were not outstanding due to the lack of especial modules for airway segmentation. 

Another critical comparison was the generalization ability on the noisy domain. Figure \ref{fig:box-plot} depicted the box 
plots of the quantitative results achieved by the top-10 algorithms on the validation set, test set, and COVID-19 set separately. It was obvious 
that TD and BD suffered a decrease in the COVID-19 set by all teams. The averages of TD and BD were 75.246\% and 64.206\%, which was far below the validation or test set. 
This performance degradation was moderate among the top 3 methods as they could still achieve more than 92\% TD and 88\% BD.  
DSC was slightly less affected than TD/BD while the variation tendency among teams was not unified. DSC decreased in the COVID-19 set among 
most teams, while T6 and T14 increased, 
which substantiated the S3 (Sec.\ref{consensus_of_effective_methods}). Both T6 and T14 designed the topology-sensitive loss functions that are beneficial 
to improving the robustness and generalization ability. The Precision had slightly increased due to the trade-off of sensitivity and specificity. 
Considering the general performance degradation in the noisy domain, especially topological completeness, further investigation of improving 
generalization is necessary.  

\begin{figure}[htbp]
\centering
\includegraphics[width=1.0\linewidth]{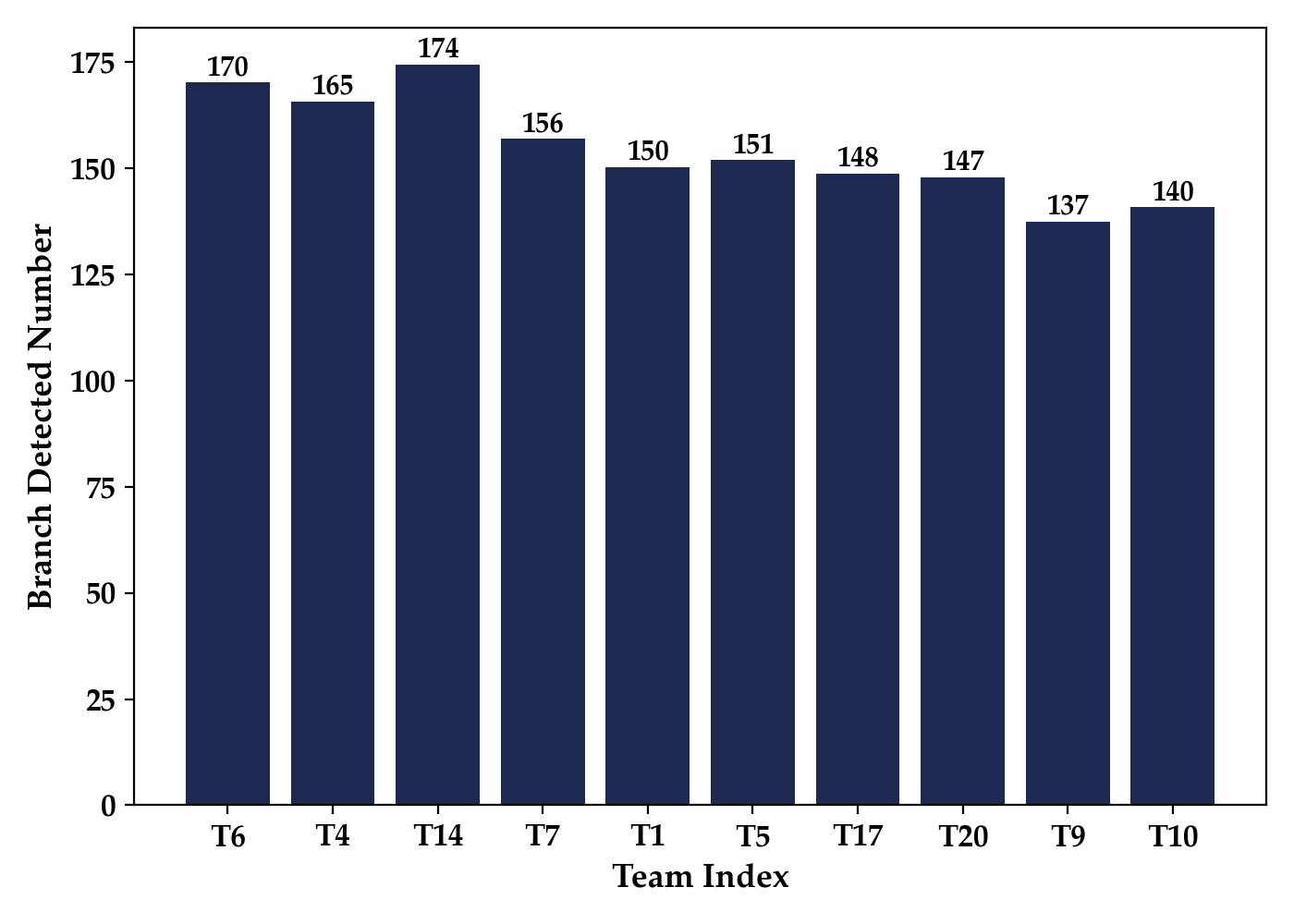}
\caption{The bar plot of the branch detected number achieved by the top 10 teams, averaged across the whole test dataset. The x-axis order (from left to right) 
follows the averaged ranking result. The averaged 154.4 of the branches were detected by all these 10 teams, 
whereas only the top 3 teams achieved more than 165 branches.}
\label{fig:branch_bar}
\end{figure}

\begin{figure*}[thbp]
  \centering
  \includegraphics[width=0.99\linewidth]{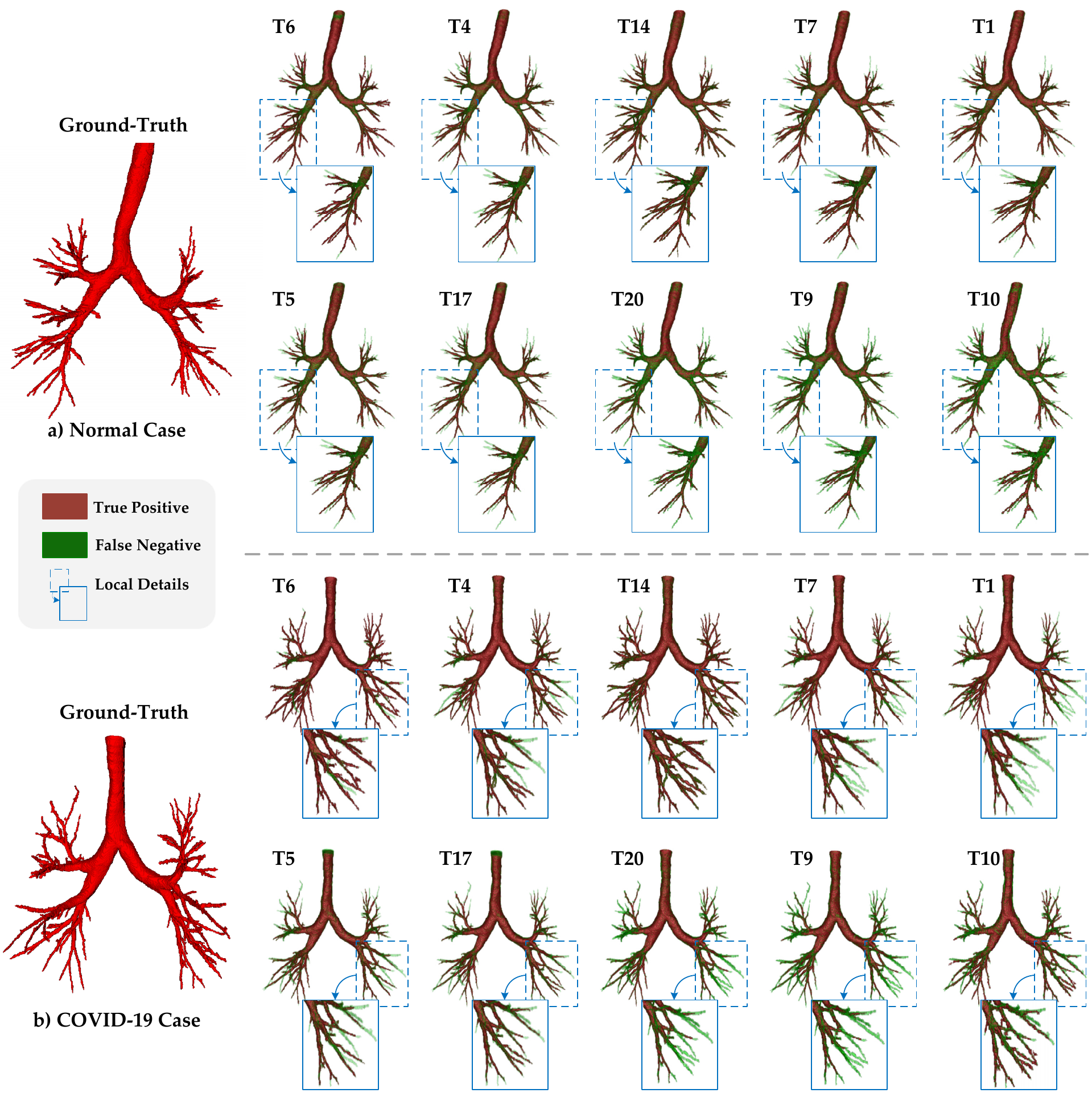}
  \caption{Visualization of the false negatives between the Top 10 teams (followed by the ranking order) of the normal case and the COVID-19 case. 
  The red part indicates the true positive and the green part denotes the false negative. Local details are highlighted in the boxes. Best viewed in color.}
  \label{fig:visualization_FN}
  \end{figure*}

\noindent \textbf{Branch-wise True Positive Analysis:} Table \ref{fig:branch_bar} described the branch detected number achieved by the top 10 teams. The 
top 10 methods obtained 154.4 branches on average across the whole test dataset. T6, T4, T14 achieved more than 165 branches, which surpassed other teams. 
To figure out the difference in detected branch numbers concretely, we conducted the visualization of the color-encoded airway prediction in Figure 
\ref{fig:airway_pred_encoded}. The airway branches are assigned a unique color from dark red (detected by only one team) to green (detected by all ten teams). 
Both the normal cases and 
the COVID-19 cases were provided for visualization. It can be explicitly observed that the disparity usually happened in the peripheral airways, 
especially the fifth/sixth generation of airways. The TD and BD exceeded 90\% by the top 3 ranked methods whereas the lowest BD was only around 50\% from the rest 
of the teams. Another interesting fact is that the top-ranked methods did perform well in each case. 
For example, Figure \ref{fig:airway_pred_encoded}.a) demonstrated that the airway in the left upper lobe was difficult to be detected, 
where red and blue dominated, revealing only a few teams detected correctly. The first-place team suffered a failure in this case. 
The underlying explanation for this observation is that the first-place algorithm shared some specific feature bias in contradiction with the failure case. 
It reminded us that pursuing the consensus of the prediction result is also significant to improve the reliability of deep learning models.

\begin{figure*}[thbp]
\centering
\includegraphics[width=1.0\linewidth]{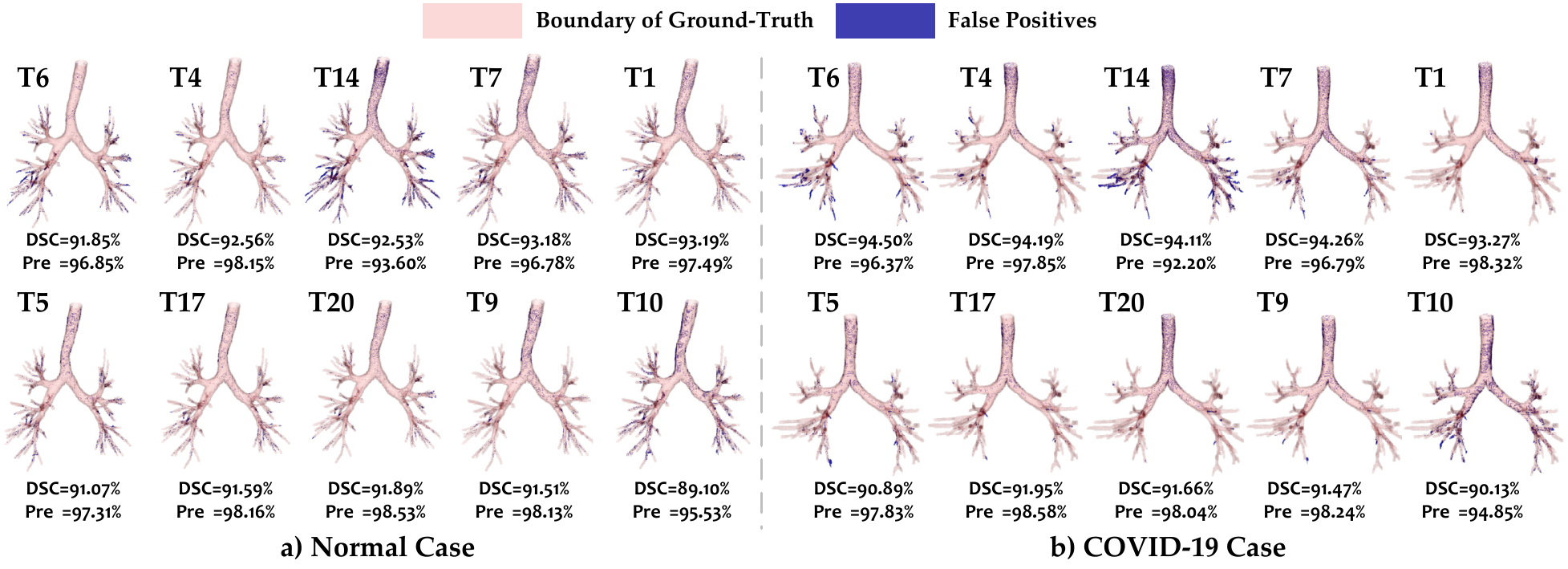}
\caption{Visualization of the false positive between the Top 10 teams (followed by the ranking order) of the normal case and the COVID-19 case, same as Figure \ref{fig:visualization_FN}. 
The red part indicates the boundary of the airway ground-truth and the blue part denotes the false positive. Best viewed in color. }
\label{fig:visualization_FP}
\end{figure*}

\noindent \textbf{False Negative Analysis:} As discussed before, the primary difference among the methods focused on the peripheral airways. Figure \ref{fig:visualization_FN} 
illustrated these areas separately for each method of the top 10 algorithms. The canonical normal case and COVID-19 case were chosen for representation, where the 
red part indicates the true positive and the green part denoted the false negative. The local details highlighted the right middle and lower lobes of the normal case, 
and the left lower lobe of the noisy COVID-19 case. In the first normal case, The top 5 methods segmented more accurate than the rest of the teams. 
For the COVID-19 case, T6 and T14 demonstrated significant improvement compared to others in detecting more bronchi and preserving better completeness 
under such noisy COVID-19 imaging characteristics. These observations confirm that the effective solutions (S1,S2,S3) are beneficial to reduce the false negatives.

\noindent \textbf{False Positive Analysis:} The increase in the false positive accounted for the dilation problem of the airway segmentation task~\citep{zheng2021alleviating}. 
Figure \ref{fig:visualization_FP} rendered the false positives (blue) and the boundary of the airway ground-truth (red). Except for the uncertainty of distal airways, 
the false positive concentrated on the trachea and main bronchi. Took T14, T7, and T10 as an example, They seemed thicker than the ground-truth and the corresponding Precision 
were also lower than other teams. The underlying reasons can be summarized as two aspects. 1) Over-emphasizing topological completeness could be harmful to the topological 
correctness (T14). 2) The airway segmentation task itself is challenging as the intensity shares large variation among the trachea, main bronchi, lobar bronchi, and 
distal segmental bronchi (T1, T10). To reduce the false positive, the promising effective solution is increasing the intra-class feature discrimination ability and 
trying to achieve a satisfactory trade-off between topological completeness and correctness.

\begin{figure*}[htbp]
\centering
\subfloat[Tree length detected rate \textit{v.s.} Dice Similarity Coefficient]{
    \includegraphics[width=0.48\linewidth]{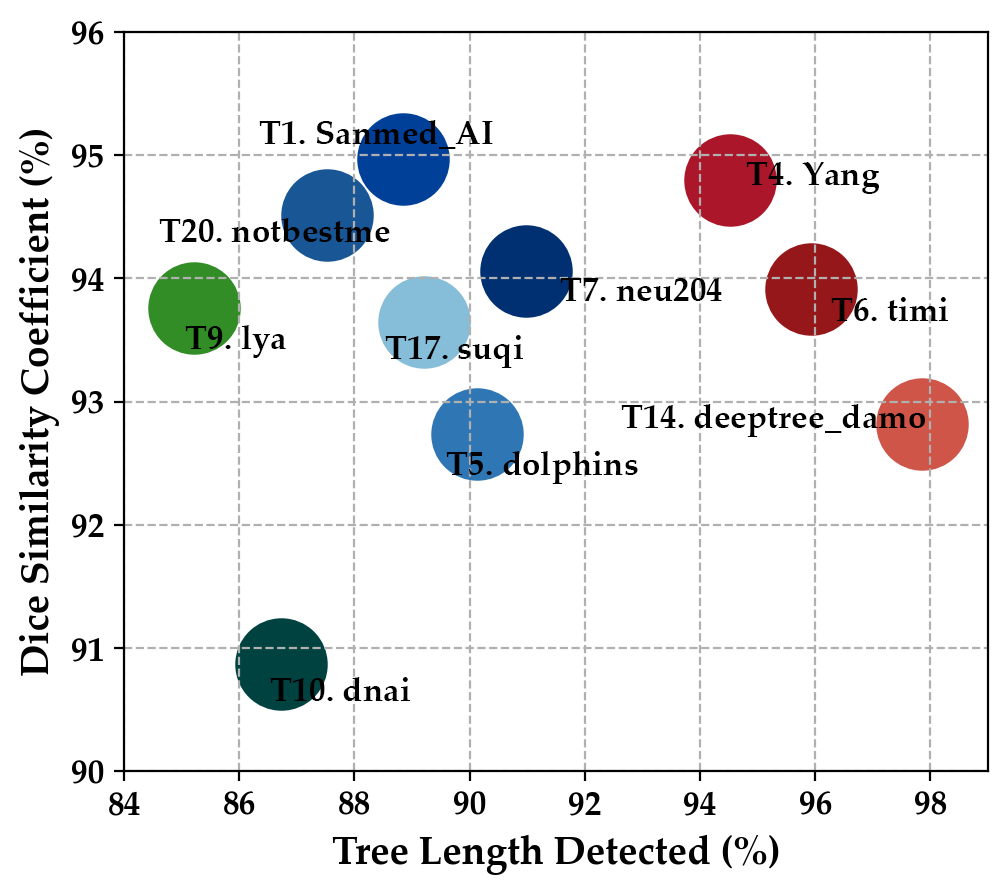}
}
\subfloat[Tree length detected rate \textit{v.s.} Precision]{
\includegraphics[width=0.48\linewidth]{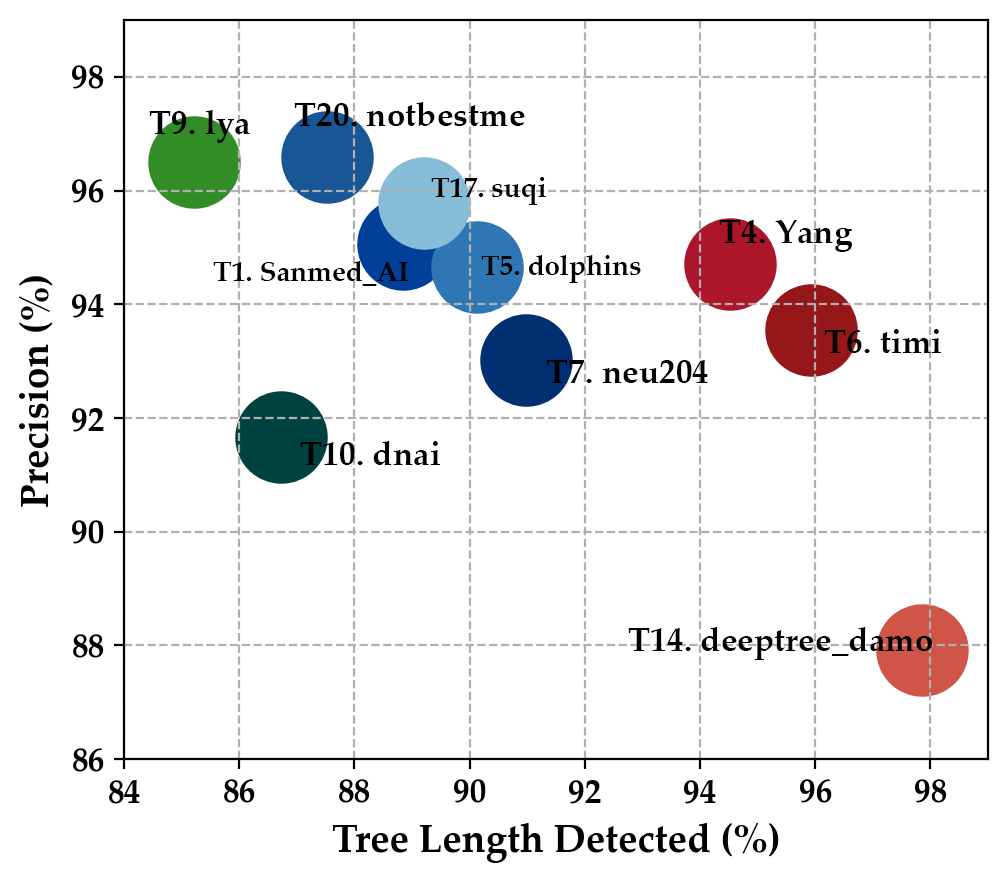}
}
\quad    
\subfloat[Branch detected rate \textit{v.s.} Dice Similarity Coefficient]{
  \includegraphics[width=0.48\linewidth]{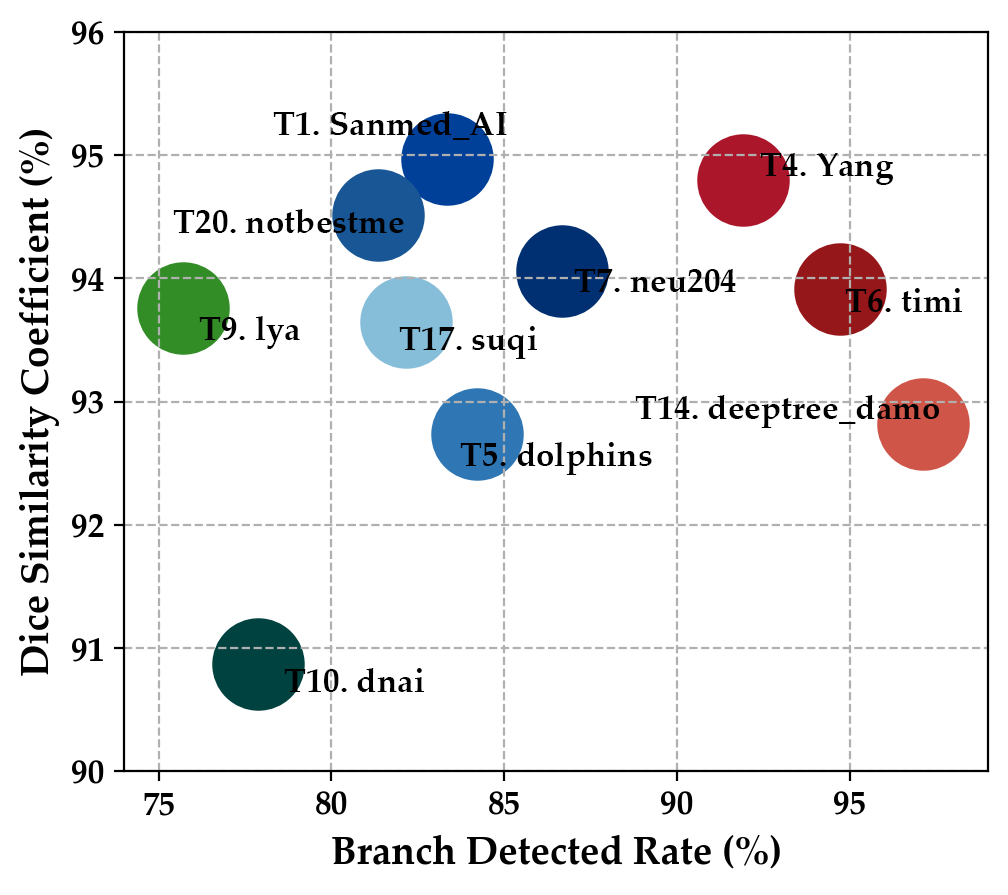}
}
\subfloat[Branch detected rate \textit{v.s.} Precision ]{
\includegraphics[width=0.48\linewidth]{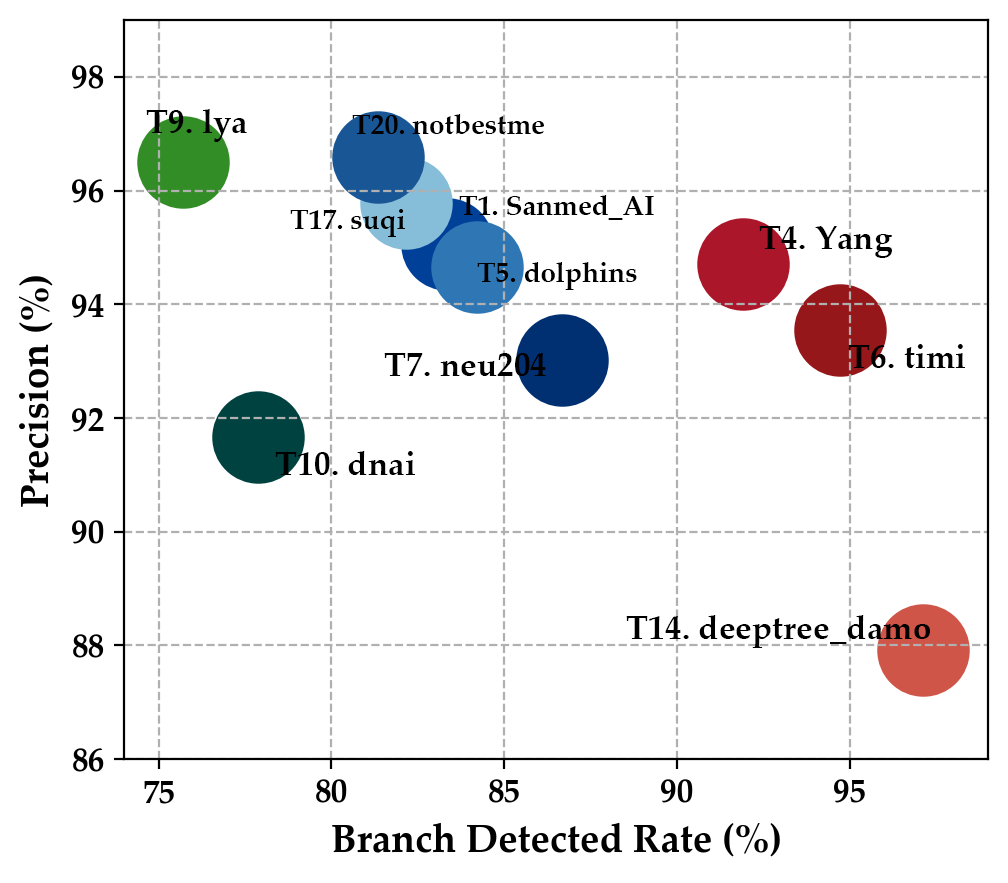}
}
\caption{The metric correlation scatter maps of the top 10 methods. Concretely, (a) the TD \textit{v.s.} DSC, (b) the TD \textit{v.s.} Precision, 
(c) the BD \textit{v.s.} DSC, (d) the BD \textit{v.s.} Precision. The team name is annotated besides the splashes. Best viewed in color.}
\label{fig:metric-correlation}
\end{figure*}

\subsubsection{Metric Correlation Analysis} The evaluation system adopted in this challenge contains two critical components, topological completeness (TD and BD) 
and topological correctness (DSC and Precision). Quantitative results showed that there was no single team achieves the highest performance on all of these metrics. 
More remarkably, T14 achieved the highest the TD (97.853\%) and BD (97.129\%) while the lower Precision (87.928\%) was the main drawback. Moreover, The high 
DSC or Precision cannot guarantee the topological completeness (T1, T7, T9). For example, T1 obtained the outstanding DSC across the validation, test, and the COVID-19 sets, 
however, their performance of TD and BD on the test phase was lower than 90\% and 85\%, respectively. Figure \ref{fig:visualization_FP} demonstrated that they did not produce 
many false positives while Figure \ref{fig:visualization_FN} corroborated the missing of substantial branches that led to the inferior TD and BD. To intuitively observe the 
inter-relation of these metrics, we conducted the metric correlation analysis, as seen in Figure \ref{fig:metric-correlation}. Four subplots were provided, the 
TD \textit{v.s.} DSC (Figure \ref{fig:metric-correlation}.a), TD \textit{v.s.} Precision (Figure \ref{fig:metric-correlation}.b), 
BD \textit{v.s.} DSC (Figure \ref{fig:metric-correlation}.c), and BD \textit{v.s.} Precision (Figure \ref{fig:metric-correlation}.d). Each metric correlation could be 
supposed to split into four quadrants. The expected method should appear in the first quadrant, where both the topological completeness and topological correctness metrics are 
high. The closest approaches (T6 and T4) so far still have a lot of room for improvement. The majority of the top 10 methods were distributed on the second quadrant, which 
demonstrated that high topological completeness is far tougher than high topological correctness to be achieved. This observation was in line with the main challenges 
encountered by the airway segmentation task (seen in Sec.\ref{sec:challenges_of_pulmonary_airway_segmentation}). The situation of T10 and T14 were quite different, T10 achieved 
high TD and BD, moderate DSC, and a low Precision. T14 was contrary to T10, Their TD, BD, and DSC were low, Precision was moderate. The above findings revealed that the 
principal challenge is improving topological completeness. Secondly, the topological correctness should be carefully handled to prevent a dramatic decrease.

\begin{table*}[htbp]
\begin{minipage}{\textwidth}
\begin{minipage}[t]{0.50\textwidth}
\renewcommand\arraystretch{1.2}
\centering
\makeatletter
\def\hlinew#1{%
\noalign{\ifnum0=`}\fi\hrule \@height #1 \futurelet
\reserved@a\@xhline}
\makeatother
\caption{\justify{Rankings for the validation phase and the test phase. Mean scores were used in ranking for all final 20 teams. Score is under percentage format.}}\label{tab:rank_mean_score_val_test}
\scalebox{0.7}{
\begin{threeparttable}
\begin{tabular}{ccccccc}
\hlinew{1.25pt}
\multicolumn{3}{l}{\textbf{The validation phase}}        &  & \multicolumn{3}{l}{\textbf{The test phase}}              \\ \cline{1-3} \cline{5-7} 
\textbf{Rank} & \textbf{\footnotesize{Team Name}} & \textbf{Mean Score} &  & \textbf{Rank} & \textbf{Team Name} & \textbf{Mean Score} \\ \hline
1             &timi                    &94.7038                     &  & 1             &timi                    & 94.5278                    \\
2             &YangLab                    &94.7035                     &  & 2             &YangLab                   & 93.9848                   \\
3             &neu204                    &93.9927                     &  & 3             &deeptree\_damo                    & 93.9323                   \\
4             &deeptree\_damo                    &93.5555                     &  & 4             &neu204                     & 91.1818                    \\
5             &Sanmed\_AI                   &91.5205                     &  & 5             &Sanmed\_AI                     & 91.1738                     \\
6             &satsuma                    &90.8943                     &  & 6             &dolphins                    & 90.4313                     \\
7             &LinkStartHao                    &90.8383                     &  & 7             &suqi                    & 90.1990                    \\
8             &CITI-SJTU                    &90.7885                     &  & 8             &notbestme                    & 89.9915                    \\
9             &lya                    &90.6783                     &  & 9             &lya                    & 87.7948                     \\
10            &blackbean                    &90.6618                     &  & 10            &dnai                    & 86.7915                    \\
11            &Median                    &90.3788                     &  & 11            &CITI-SJTU                    & 85.9390                     \\
12            &notbestme                    &88.9638                     &  & 12            &blackbean                   & 85.7050                    \\
13            &dolphins                    &87.5408                     &  & 13            &LinkStartHao                     & 85.4848                    \\
14            &dnai                    &87.4095                     &  & 14            &satsuma                    & 85.4680                    \\
15            &miclab                    &87.2865                     &  & 15            &Median                    & 84.0613                    \\
16            &suqi                    &85.5348                     &  & 16            &miclab                    & 82.8338                     \\
17            &CBT\_IITDELHI            &82.7658                     &  & 17            &bwhacil                    & 76.3725                    \\
18            &fme                    &76.3648                     &  & 18            &CBT\_IITDELHI                    & 75.4443                    \\
19            &bwhacil                    &75.4048                     &  & 19            &fme                     & 75.1083                    \\
20            &biomedia                    &69.6055                     &  & 20            &biomedia                    & 73.0363                    \\\hlinew{1.25pt}
\end{tabular}
\end{threeparttable}}
\end{minipage}
\begin{minipage}[t]{0.50\textwidth}
\renewcommand\arraystretch{1.2}
\centering
\makeatletter
\def\hlinew#1{%
\noalign{\ifnum0=`}\fi\hrule \@height #1 \futurelet
\reserved@a\@xhline}
\makeatother
\caption{\justify{Rankings stability analysis by different scores calculation methods on the test phase for all final 20 teams. Score is under percentage format.}}\label{tab:rank_weighted_mean_test}
\scalebox{0.7}{
\begin{threeparttable}
\begin{tabular}{ccccccc}
\hlinew{1.25pt}
\multicolumn{3}{l}{\textbf{The test phase by weighted score}}        &  & \multicolumn{3}{l}{\textbf{The test phase by mean score}}              \\ \cline{1-3} \cline{5-7} 
\textbf{Rank} & \textbf{\footnotesize{Team Name}} & \textbf{Weighted Score} &  & \textbf{Rank} & \textbf{Team Name} & \textbf{Mean Score} \\ \hline
1             & deeptree\_damo                   &95.3558                    &  & 1             &timi                    & 94.5278                    \\
2             & timi                  & 94.8463                  &  & 2             &YangLab                   & 93.9848                   \\
3             & YangLab                   & 93.6773                   &  & 3             &deeptree\_damo                    & 93.9323                   \\
4             & neu204                  & 90.2379                    &  & 4             &neu204                     & 91.1818                    \\
5             & Sanmed\_AI                & 89.1429                   &  & 5             &Sanmed\_AI                     & 91.1738                     \\
6             & dolphins                  & 89.1258                   &  & 6             &dolphins                    & 90.4313                     \\
7             & suqi                 & 88.3940                  &  & 7             &suqi                    & 90.1990                    \\
8             & notbestme                 & 87.7671                &  & 8             &notbestme                    & 89.9915                    \\
9             & dnai                & 84.9991                  &  & 9             &lya                    & 87.7948                     \\
10            & lya                & 84.8609                 &  & 10            &dnai                    & 86.7915                    \\
11            & CITI-SJTU            & 82.8748                &  & 11            &CITI-SJTU                    & 85.9390                     \\
12            & blackbean            & 82.1272                  &  & 12            &blackbean                   & 85.7050                    \\
13            & LinkStartHao             & 81.721                  &  & 13            &LinkStartHao                     & 85.4848                    \\
14            & satsuma                &  81.7576                  &  & 14            &satsuma                    & 85.4680                    \\
15            & Median               & 79.8302                     &  & 15            &Median                    & 84.0613                    \\
16            & miclab         & 77.9807                    &  & 16            &miclab                    & 82.8338                     \\
17            & bwhacil        & 74.6303                   &  & 17            &bwhacil                    & 76.3725                    \\
18            & CBT\_IITDELHI           & 70.3930                &  & 18            &CBT\_IITDELHI                    & 75.4443                    \\
19            & fme                 & 70.1270                 &  & 19            &fme                     & 75.1083                    \\
20            & biomedia               & 67.4702                 &  & 20            &biomedia                    & 73.0363                    \\\hlinew{1.25pt}
\end{tabular}
\end{threeparttable}}
\end{minipage}
\end{minipage}
\end{table*}
  
\subsubsection{Ranking Stability Analysis}\label{sec:ranking_stability_analysis}
As defined in Sec.\ref{sec:evaluation_metrics}, we adopted the mean score calculation as the ranking criterion. The rankings included all successful participants 
(i.e., who took part in both the validation and test phase). Table \ref{tab:rank_mean_score_val_test} reported the rankings of 20 teams in the validation phase and test phase. 
It is observed that the top five teams in the validation phase also occupied the top five positions, only neu204 (T7) and deeptree\_damo (T14) interchanged the order. 
However, From the sixth position to the twentieth position, an obvious variation arose between the ranking of the validation and test phase. 

Kendall's $\tau$ ~\citep{kendall1938new} was adopted to determine the variability of the rankings. In Table \ref{tab:rank_mean_score_val_test}, 
the Kendall's $\tau$ is 0.607 with p-value of 0.000189, which also implied the fluctuation of the ranking. 
The generalization ability of the methods accounts for this phenomenon, as the top five methods performed better than the rest of the teams with regard to the 
generalization ability. To measure the sensitivity of our score calculation, we modified it to a weighted formulation that slightly emphasizes the 
geometry of airway prediction results via adjusting the weighting coefficients:
\begin{align}
  \text{Weighted Score} = 0.30 * TD + 0.30 * BD \notag \\  + 0.15 * DSC + 0.15 * Precision.
\end{align}
Table \ref{tab:rank_weighted_mean_test} reported the ranking results by the mean score and weighted score separately. The result has shown that only 
the team deeptree\_damo (T14) moved to first place while the relative orders of other teams remained unchanged. As discussed before, T14 put much  
emphasis on topological completeness, hence, they were sensitive to the weighting coefficients. In general, the ranking order was close to identical as 
the corresponding Kendall's $\tau$ is 0.979 with p-value of 1.87e-09. The above experimental results demonstrated that the ranking criterion was reasonable.

\subsubsection{Model Complexity Analysis}
\begin{figure*}[thbp]
\centering
\subfloat[Tree length detected rate]{
    \includegraphics[width=0.50\linewidth]{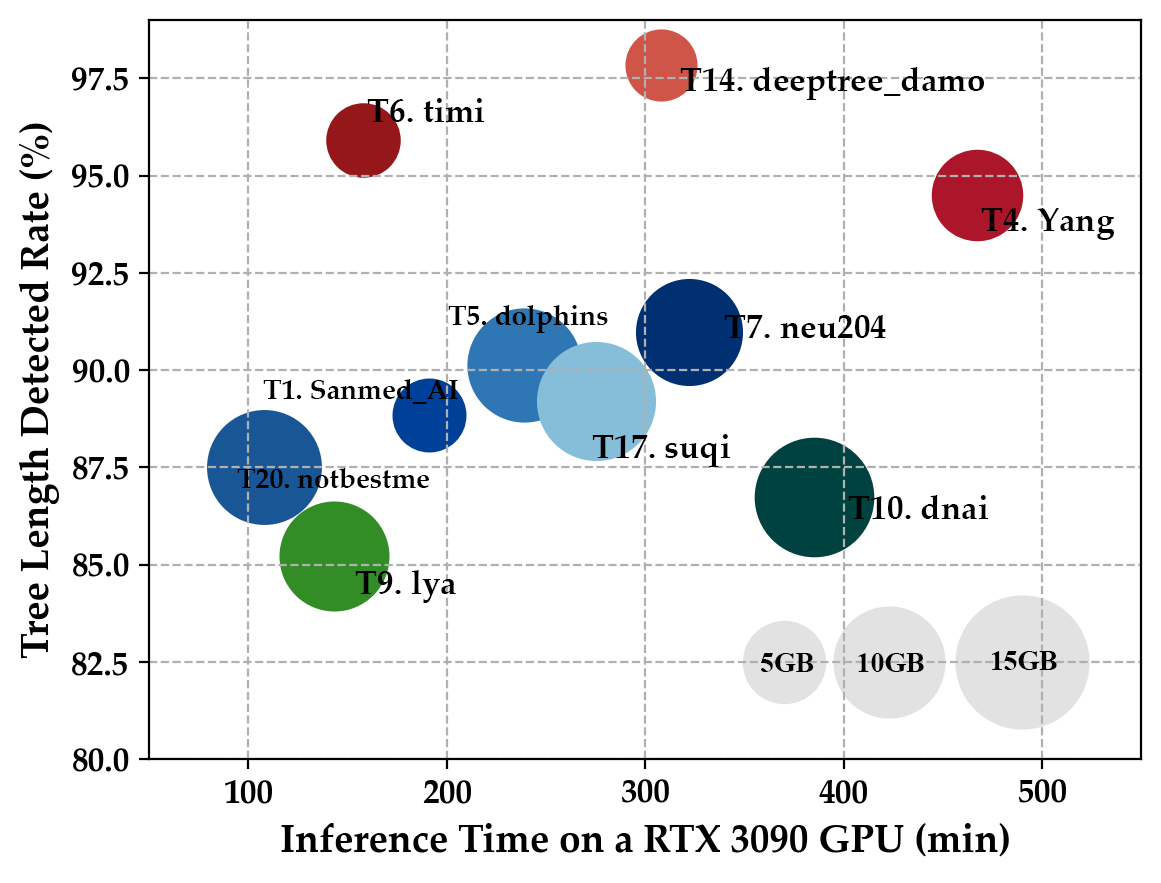}
}
\subfloat[Branch detected rate]{
\includegraphics[width=0.50\linewidth]{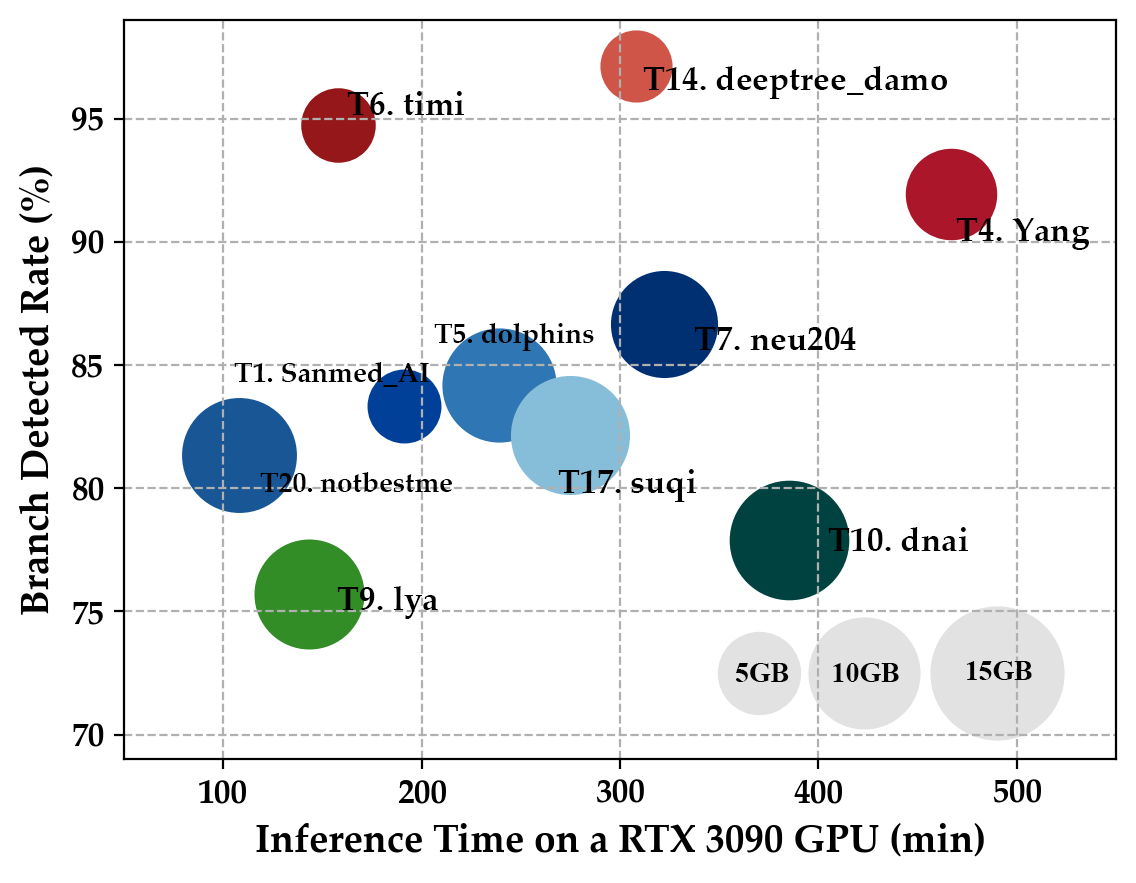}
}
\quad    
\subfloat[Dice Similarity Coefficient]{
  \includegraphics[width=0.50\linewidth]{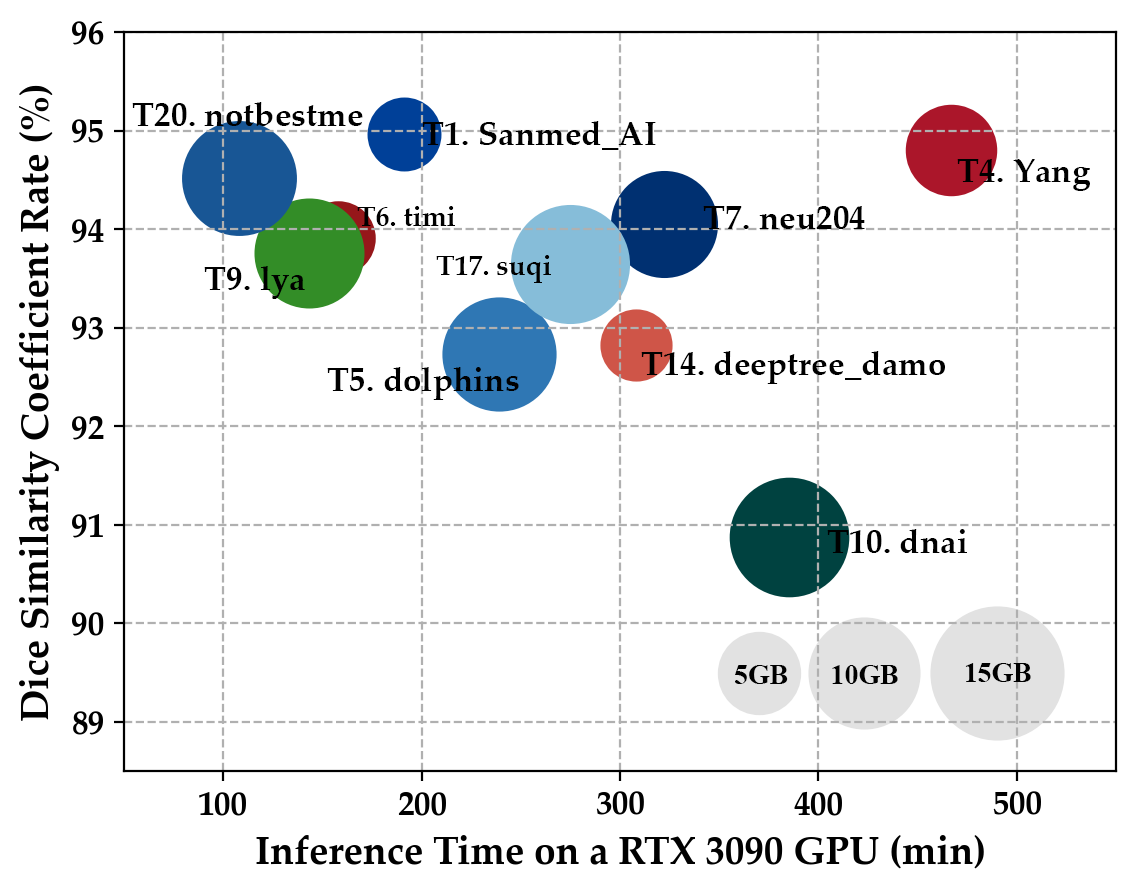}
}
\subfloat[Precision]{
\includegraphics[width=0.50\linewidth]{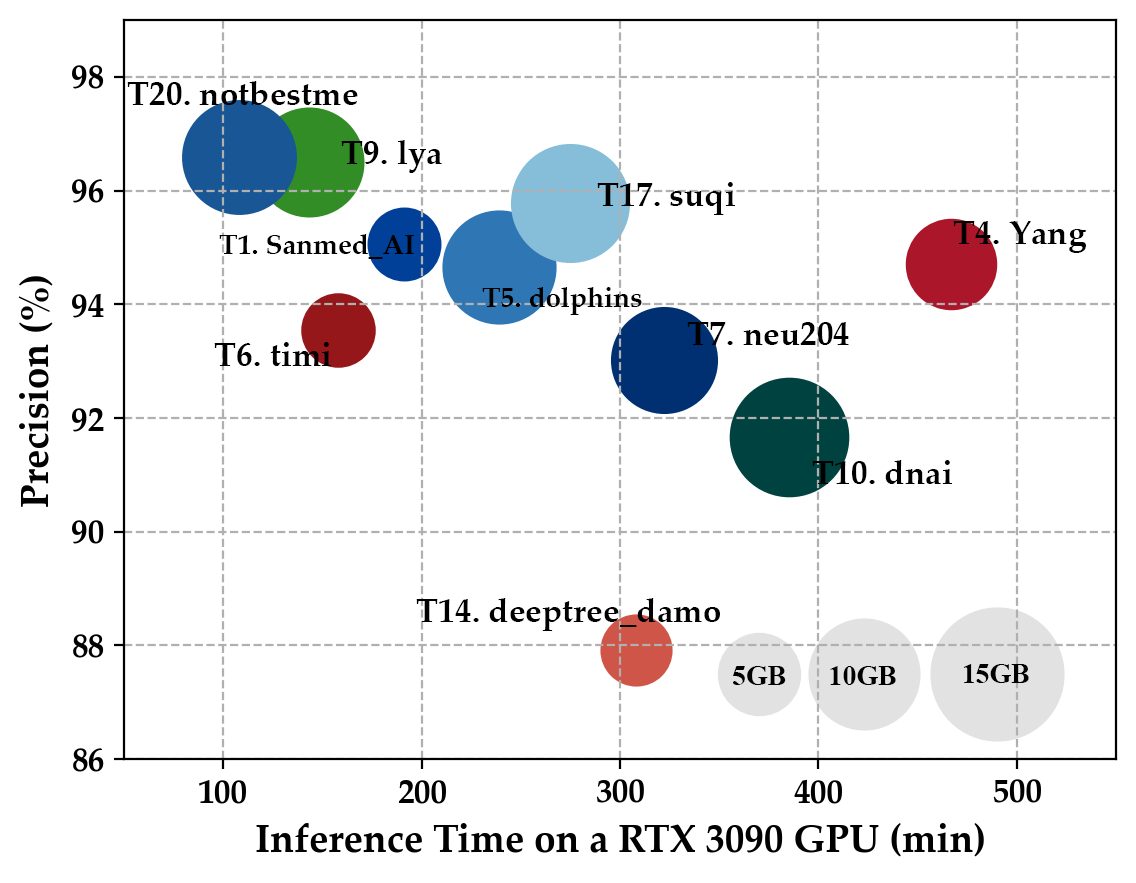}
}
\caption{Comparison of different models on the inference time and metrics. The larger markers indicate that the models share more parameters.}
\label{fig:gpu-infertime-metrics}
\end{figure*}

The efficiency of the method is also critical in medical applications. For example, the excellent efficiency provides the potential to 
reconstruct the airway in real-time from the mobile CT, which is helpful in guidance for thoracic surgery. Recently, efficiency also raised 
attention in biomedical challenges. In previous challenge settings, only the prediction results were required to submit, consequently, the efficiency 
behind the methods was non-transparent. As the competition standard rose, submitting the docker of the algorithm is preferred as a primary choice. 
FLARE'21~\citep{ma2022fast} challenge considered the running time and the maximum GPU memory consumption as a part of the ranking score calculation. 
Although efficiency was not involved in the ranking calculation, we conducted the model complexity analysis for the complementary measurement. 
The test dockers submitted by the participants were executed on the same Linux workstation with Intel Xeon Gold 5119T CPU @ 1.90GHz, 128 GB RAM, 
and 2 NVIDIA Geforce RTX 3090 GPUs. The maximum GPU memory consumption and inference time cost of each team were recorded in Table~\ref{tab:test-model-complexity}. 
We also compared the metrics conditioned on the efficiency in Figure \ref{fig:gpu-infertime-metrics}. 
T6 and T14 achieved an overall high performance, meanwhile, they maintained the competitive efficiency. 
Compared to T6 and T14, T4 was time-consuming. The post-process procedure used by T4 may increase the inference time cost. In the future, the 
model complexity is likely to be added to the evaluation to encourage the creation of effective methods with high efficiency.

\makeatletter
\def\hlinew#1{%
\noalign{\ifnum0=`}\fi\hrule \@height #1 \futurelet
\reserved@a\@xhline}
\makeatother
\begin{table}[thbp]
\renewcommand\arraystretch{1.2}
\centering
\caption{\justify{The maximum GPU memory consumption and inference time used by the participants on the test phase, 150 cases in total. 
All the test dockers were executed on the same device.}}\label{tab:test-model-complexity}
\scalebox{0.9}{
\begin{threeparttable}
\begin{tabular}{lcc}
\hlinew{1.25pt}
\textbf{Team name}   & \textbf{GPU memory} & \textbf{Inference time} \\ \hline
Sanmed\_AI (T1)    &  3.74 GB &  3h 11min\\
fme (T2)    &  17.89 GB &  2h 18min\\
LinkStartHao (T3)     &  9.64 GB &  4h 39min\\
YangLab (T4)   &  6.24 GB &  7h 47min\\
dolphins (T5)  &  10.43 GB &  3h 59min\\
timi (T6)   &  3.81 GB &  2h 38min\\
neu204 (T7)   &  9.00 GB &  5h 11min\\
blackbean (T8)   &  3.72 GB &  12h 13min\\
lya (T9) & 9.58 GB & 2h 23min\\
dnai (T10) & 11.59 GB & 6h 25min\\
bms410 (T11) & 23.39 GB &11h 01min\\
miclab (T12) & 11.54 GB &2h 05min\\
CITI-SJTU (T13) & 3.24 GB &5h 41min\\
deeptree\_damo (T14) & 3.52 GB &5h 08min\\
CBT\_IITDELHI (T15) & 7.13 GB &3h 35min\\
bwhacil (T16) & 23.39 GB &1h 56min\\
suqi (T17) & 11.47GB & 4h 35min\\
Median (T18) & 5.53GB & 6h 06min\\
notbestme (T19) & 10.58GB & 1h 48min\\
satsuma (T20) & 	3.71GB & 	17h 42min\\
biomedia (T21) & 11.52GB & 9h 12min\\
\hlinew{1.25pt}
\end{tabular}
\end{threeparttable}}
\end{table}

\section{Discussions}\label{sec:discussion}
\subsection{Clinical Applications}
Our challenge emphasizes the topological correctness and completeness of the airway. Both are clinically significant because pulmonary disease assessment and 
endobronchial intervention require accurate airway segmentation for quantitative measurements of bronchial features. 
With regard to topological correctness, the performance of the voxel-wise segmentation determines the accuracy of the quantitative measurements. 
The measurements of bronchial morphometric parameters such as wall thickness, total airway count, and lumen diameter
can be used in the diagnosis of cystic fibrosis~\citep{wielputz2013automatic} chronic obstructive pulmonary disease (COPD)~\citep{kirby2018total}, and asthma~\citep{eddy2020computed}.
The accurate segmentation of airways can relieve the burden of clinicians and reduce the large variability of a higher order of the branches. The 
high topological correctness could facilitate better quantification of airway pathologies, and then improve the comprehension of the 
mechanisms of disease progression.
As for topological completeness, it is quite important for the navigation of endobronchial interventions. 
Detailed pulmonary airway segmentation, which traditionally works on the level of trachea and bronchi and ideally reached the granularity of alveoli is
required for navigation in bronchoscopic-assisted surgery since it shows great advantages in the treatment of lung cancer~\citep{reynisson2014navigated}, 
chronic obstructive pulmonary disease (COPD) ~\citep{wan2006bronchoscopic} and the recent COVID-19 ~\citep{luo2020performing}. 
As claimed in ~\citep{gu2022vision}, the outer diameter of current flexible bronchoscopes is smaller than 5\textit{mm}, which allows the direct exploration of distal 
small bronchi. Hence, a detailed modeling of the bronchial tree is demanded to build the virtual lung model for preoperative path planning and 
intraoperative navigation. In conclusion, both topological correctness and completeness of the airway are critical indicators for the diagnosis and treatment of 
pulmonary diseases. Hence, the aim of our challenge encourages the development of airway segmentation algorithms that consider both correctness and completeness.

\subsection{Rethinking the Evaluation of Tubular Structures}
Although the overlap-wise metrics (DSC, IOU, etc.) and the distance-based metrics (Chamfer distance, Average symmetric surface distance, etc.) are the standard 
metrics adopted by the computer vision community, biomedical tasks often have special domain-specific requirements. In this organized ATM'22 challenge, The topological 
completeness (i.e., connectivity) is of particular interest, hence, the specific metrics should be taken into consideration. Deep distance transform ~\citep{wang2020deep} 
was designed for tubular structure segmentation in CT scans while they merely employ the DSC and mean surface distance for evaluation. 
They particularly formulated the tubular structure as the envelope of spheres with continuously changing center points and radii to be distinct from other general 
object segmentation, however, no specific metrics were adopted for evaluation. The situation is opposite to the EXACT'09 challenge~\citep{EXACT09}. 
It provided the specific metrics (tree length detected and branches detected) to measure topological completeness of the airway extracted by the segmentation algorithms. 
However, in that period, the methods were mainly intensity-based, which cannot achieve the high performance of the topological metrics due to the lack of airway tree prior. 
Fortunately, our organized ATM'22 challenge provided a large dataset with full airway annotation. In addition, we emphasized the specific tree-like topology in the 
airway segmentation task. Equipped with the deep-learning technique and specified modules that 
emphasize the topological knowledge, it is promising to explore the intrinsic knowledge of the tubular structure and narrow the gap between the methodology and evaluation.

As demonstrated before, we adopted the TD and BD to measure the topological completeness, and DSC and Precision were used to evaluate the topological correctness. 
The \textit{Betti numbers} are crucial topological invariants, and the airway of the golden standard shares the constant topological feature of $\beta_{0} = 1$ 
(i.e., owns only a single connected component). The Betti error is hard to directly minimize due to the non-differentiable property. Moreover, the Betti error alone 
cannot always reflect the real topological completeness. It will fall into such a pitfall: For example, the region growing method definitely
produces a single connected airway prediction due to the growth rule. However, the peripheral airways are not detected. Under such circumstances, the error of $\beta_{0}$ 
is zero while the TD/BD will be quite low. 

Airway thickness~\citep{orlandi2005chronic, achenbach2008mdct} had demonstrated a strong correlation with reduced air flow, and the Airway Fractal Dimension (AFD) was 
proposed to conduct an auxiliary diagnosis of respiratory morbidity and mortality in COPD~\citep{bodduluri2018airway}. These metrics are 
biased toward the clinical analysis of the morphological parameters, which may be involved in our future works.
The most relevant metric to our task is clDice~\citep{shit2021cldice}. It emphasizes the connectivity to evaluate tubular structure segmentation 
based on extracted soft skeletons and masks. The clDice is defined as the harmonic mean of Topology Precision and Topology Sensitivity. It handles the FP and FN samples 
simultaneously, reinforcing the network to be connectivity-aware. The advantage of clDice is that it can be designed as a loss function due to the differentiable property.
However, the clDice loss function cannot guarantee the topology of the airway because the TD/BD is calculated after the 
largest component extraction of the prediction while clDice does not. To our best knowledge, the operation of the largest component extraction is non-differentiable. 
The clDice is also affected by the structure size, even if the predictions of the two algorithms differ in only a single pixel, the clDice varies remarkably. 
In addition, our preliminary experiments observed the soft-skeleton extracted by clDice on the airway data was of poor quality. In conclusion, the 
non-differentiable topological metrics (TD/BD) are unfriendly to the pipeline of the deep learning models. The implicit modules, such as novel objective 
functions, are encouraged to be investigated since they had shown the ability to improve topology performance. The best supervision signals to characterize the topology
priors are still far away from being solved, hence further research is necessary.

\subsection{Limitations and Future of ATM'22}
ATM'22 is primarily aimed at establishing the new standard norm of the airway segmentation field in the deep learning era. 
Our future work can be roughly concentrated on three aspects. 
Firstly, more cases with diverse diseases and low resolution would be introduced to strengthen the evaluation of the generalization ability. ATM'22 only 
introduced the COVID-19 disease due to the extreme difficulty of annotating these noisy CT scans. 
Benefiting from the ensemble of the top-ranked algorithms from the outcome of ATM'22, it is feasible to introduce more diseased cases. These diseased cases can be first 
pre-segmented by a strong ensemble model to relieve the burden of the clinician. 
Secondly, we would extend our challenge. ATM'22 is currently focused on binary airway segmentation. There are several recent works~\citep{tan2021sgnet,xie2022structure,yu2022tnn} 
that start the airway anatomical labeling (i.e., branch-wise airway classification or segmentation). The qualified binary segmentation is the foundation of the assignment
of anatomical names to the corresponding branches of the airway tree. The assignment task is a promising direction to promote our challenge as fine-grained labels
provide a detailed map for bronchoscopic navigation and the morphological changes will be position-aware.
Thirdly, we would expand our challenge to include more tubular structures. The topology performance is also crucial in other tubular structures, such as the 
fundus blood vessel, hepatic vessels, and coronary artery. The generalization ability could not only be tested on the unseen domain of the same class, but also on 
universal tasks.

\section{Conclusion}\label{sec:conclusion}
In this paper, we presented the Multi-site, Multi-domain Airway Tree Modeling (ATM'22) benchmark. 
The largest chest CT scans (500 scans in total) with full pulmonary airway annotation and the 
most comprehensive evaluation system were provided for this task. We summarized four typical challenges 
in airway segmentation, among which achieving both the high performance of topological completeness and 
correctness was the most crucial. Generally speaking, most teams performed better on the topological correctness (91.277\% DSC and 93.669\% Precision in average)
than the topological completeness (83.350\% TD and 75.596\% BD on average). Experimental results also demonstrated that 
high performance of topological correctness can not consistently guarantee the topological completeness of the airways and vice versa. 
Several consensuses of effective methods were derived to deal with the challenges of the airway segmentation. 
Improve intra-class discrimination and design novel objective functions were recognized as promising directions to 
achieve the outstanding trade-of between topological completeness and correctness. 
Moreover, the non-differentiable topological metrics(TD/BD) are unfriendly to the deep learning models. Hence, the 
best supervision signals to characterize the topology priors are still needed further research.

\section*{Acknowledgments}
This work is supported in part by the Open Funding of Zhejiang Laboratory under Grant 2021KH0AB03, in part by the Shanghai Sailing Program
under Grant 20YF1420800, and in part by NSFC under Grant 62003208, and in part by Shanghai Municipal of Science and Technology Project, 
under Grant 20JC1419500 and Grant 20DZ2220400. The authors would like to thank the MICCAI challenge society and the support of the Amazon Web Services and Grand-Challenge.org.\\

\section*{Author contributions}
\noindent
\textbf{M.Z:} Conceptualization, Methodology, Data curation, Software, Formal analysis, Writing- original draft; 
\textbf{Y.W:} Data curation, Software, Writing- original draft;
\textbf{H.Z:} Data curation, Writing- review and editing;
\textbf{Y.Q:} Data curation, Software;
\textbf{H.Z:} Data curation, Software;
\textbf{J.S:} Conceptualization, Data curation;
\textbf{G.Z.Y:} Conceptualization, Supervision, Writing- review and editing;
\textbf{Y.G:} Project administration, Conceptualization, Methodology, Supervision, Writing- review and editing.

W.T, C.A, C.P, P.Y, Y.N, G.Y, S.W, D.C.M, M.K, P.W, D.G, D.J, Y.W, S.Z, R.C, B.Z, X.L, A.Q, M.M, Q.S, Y.W, Y.L, Y.Z, J.Y, A.P, B.R, R.S.J.E, C.C.E were participants of the ATM'22 challenge, and provided their results for evaluation and the description of their algorithms. The final manuscript was approved by all authors.

\section*{Declaration of Competing Interest}
The authors declare that they have no known competing financial interests or personal relationships that could be appeared to influence the work reported in this paper.

\bibliographystyle{model2-names.bst}\biboptions{authoryear}
\bibliography{refs}

\end{document}